\documentclass[%
 reprint,
 amsmath,amssymb,
 aps,superscriptaddress,
]{revtex4-1}

\usepackage{color} 
\usepackage{amsmath}
\usepackage{natbib}
\usepackage{graphicx}
\usepackage{dcolumn}
\usepackage{bm}
\usepackage{subfigure}

\begin{document}

\title{Mechanisms and dynamics of the NH$_2^{+}$ + H$^{+}$ and NH$^{+}$ + H$^{+}$ + H fragmentation channels upon single-photon double ionization of NH$_3$}

\author{Kirk A. Larsen}
\email{klarsen@lbl.gov}
\affiliation{%
 Graduate Group in Applied Science and Technology, University of California, Berkeley, CA 94720, USA}
\affiliation{%
 Chemical Sciences Division, Lawrence Berkeley National Laboratory, Berkeley, CA 94720, USA}%

\author{Thomas N. Rescigno}
\email{tnrescigno@lbl.gov}
\affiliation{%
 Chemical Sciences Division, Lawrence Berkeley National Laboratory, Berkeley, CA 94720, USA}%

\author{Zachary L. Streeter}
\affiliation{%
 Chemical Sciences Division, Lawrence Berkeley National Laboratory, Berkeley, CA 94720, USA}%
\affiliation{%
 Department of Chemistry, University of California, Davis, CA 95616, USA}%

\author{Wael Iskandar}
\affiliation{%
 Chemical Sciences Division, Lawrence Berkeley National Laboratory, Berkeley, CA 94720, USA}%
 
\author{Saijoscha Heck}
\affiliation{%
 Chemical Sciences Division, Lawrence Berkeley National Laboratory, Berkeley, CA 94720, USA}%
\affiliation{%
 Max-Planck-Institut f\"{u}r Kernphysik, Saupfercheckweg 1, 69117 Heidelberg, Germany}%
\affiliation{%
 J.W. Goethe Universit\"{a}t, Institut f\"{u}r Kernphysik, Max-von-Laue-Str. 1, 60438 Frankfurt, Germany}%

\author{Averell Gatton}
\affiliation{%
 Chemical Sciences Division, Lawrence Berkeley National Laboratory, Berkeley, CA 94720, USA}%
\affiliation{%
 Department of Physics, Auburn University, Alabama 36849, USA}%
 
\author{Elio G. Champenois}
\affiliation{%
 Graduate Group in Applied Science and Technology, University of California, Berkeley, CA 94720, USA}
\affiliation{%
 Chemical Sciences Division, Lawrence Berkeley National Laboratory, Berkeley, CA 94720, USA}%
 
\author{Travis Severt}
\affiliation{%
 J.R. Macdonald Laboratory, Physics Department, Kansas State University, Manhattan, Kansas 66506, USA}

\author{Richard Strom}
\affiliation{%
 Department of Physics, Auburn University, Alabama 36849, USA}%
 
\author{Bethany Jochim}
\affiliation{%
 J.R. Macdonald Laboratory, Physics Department, Kansas State University, Manhattan, Kansas 66506, USA}
 
\author{Dylan Reedy}
\affiliation{%
  Department of Physics, University of Nevada Reno, Reno, Nevada 89557, USA}
  
\author{Demitri Call}
\affiliation{%
  Department of Physics, University of Nevada Reno, Reno, Nevada 89557, USA}

\author{Robert Moshammer}
\affiliation{%
  Max-Planck-Institut f\"{u}r Kernphysik, Saupfercheckweg 1, 69117 Heidelberg, Germany}
  
\author{Reinhard D\"{o}rner}
\affiliation{%
  J.W. Goethe Universit\"{a}t, Institut f\"{u}r Kernphysik, Max-von-Laue-Str. 1, 60438 Frankfurt, Germany}
 
\author{Allen L. Landers}
\affiliation{%
  Department of Physics, Auburn University, Alabama 36849, USA}
  
\author{Joshua B. Williams}
\affiliation{%
  Department of Physics, University of Nevada Reno, Reno, Nevada 89557, USA}
  
\author{C. William McCurdy}
\affiliation{%
 Chemical Sciences Division, Lawrence Berkeley National Laboratory, Berkeley, CA 94720, USA}%
\affiliation{%
 Department of Chemistry, University of California, Davis, CA 95616, USA}%
 
\author{Robert R. Lucchese}
\affiliation{%
 Chemical Sciences Division, Lawrence Berkeley National Laboratory, Berkeley, CA 94720, USA}%

\author{Itzik Ben-Itzhak}
\affiliation{%
 J.R. Macdonald Laboratory, Physics Department, Kansas State University, Manhattan, Kansas 66506, USA}
 
\author{Daniel S. Slaughter}
\affiliation{%
 Chemical Sciences Division, Lawrence Berkeley National Laboratory, Berkeley, CA 94720, USA}%
 
\author{Thorsten Weber}
\email{tweber@lbl.gov}
\affiliation{%
 Chemical Sciences Division, Lawrence Berkeley National Laboratory, Berkeley, CA 94720, USA}%

\date{\today}

\begin{abstract}
We present state-selective measurements on the NH$_2^{+}$ + H$^{+}$ and NH$^{+}$ + H$^{+}$ + H dissociation channels following single-photon double ionization at 61.5~eV of neutral NH$_{3}$, where the two photoelectrons and two cations are measured in coincidence using 3-D momentum imaging. Three dication electronic states are identified to contribute to the NH$_2^{+}$ + H$^{+}$ dissociation channel, where the excitation in one of the three states undergoes intersystem crossing prior to dissociation, producing a cold NH$_2^+$ fragment. In contrast, the other two states directly dissociate, producing a ro-vibrationally excited NH$_2^+$ fragment with roughly 1~eV of internal energy. The NH$^{+}$ + H$^{+}$ + H channel is fed by direct dissociation from three intermediate dication states, one of which is shared with the NH$_2^{+}$ + H$^{+}$ channel. We find evidence of autoionization contributing to each of the double ionization channels. The distributions of the relative emission angle between the two photoelectrons, as well as the relative angle between the recoil axis of the molecular breakup and the polarization vector of the ionizing field, are also presented to provide insight on both the photoionization and photodissociation mechanisms for the different dication states.
\end{abstract}

\pacs{Valid PACS appear here}

\maketitle

\section{\label{sec:level1}Introduction}

Molecular dissociation of polyatomic systems that follows single Photon Double Ionization (PDI) can involve an excitation to a given dication electronic state reaching its own respective adiabatic limit or undergoing non-adiabatic transitions between states preceding the fragmentation \cite{Gaire,Gaire1,Gaire2}. Non-adiabatic dynamics, e.g. ultrafast state coupling via Conical Intersections (CIs) in polyatomic molecules, occur often. CIs are degeneracies between Born-Oppenheimer Potential Energy Surfaces (PESs) that result in non-adiabatic transitions between excited states. They can be facilitated via (a) internal conversion, where two interacting hypersurfaces have the same multiplicity, or (b) intersystem crossing processes, where the multiplicity is different and spin-orbit coupling is required. Both flavors (a) and (b) represent a breakdown of the Born-Oppenheimer approximation \cite{Yarkony}. CIs have been observed in numerous instances to significantly shape the molecular fragmentation processes occurring in the molecular dication, influencing the branching ratios, energy dispersion, breakup kinematics, and timescales \cite{Gaire,Gaire1,Gaire2}. However, identifying the flavor of the CI, i.e. distinguishing between type (a) and (b), by precisely tracing the electron and nuclear dynamics between the instant of PDI and when the dissociation has set in, is very challenging. This is because polyatomic molecules can break up in many different ways after PDI, and each reaction channel inherently carries different coupled rovibrational degrees of freedom that can be excited. Some of these non-adiabatic fragmentation dynamics leave their fingerprint in the energy or momentum domain of the emitted particles, and they can be determined if the 3-D momenta  of the photoelectron-pair and the recoiling fragments can be measured in coincidence to produce highly differential observables, which is the aim of this work. In this paper we report on the PDI of NH$_3$ and reveal the role and flavor of a non-adiabatic transition occurring in a select two-body breakup channel, while identifying the other two- and three-body breakup channels as direct, in a combined effort of experiment and theory.

The electronic states of NH$_3^{2+}$ and its dissociation channels have been studied in detail experimentally following PDI, electron impact double ionization, and double-charge-transfer spectroscopy, as well as theoretically \cite{Winkoun,Stankiewicz,Locht1,Locht2,Eland,Tarantelli,Samson,Appell,Cheret,Langford,Griffiths,Locht3,White,Camilloni,Okland,Jennison,Boyd}. These studies have largely focused on establishing the appearance energies of the various dissocation channels and the relative energies of the dication electronic states. Early experimental studies on the two-body NH$_2^+$ + H$^+$ dissociation channel in a narrow photon energy range near the PDI threshold revealed that intersystem crossing from the X ($3a_1^{-2}$) $^1A_1$ to the A ($3a_1^{-1},1e^{-1}$) $^3E$ state can facilitate fragmentation \cite{Boyd}. However, non-adiabatic effects have remained unobserved in any of the other fragmentation channels, whether near or well above the double ionization threshold. Additionally, to our knowledge, no study to date has examined the exact nature of the PDI mechanisms in NH$_3$ in detail, e.g. the level at which autoionization contributes to the PDI, if at all, or if the PDI is primarily based on an electron-electron knock-out process (so called Two-Step 1 interaction \cite{Ishihara}).

In our recent study~\cite{Larsen}, hereafter referred to as [I], we reported the photoionization mechanisms and photodissociation dynamics of the H$^+$ + H$^+$ fragmentation channels of NH$_3^{2+}$ following PDI of neutral NH$_{3}$ molecules at 61.5~eV. In that study we observed non-adiabatic dynamics that enables both a sequential dissociation mechanism and a charge transfer process. In this report we extend our investigation to the NH$_2^{+}$ + H$^{+}$ and NH$^{+}$ + H$^{+}$ + H dissociation channels of NH$_3^{2+}$ following PDI of NH$_{3}$ molecules at 61.5~eV, approximately 27~eV above the PDI threshold, where both the photoelectron- and cation-pair are measured in coincidence using charged particle 3-D momentum imaging.

\section{\label{sec:level2}Experiment}

Both the NH$_2^{+}$ + H$^{+}$ and NH$^{+}$ + H$^{+}$ + H fragmentation channels of NH$_3^{2+}$ following PDI of neutral NH$_{3}$ molecules at 61.5~eV were investigated using COLd Target Recoil Ion Momentum Spectroscopy (COLTRIMS) \cite{Dorner,Ullrich}, where the two photoelectrons and two cations produced by PDI are detected with full $4\pi$ solid angle, and their 3-D momenta are measured in coincidence on an event-by-event basis. The photoelectron- and cation-pair were guided using static parallel electric and magnetic fields, 11.4~V/cm and 10.0~G, respectively, to multi-hit capable position- and time-sensitive detectors. The detectors comprised a Multi-Channel Plate (MCP) stack in chevron configuration, backed by a delay-line anode readout, each at opposite ends of the spectrometer. The electron and ion delay-line detectors were a three-layer hex-anode with an 80~mm MCP stack and a two-layer quad-anode with a 120~mm MCP stack, respectively. This system encodes a charged particle's 3-D momentum into its hit position on the detector and Time-of-Flight (TOF) relative to the incoming XUV light pulses. These detectors are subject to multi-hit dead-time effects that are most prominent in the electron pair detection, due to the small variation in the electron's arrival times and hit positions on the detector \cite{Jagutzki}, whereas the dead-time effects play a negligible role for the detection of the cation pair. This dead-time effect can influence the measured relative electron-electron angular distribution, hence it is important to quantify this deficiency, in order to distinguish real features from those which originate from the underperforming detection scheme. The electron-pair resolution is estimated by simulating the relative motion of the electron pair in the spectrometer fields with various electron sum kinetic energies and in various electron energy sharing conditions. For each pair of electron trajectories, the relative hit position and TOF is computed, which is used to determine the fraction of simulated electron-pair events lost due to an estimated detector response, and thus approximate the fraction of actual losses.

PDI was performed using a linearly polarized tunable monochromatic beam of extreme ultraviolet (XUV) photons produced at beamline 10.0.1.3. at the Advanced Light Source (ALS), located at Lawrence Berkeley National Laboratory. The beamline monochromator was configured to provide 61.5~eV photons to the experiment with an energy resolution of less than $\pm$50~meV. The photon energy of 61.5~eV was chosen to be near the maximum of the PDI cross section of NH$_3$, while at the same time providing electron kinetic energies that can be detected with full solid angle and adequate energy resolution (around 1:10). Moreover, it is beneficial to keep the electron sum energy greater than $\sim5$~eV in order to utilize a large region of the 3D electron pair detection phase space, minimizing losses due to the electron detector dead-time.

A beam of rotationally and vibrationally cold ($\sim$80~K) NH$_{3}$ molecules (Praxair, anhydrous ammonia $>$99\% purity) was produced by an adiabatic expansion of pressurized gas through a 50~$\mu$m nozzle and collimated by a pair of downstream skimmers. The first skimmer had a diameter of 0.3~mm, and the second skimmer had a diameter of 0.5~mm. The first skimmer was placed 8~mm downstream of the nozzle and in the zone of silence of the supersonic expansion. The second skimmer was 10~mm downstream of the first skimming stage. This supersonic gas jet propagated perpendicular to the photon beam, where the two beams crossed at the interaction region ($\sim0.15 \times 0.15 \times 1.0$~mm$^3$) inside the 3-D momentum imaging spectrometer, resulting in the PDI of neutral ammonia in its ground state at an average rate of less than 0.01 events per XUV pulse, assuring unambiguous coincidence conditions. The pressure in the target chamber was $\sim 2 \times 10^{-8}$ Torr with the supersonic beam running, and was approximately a factor of 2 lower without the jet. The target gas itself was delivered through a room temperature injection line using Swagelok connections. Background water in the target chamber was the largest contaminating species, and was minimized through the use of a cold trap filled with liquid nitrogen.

The TOF and hit position of charge particles produced by the PDI were recorded in list mode on an event-by-event basis, enabling relevant events to be captured and examined in a detailed off-line analysis, using the LMF2Root software package \cite{lmf2root} described in Ref.~\cite{jahnkelmf2root}. After cleaning, calibrating, and sorting the data set, for each PDI event the photoelectron kinetic energy was determined from the 3-D photoelectron momentum, while the Kinetic Energy Release (KER) of the fragmentation was determined from the 3-D momenta of the two cations. We inferred the momentum of the neutral H fragment in the three-body dissociation channel from momentum conservation. 

\section{\label{sec:level3}Theory}

\begin{figure}[h!]
    \includegraphics[width=8.35cm]{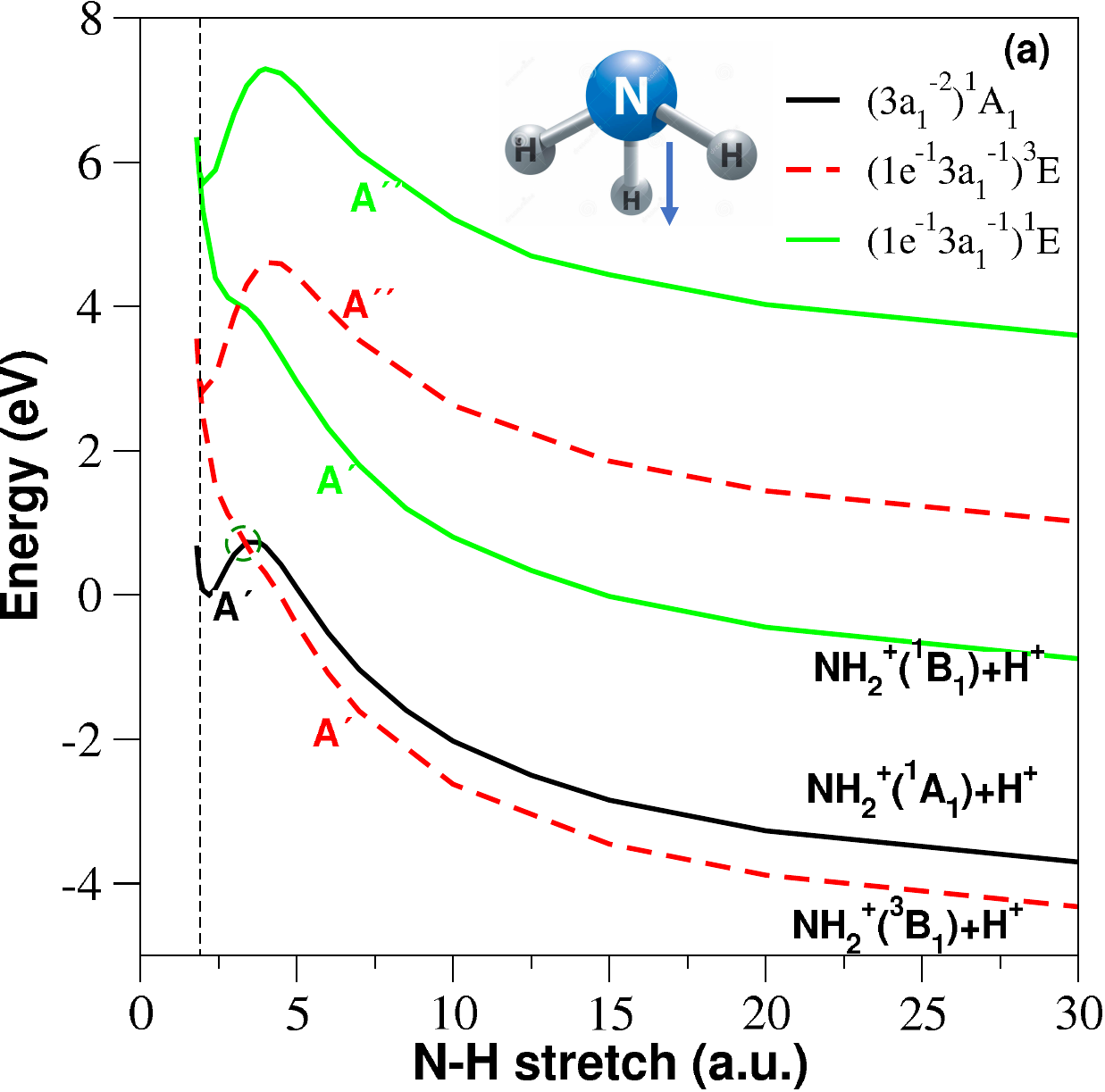}
    \includegraphics[width=8.35cm]{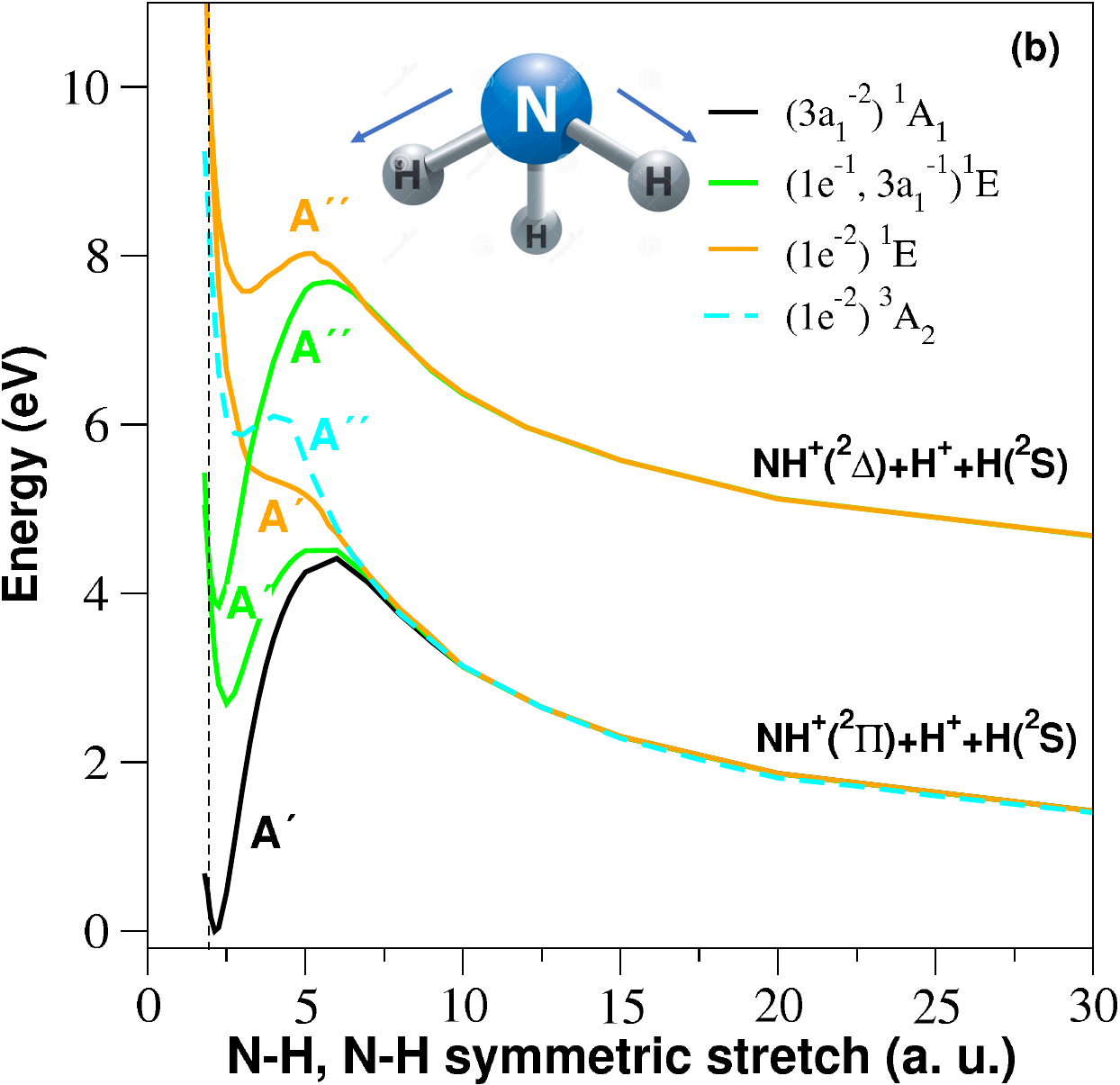}
\caption{PES cuts of the experimentally identified states of the NH$_3$ dication generated from MRCI calculations as described in the text. In panel (a) one hydrogen is stretched while the other two hydrogens remain fixed, with all internal angles frozen at the geometry of neutral ammonia. In panel (b) two hydrogens are symmetrically stretched while the third hydrogen remains fixed, with all internal angles frozen at the geometry of neutral ammonia. The dissociation limits are given in Table~\ref{table:asymptotes}. The zero of energy is set to the ground-state ($^1A_1$) of the ammonia dication at the geometry of the neutral ammonia molecule, which lies 34.8 eV below the dication ground state~\cite{Locht2}. On this energy scale, the 61.5 eV photon energy falls at 26.74. The green dashed circle in panel (a) indicates the region where intersystem crossing may occur. The broken vertical lines indicate the equilibrium geometry of neutral NH$_3$.}
\label{fig:PEC_NH3}
\end{figure}

The electron configuration of neutral NH$_3$ in its ground-state is $(1a_{1})^{2}(2a_{1})^{2}(1e)^{4}(3a_{1})^{2}$. Nine low-lying singlet and triplet states of the ammonia dication can be formed by distributing six electrons over the outer 1e and/or 3a$_1$ orbitals, all of which are accessible by single photon absorption at an energy of $61.5$~eV. In order to determine which of these states correlate with the two-body NH$_2^+$ + H$^+$ dissociation channel and three-body NH$^+$ + H$^+$ + H fragmentation channel, we carried out a series of electronic structure calculations, analogous to those described in [I]. As in our recent work, the molecular orbitals at each geometry considered were generated from state-averaged, Complete Active Space (CAS) Multi-Configuration Self-Consistent Field (MCSCF) calculations on the two lowest triplet ($^3$E) states of the dication, keeping one orbital (N 1s) frozen and including seven orbitals in the CAS space. These were followed by Multi-Reference Configuration-Interaction (MRCI) calculations, including all single and double excitations from the CAS reference space, to generate cuts through the calculated Potential Energy Surfaces (PESs). As in [I], all bond angles were frozen at the equilibrium geometry of neutral ammonia (107$^o$), as was one hydrogen bond length (1.9138 Bohr), while either one or two hydrogen bonds were stretched. The results of the calculations are shown in Fig.~\ref{fig:PEC_NH3}(a) and (b), respectively. The electron configuration and state labels of each dication PES are given in the legend. The vertical energies at the neutral NH$_3$ geometry and the energies at the asymptotic limits (extrapolated from 30 Bohr to infinity under the assumption of a purely repulsive Coulomb interaction between the positively charged fragments) are given in Table~\ref{table:asymptotes}.

\begin{table*}
\centering
\begin{tabular}{  c  c  c  c  } 
 \hline\hline
 State & Vertical Energy (eV) & Asymptote & Adiabatic Limit Energy (eV)\\
 \hline
 &&Two-Body Channels&\\
 \hline
 (3a$_1^{-2}$)$^1$A$_1$ (black) & 0.44 & NH$_2^+$($^1$A$_1$)+H$^+$ & -4.61  \\
(1e$^{-1}$,3a$_1^{-1}$)$^3$E(red) & 2.97 & NH$_2^+$($^3$B$_1$)+H$^+$ &  -5.23\\ 
(1e$^{-2}$)$^1$E (gold) & 5.74 & NH$_2^+$($^1$B$_1$)+H$^+$ & -3.09 \\
 \hline
  &&Three-Body Channels&\\
 \hline
 (3a$_1^{-2}$)$^1$A$_1$ (black) & 0.44 & NH$^+$($^2\Pi$)+H$^+$+H & 0.52  \\
 (1e$^{-1}$,3a$_1^{-1}$)$^1$E[A$'$] (green) & 5.74 & NH$^+$($^2\Pi$)+H$^+$+H & 0.52  \\
  (1e$^{-1}$,3a$_1^{-1}$)$^1$E[A$''$] (green) & 5.74 & NH$^+$($^2\Delta$)+H$^+$+H & 3.78  \\
 (1e$^{-1}$,3a$_1^{-1}$)$^3$A$_2$ (cyan) & 8.64 & NH$^+$($^2\Pi$)+H$^+$+H & 0.52  \\
 (1e$^{-2}$)$^1$E[A$'$] (gold) & 10.39 & NH$^+$($^2\Pi$)+H$^+$+H & 0.52  \\
  (1e$^{-2}$)$^1$E[A$''$] (gold) & 10.39 & NH$^+$($^2\Delta$)+H$^+$+H & 3.78  \\
 \hline
\end{tabular}
\caption{Ammonia dication vertical energies at neutral NH$_3$ geometry and asymptotic two- and three-body dissociation limits extrapolated from {\em ab initio} calculations at N-H distances of 30.0 bohr (see text). The zero of energy is set to the ground-state ($^1A_1$) of the ammonia dication at the geometry of the neutral ammonia molecule.}
\label{table:asymptotes}
\end{table*}

Our calculations reveal that there are five relevant dication electronic states accessible in the Franck-Condon (FC) region. Three of these states are singlets, ($3a_1^{-2}$) $^1A_1$, ($3a_1^{-1},1e^{-1}$) $^1E$, and ($1e^{-2}$) $^1E$, shown as solid curves (black, green, and gold), while two are triplets, ($3a_1^{-1},1e^{-1}$) $^3E$, and ($1e^{-2}$) $^3A_2$ shown as dashed curves (red and cyan).

In the case of stretching a single proton (asymmetric stretch), Fig.~\ref{fig:PEC_NH3}(a) shows that the three NH$_3^{2+}$ states ($3a_1^{-2}$) $^1A_1$, ($3a_1^{-1},1e^{-1}$) $^3E$, and ($3a_1^{-1},1e^{-1}$) $^1E$ correlate with the $^1$A$_1$, $^3$B$_1$, and $^1$B$_1$ states of NH$_2^+$, respectively. The ($3a_1^{-2}$) $^1A_1$ state is evidently predissociated by the A$'$ component of the $^3$E state, which can lead to a non-adiabatic transition between these states via intersystem crossing.

In the case of symmetric stretch of two hydrogen bonds, it can be seen in Fig.~\ref{fig:PEC_NH3}(b) that the $^1$A$'$ components of the ($3a_1^{-1},1e^{-1}$) $^1E$, ($1e^{-2}$) $^1E$ states, as well as the $^3$A$''
($($1e^{-2}$) $^3$A$_2$) state, all dissociate to the three-body channel NH$^{+}$ ($^2\Pi$) + H$^{+}$ + H. The two $A''$ symmetry components of the ($3a_1^{-1},1e^{-1}$) $^1E$ and ($1e^{-2}$) $^1E$ dication states dissociate to NH$^{+}$ ($^2\Delta$) + H$^{+}$ + H. Previous measurements have also indicated that the ($1e^{-2}$) $^1E$ dication state dissociates to the NH$^{+}$ + H$^{+}$ + H channel \cite{Stankiewicz}.

To get theoretical estimates of the expected spread of the observed photoelectron energies for the various dication states, we use a variant of the so-called reflection approximation~\cite{Rescigno88}. The range of detectable KERs is determined by the FC envelope of the initial (neutral) vibrational state reflected onto the final dication PESs. For the initial vibrational wavefunction we use a harmonic oscillator function $\chi_0$, obtained from a fit of the ground state energy of ammonia as a function of the symmetric stretch. If we assume that the PDI cross section varies little over the FC region (which is a good approximation, since there are no resonances or near-threshold effects at the chosen photon energy) and that the final continuum vibrational wavefunctions can be approximated by delta functions about the classical turning points on the dication PESs~\cite{Vanroose}, then the envelope of the expected photoelectron energies is given by the values of the vertical PDI energies as a function of the symmetric-stretch coordinate, weighted by the square of the symmetric-stretch vibrational wavefunction. We find that $|\chi_0|^2$ reaches half its maximum value at a symmetric-stretch displacement of approximately $\pm$0.11~Bohr from  equilibrium, and we have used these values to calculate the FWHM of the photoelectron distributions below.

\section{\label{sec:level4}Results}

\begin{figure}[h!]
        \includegraphics[width=8.5cm, trim=2.5cm 0.25cm 2.5cm 1.0cm, clip]{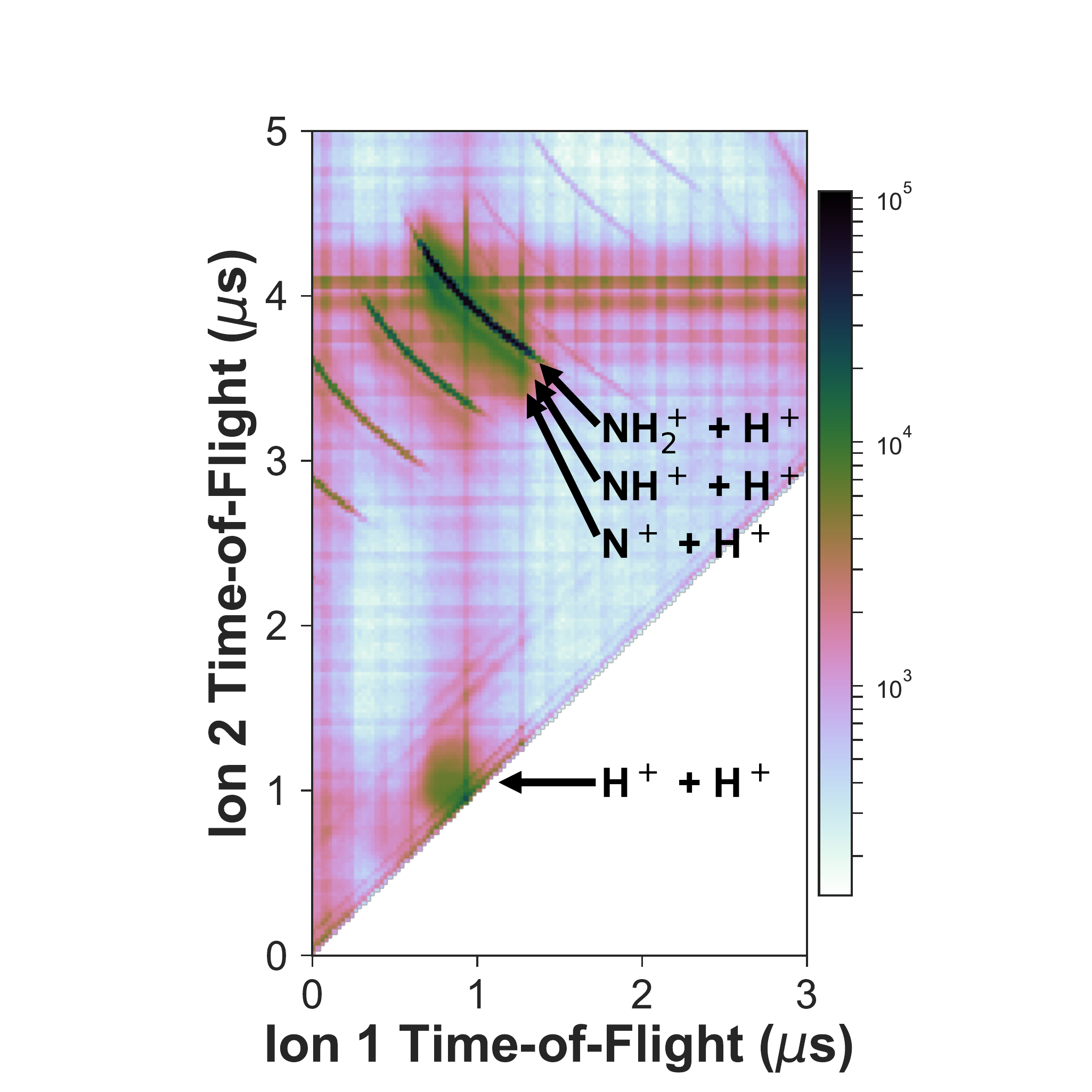}
\caption{The photoion-photoion time-of-flight coincidence map (PIPICO spectrum), shown on a logarithmic scale. The observed photoion-photoion coincidences from the respective breakup channels following PDI of NH$_3$ at 61.5~eV are indicated by the black arrows and text. Two of these channels (H$^+$ + H$^+$ and N$^+$ + H$^+$) are the topic of paper [I].}
\label{fig:PIPICO}
\end{figure}

The NH$_2^{+}$ + H$^{+}$ two-body and NH$^{+}$ + H$^{+}$ + H three-body dissociation channels of NH$_3^{2+}$, following PDI of NH$_3$ at 61.5~eV, $\sim$27~eV above the PDI threshold, are identified and isolated by selecting the two charged fragments in the PhotoIon-PhotoIon COincidence (PIPICO) TOF spectrum and then in momentum space. Moreover, we also enforce that two electrons are measured in coincidence with the two ionic fragments. We show the PIPICO spectrum in Fig.~\ref{fig:PIPICO}, where the photoion-photoion coincidence yield is shown on a logarithmic scale. Here we observe four photoion-photoion coincidence features following PDI of NH$_3$ molecules, two of which are addressed in this report, NH$_2^+$ + H$^+$ and NH$^+$ + H$^+$, while the other two coincidence features, H$^+$ + H$^+$ and N$^+$ + H$^+$, are the topic of manuscript [I]. The N$^+$ + H$^+$ channel is very faint and diffuse, which renders it difficult to visually identify in the PIPICO spectrum alone, however it emerges upon further analysis (examined in [I]). The vertical and horizontal features, as well as the periodically repeating features, are the result of false coincidences, which are removed later in our analysis. We first begin with a discussion of the NH$_2^+$ + H$^+$ two-body fragmentation channel before turning to the NH$^+$ + H$^+$ + H three-body channel. By inspecting the yield of the two features in the PIPICO corresponding with NH$_2^+$ + H$^+$ and NH$^+$ + H$^+$ coincidences, we find the approximate branching ratio between the NH$_2^+$ + H$^+$ two-body and the NH$^+$ + H$^+$ + H three-body dissociation channels to be 86:14. 

\subsection{\label{}Two-body breakup channel: NH$_2^{+}$ + H$^{+}$}

\begin{figure}[h!]
        \includegraphics[width=8.5cm]{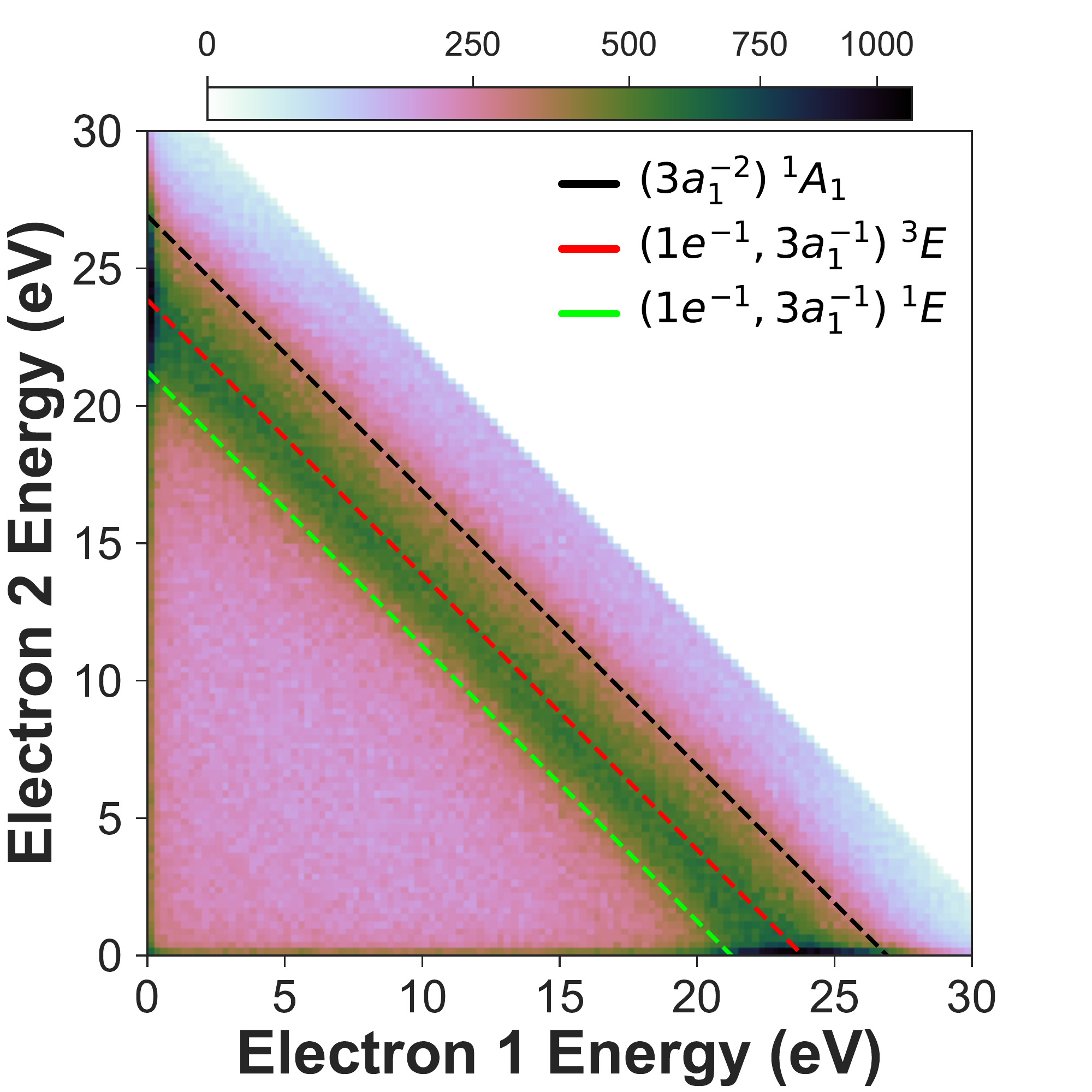}
\caption{The NH$_2^+$ + H$^+$ yield as a function of the kinetic energy of the first and second photoelectron after PDI of NH$_3$ at 61.5~eV. The three features indicating the active dication states are color-coded (black, red, and green) and shown as diagonal lines to guide the eye. Electrons with energy sums beyond 35~eV were not detected with full 4$\pi$ solid angle and are hence omitted.}
\label{fig:Ee_2}
\end{figure}

We plot the PDI yield of the NH$_2^+$ + H$^+$ fragmentation channel of NH$_3^{2+}$ as a function of the kinetic energy of the first and second detected electrons, to produce an electron-electron energy correlation map, as shown in Fig.~\ref{fig:Ee_2}. Since the two electrons are indistinguishable particles, the labeling (as 1 and 2) is arbitrary, and the figure has been symmetrized across the diagonal (the line E$_1$ = E$_2$) to account for this. With the guidance of the calculated PES cuts, we identify three features corresponding with three different NH$_3^{2+}$ electronic states that feed into the NH$_2^{+}$ + H$^{+}$ two-body fragmentation channel. Although difficult to visually identify here, these three features are better separated in different spectra that are shown below in Fig.~\ref{fig:EeKER_2}. Each of these three features correspond to two photoelectrons with energy sums centered around 26.9~eV, 23.9~eV, and 21.4~eV, with a Full Width at Half Maximum (FWHM) of roughly 1.5~eV, 2.0~eV, and 1.7~eV, respectively. These features, indicated as diagonal lines (which take the form E$_2$ = -E$_1$ + E$_{sum}$, where E$_{sum}$ is the photoelectron energy sum corresponding to that feature), have been color-coded as black, red, and green to guide the eye. In the offline analysis we choose each of these three features by selecting carefully around the center of each feature in Fig.~\ref{fig:EeKER_2}. Enforcing conditions in a multitude of observables and dimensions (particle energy and momenta) in this fashion enables us to separate these three features for subsequent in-depth analysis. 

The FWHM of the electron energy sum of each dication state roughly corresponds with the magnitude of the gradient of the PES in the FC region (convoluted with the energy resolution of the electron detector). The characteristics of the three features suggest that the three dication states are accessed via direct PDI, as indicated by the uniformity of the negatively sloped diagonal features in Fig.~\ref{fig:Ee_2}, and also appear to be populated through autoionization, the signature being the two sharp features located at the end of the diagonals, where one of the electrons possesses nearly zero kinetic energy. The branching ratio between these three measured channels that correspond with the three dication states shown in Table~\ref{table:2body_EeKER} is estimated from the relative yield of these three features. The branching ratios and the method used to estimate them are discussed below. 

\begin{table*}
\centering
\begin{tabular}{  c  c  c  c  c  c  c  } 
 \hline\hline
 State & \multicolumn{2}{c}{Photoelectron Energy Sum (eV)} & \multicolumn{2}{c}{KER (eV)} & Branching Fraction & $\beta_2$ \\
  & Experiment & Theory$^{\rm{a}}$ & Experiment & Theory$^{\rm{a,b}}$ & & \\
 \hline
 ($3a_1^{-2}$) $^1A_1$ (black) & 26.9 (1.5) & 26.5 (1.5) & 5.7 (1.0) & 5.5 (1.5) & 13\% $\pm$ 3\% & -0.30 $\pm$ 0.01 \\
 ($3a_1^{-1},1e^{-1}$) $^3E$ (red) & 23.9 (2.0) & 23.5 (2.0)& 6.7 (1.3) & 8.2 (2.0) & 44\% $\pm$ 3\% & -0.12 $\pm$ 0.01 \\
 ($3a_1^{-1},1e^{-1}$) $^1E$ (green) & 21.4 (1.7)  & 
 21.0 (2.4) & 6.7 (1.2) & 7.5 (2.4) & 43\% $\pm$ 3\% & -0.18 $\pm$ 0.01 \\
 \hline
\end{tabular}
\caption{The measured and calculated photoelectron energy sum and KER for each of the three identified features from NH$_2^{+}$ + H$^{+}$ fragmentation following PDI of NH$_3$ at 61.5~eV, as well as the estimated branching ratios and $\beta_2$ anisotropy parameter (see text). The energy widths (FWHM) are in parentheses. $^{\rm{a}}$Theoretical FWHM values estimated from the square of the symmetric stretch vibrational wavefunction of NH$_3$ projected onto the respective dication state  (see text). $^{\rm{b}}$Theoretical KER values are calculated assuming ro-vibrational ground state fragments, i.e. assuming maximum KER with no energy channeled into internal excitations.}
\label{table:2body_EeKER}
\end{table*}

Next we plot the yield of the NH$_2^+$ + H$^+$ dissociation channel as a function of the KER and the kinetic energy sum of the photoelectron pair, in order to generate an electron-ion energy correlation map. Three features corresponding with the three color-coded diagonals in Fig.~\ref{fig:Ee_2} are present in the electron-nuclei energy correlation map, shown in Fig.~\ref{fig:EeKER_2}. These three features are marked by ellipses in their respective color-codes to guide the eye (note: these ellipses do not reflect the actual momentum gates of the analysis). The feature circled by the green ellipse is comparatively faint and can be mistaken as a part of the feature circled by the red ellipse, however it is identified as a state with the assistance of the calculated PES cuts. Here we see that each dication state possesses a different distribution of KER centered around 5.7~eV, 6.7~eV, and 6.7~eV, each with a FWHM of roughly 1.0~eV, 1.3~eV, and 1.2~eV, respectively. The FWHM of the KER of each dication state carries similar information as the electron sum energy FWHM, indicating the steepness of the PESs in the FC region (convoluted with the energy resolution of the ion spectrometer).

\begin{figure}[h!]
        \includegraphics[width=8.5cm]{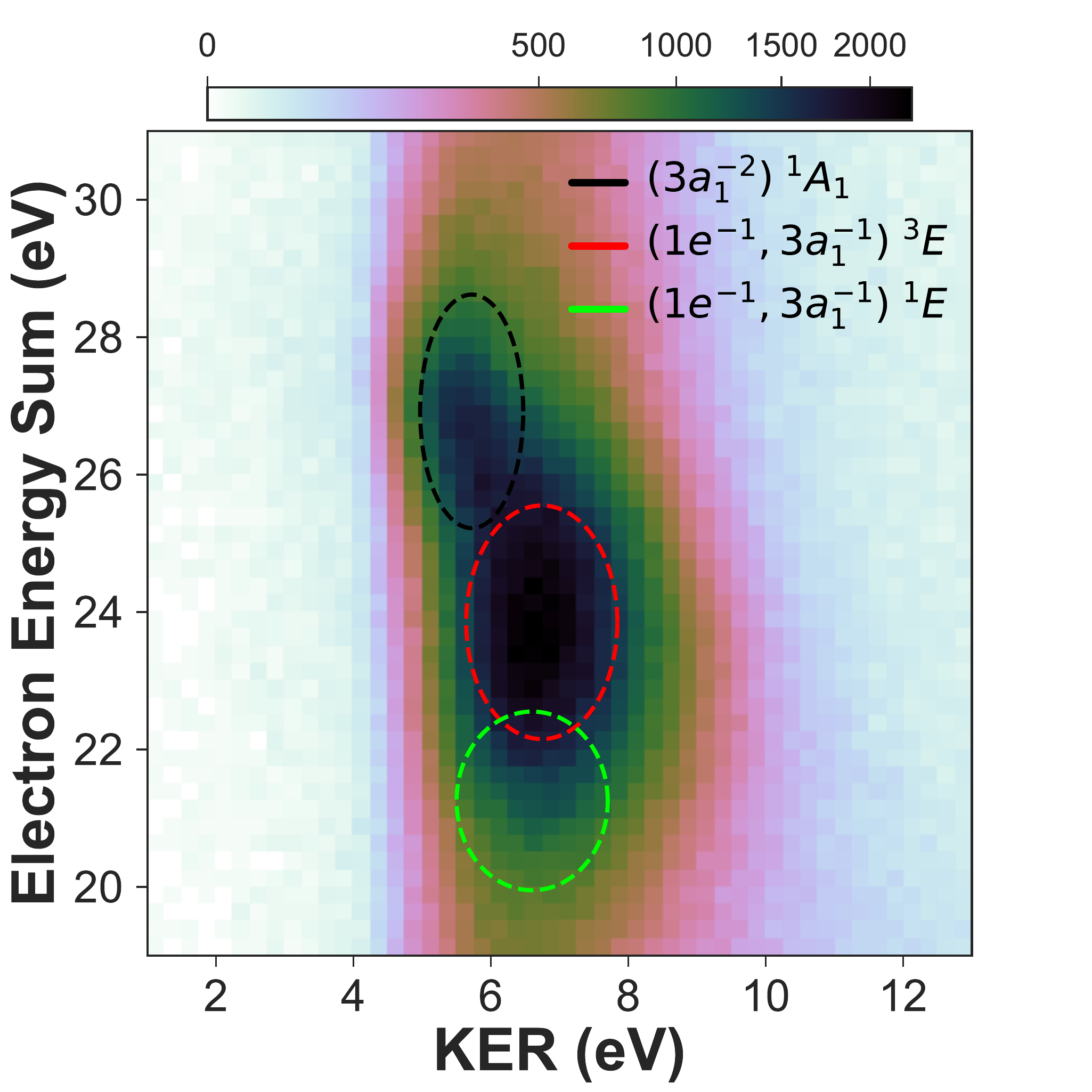}
\caption{The NH$_2^+$ + H$^+$ yield as a function of the KER and the kinetic energy sum of the photoelectron pair after PDI of NH$_3$ at 61.5~eV. The three features indicating the active dication states are color-coded (black, red, and green) and shown as ellipses to guide the eye (they only approximately represent the software gates).}
\label{fig:EeKER_2}
\end{figure}

\begin{figure}[h!]
        \includegraphics[width=8.5cm]{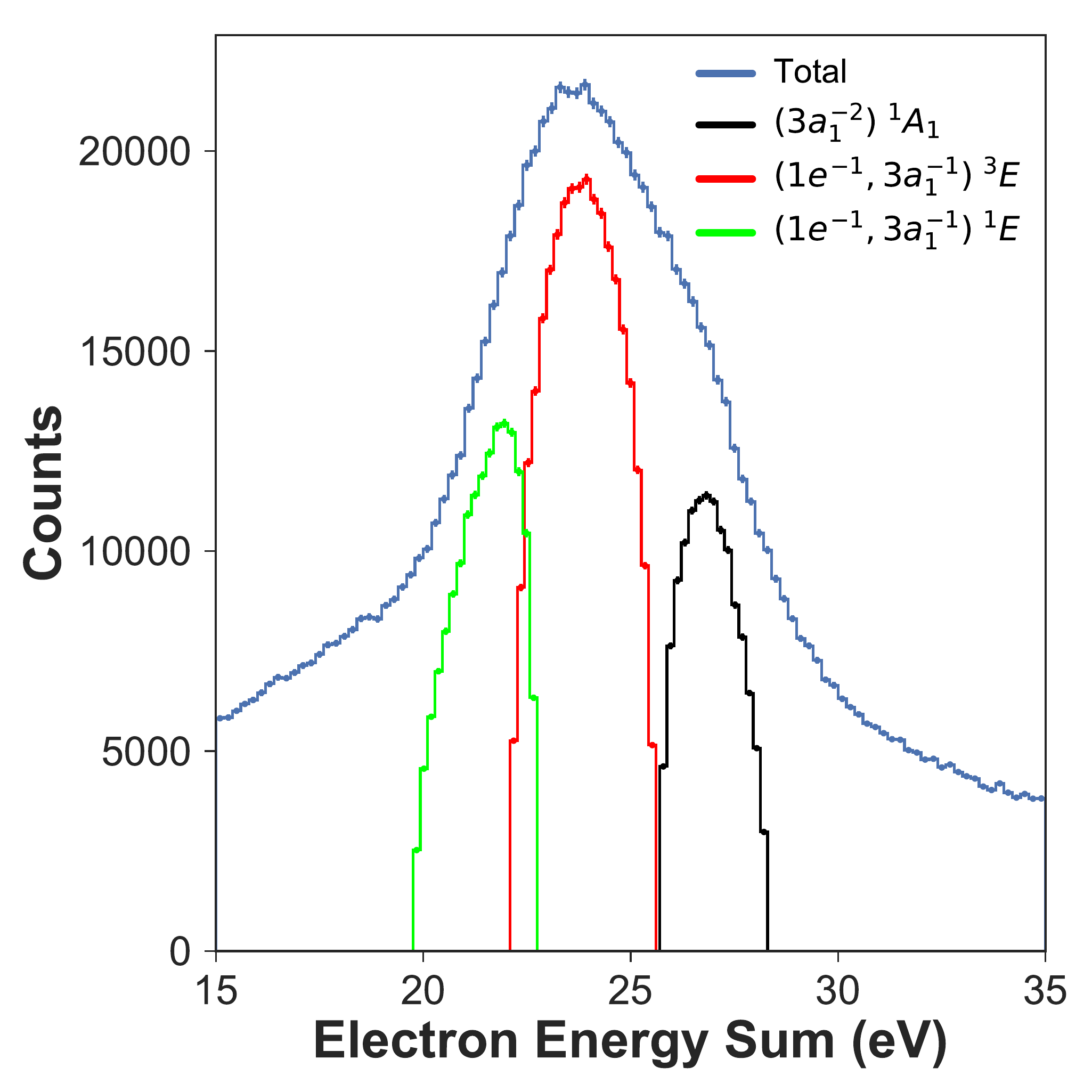}
\caption{The NH$_2^+$ + H$^+$ yield as a function of the kinetic energy sum of the photoelectron pair after PDI of NH$_3$ at 61.5~eV (shown in blue). The distributions for the three contributing dication states are shown in their respective color-codes (shown in black, red, and green, and multiplied by a factor of 1.6 for improved visibility). Contributions from individual dication states are extracted with gates as indicated in Fig.~\ref{fig:EeKER_2}, and hence their sum does not reflect the total (blue) yield (see text). The statistical error bars are on the order of the line width.}
\label{fig:Eesum_2}
\end{figure}

We present the NH$_2^+$ + H$^+$ yield as a function of the photoelectron energy sum in Fig.~\ref{fig:Eesum_2}, where each active dication state we identified in Fig.~\ref{fig:EeKER_2} has been indicated in Fig.~\ref{fig:Eesum_2} by a distribution in its corresponding color. In the total yield we observe an asymmetric monomodal distribution, exhibiting a rapid rise in yield on the low energy side of the peak and a slower decay in yield on the high energy side of the peak. The wings on the distribution originate from false coincidences and background events that are challenging to completely eliminate in the analysis, resulting in a near-uniform background (clearly visible in Fig.~\ref{fig:Ee_2}) underlying the spectrum that causes exaggerated wings.

\begin{figure}[h!]
        \includegraphics[width=8.5cm]{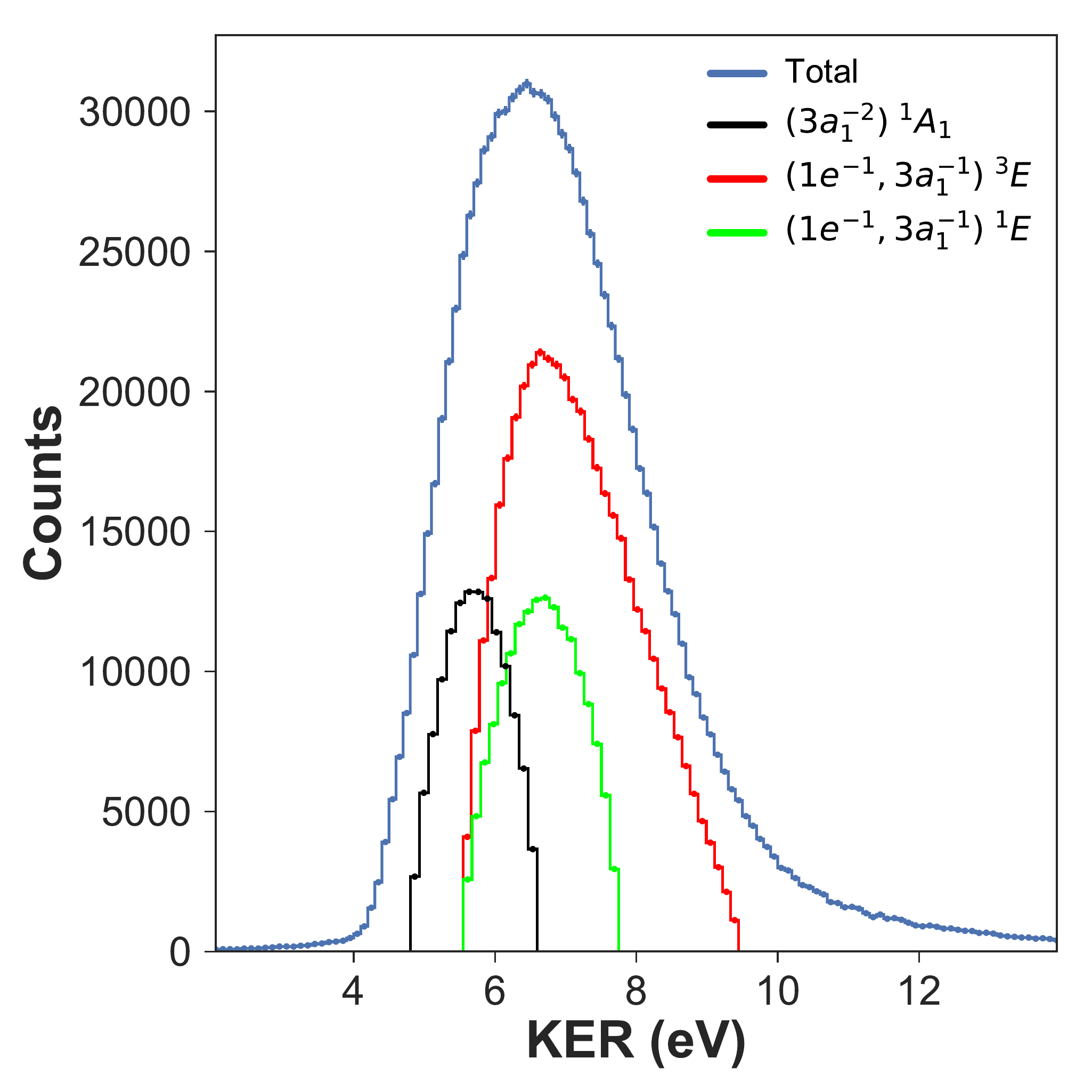}
\caption{The NH$_2^+$ + H$^+$ yield as a function of the KER after PDI of NH$_3$ at 61.5~eV (shown in blue). The distributions for the three contributing dication states are shown in their respective color-codes (shown in black, red, and green, and multiplied by a factor of 1.6 for improved visibility). Contributions from the individual dication states are extracted with gates as indicated in Fig.~\ref{fig:EeKER_2}, and hence their sum does not reflect the total (blue) yield (see text). The statistical error bars are on the order of the line width.}
\label{fig:KER_2}
\end{figure}

We show the NH$_2^+$ + H$^+$ yield as a function of KER in Fig.~\ref{fig:KER_2}, where each dication state we identified in Fig.~\ref{fig:EeKER_2} has been indicated by a distribution in its corresponding color. In the total yield we observe another broad asymmetric monomodal KER distribution with a rapid increase in yield on the low energy side of the peak and a slower decay in yield towards high energy. Both the experimental and theoretically calculated photoelectron energy sums and KERs are shown in Table~\ref{table:2body_EeKER}, which show good agreement.

These three corresponding dication states were identified using MRCI calculations, as discussed in Section~\ref{sec:level3}, and are consistently color-coded throughout the paper. The ($3a_1^{-2}$) $^1A_1$ state is shown in black, the ($3a_1^{-1},1e^{-1}$) $^3E$ state in red, and the ($3a_1^{-1},1e^{-1}$) $^1E$ state in green. Our ion yield measurements suggest that the branching ratios for these three dication states, shown in Table~\ref{table:2body_EeKER}, are approximately $13\% \pm 3\%$ for the ($3a_1^{-2}$) $^1A_1$ state, $44\% \pm 3\%$ for the ($3a_1^{-1},1e^{-1}$) $^3E$ state, and $43\% \pm 3\%$ for the ($3a_1^{-1},1e^{-1}$) $^1E$ state. These branching ratios and errors (plus/minus one standard deviation) are estimated by simultaneously fitting each feature in Fig.~\ref{fig:EeKER_2} with separate 2-D Gaussian distributions (although the distributions may not be explicitly Gaussian, this is nonetheless a good approximation). The fitting procedure varied the widths along each dimension independently, while also including a varying constant background offset. Following this fitting procedure, we integrate the fit for each dication state to estimate its contribution to the total NH$_2^+$ + H$^+$ yield. The main contribution to the uncertainty of the branching ratio is rooted in the aforementioned electron pair dead-time, which influences the detection yield of the electron-ion coincidences for each dication state as a function of the electron sum energy. Applying the simulation mentioned above, we estimate the total possible loss in PDI yield for electron sum energies of 26.9~eV (($3a_1^{-2}$) $^1A_1$), 23.9~eV (($3a_1^{-1},1e^{-1}$) $^3E$), and 21.4~eV (($3a_1^{-1},1e^{-1}$) $^1E$) to be 4.5\%, 5.2\%, and 5.9\%, respectively. This translates to an error of up to 3\% in the branching ratio. Errors due to deviations from the assumed Gaussian shape of each feature in the fitting process and the quality of the fit are estimated to be small (both $<$1\%).

\begin{figure}[h!]
    \includegraphics[width=4.225cm, trim=0.4cm 0.6cm 0.5cm 0.45cm, clip]{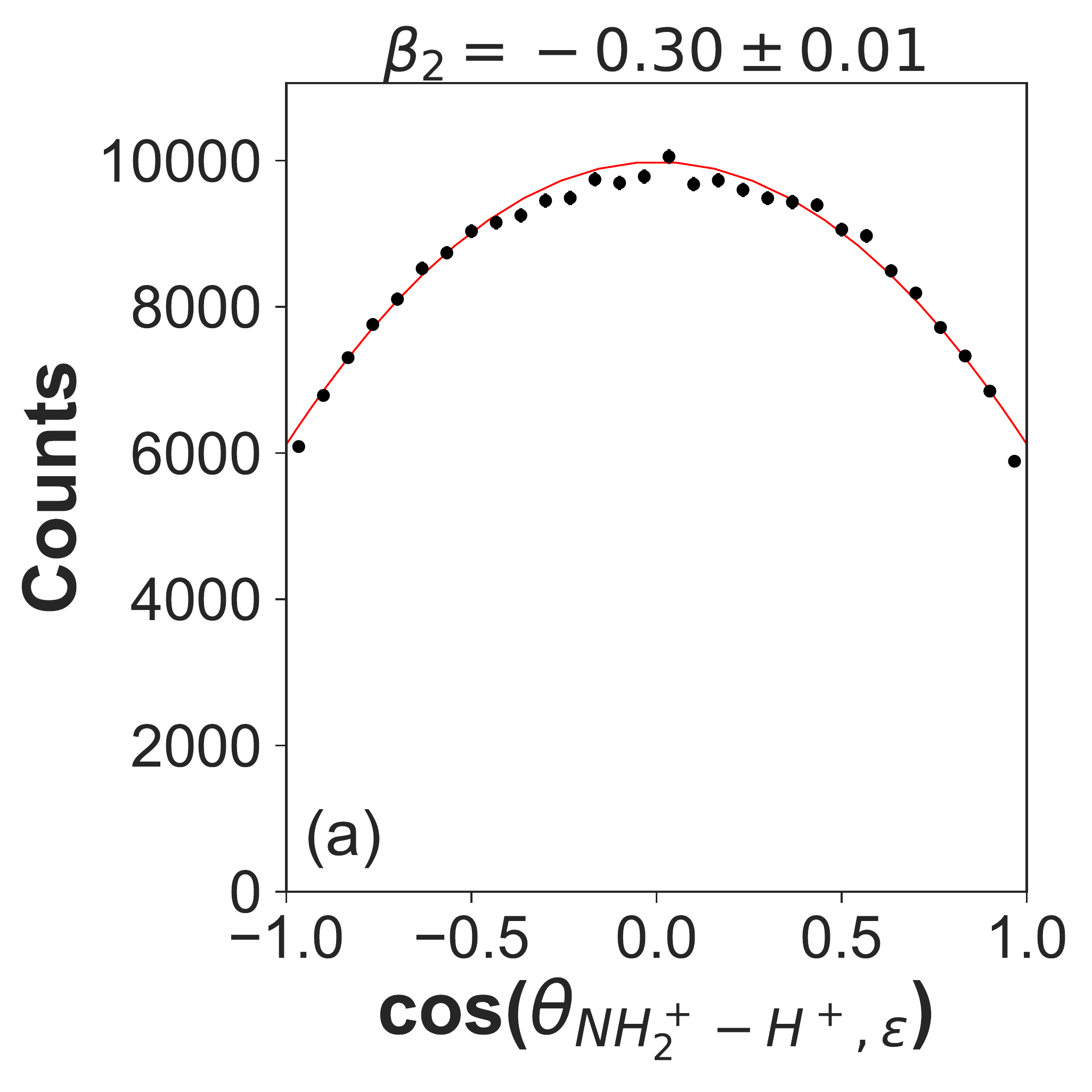}
    \includegraphics[width=4.225cm, trim=0.4cm 0.6cm 0.5cm 0.45cm, clip]{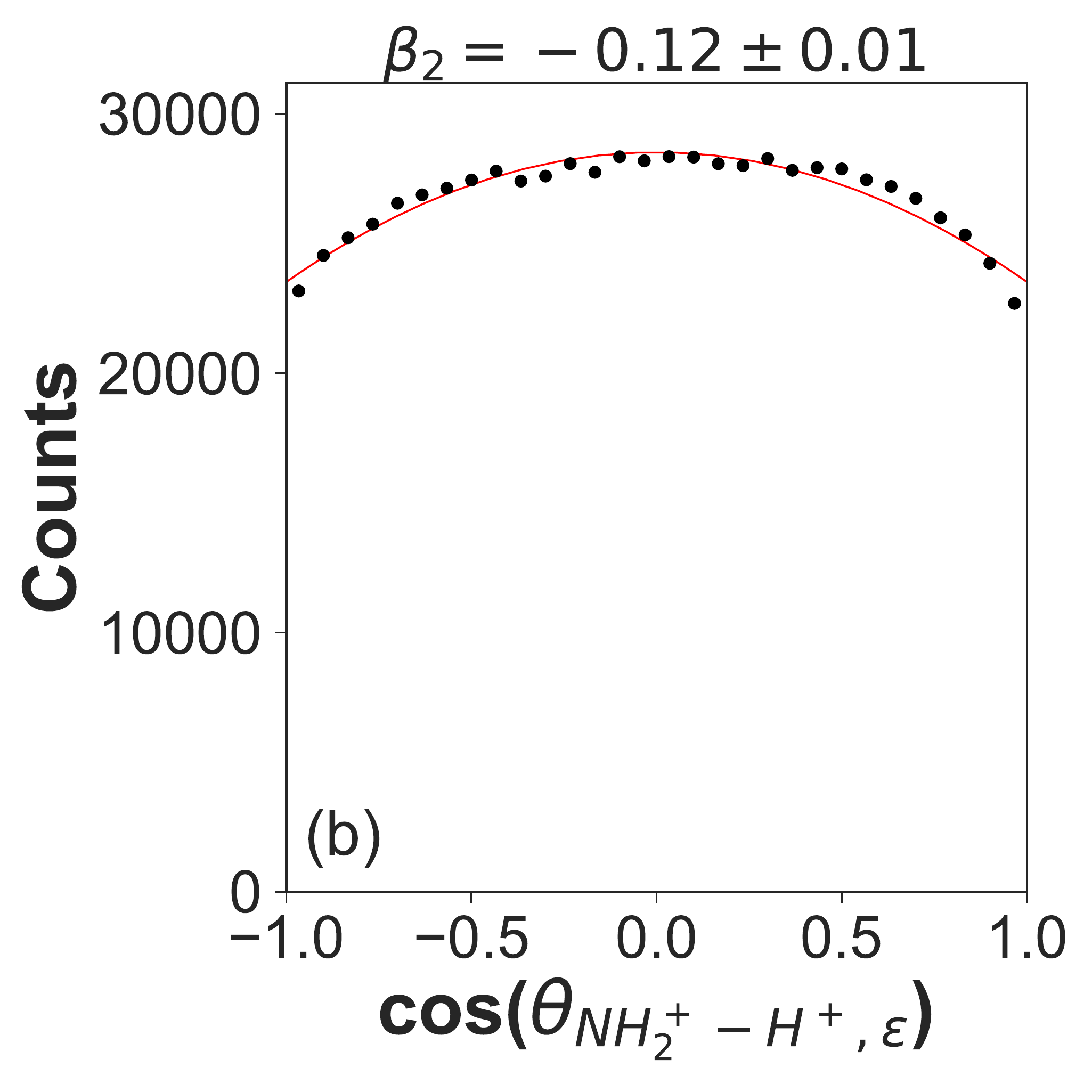}
    \includegraphics[width=4.225cm, trim=0.4cm 0.6cm 0.5cm 0.45cm, clip]{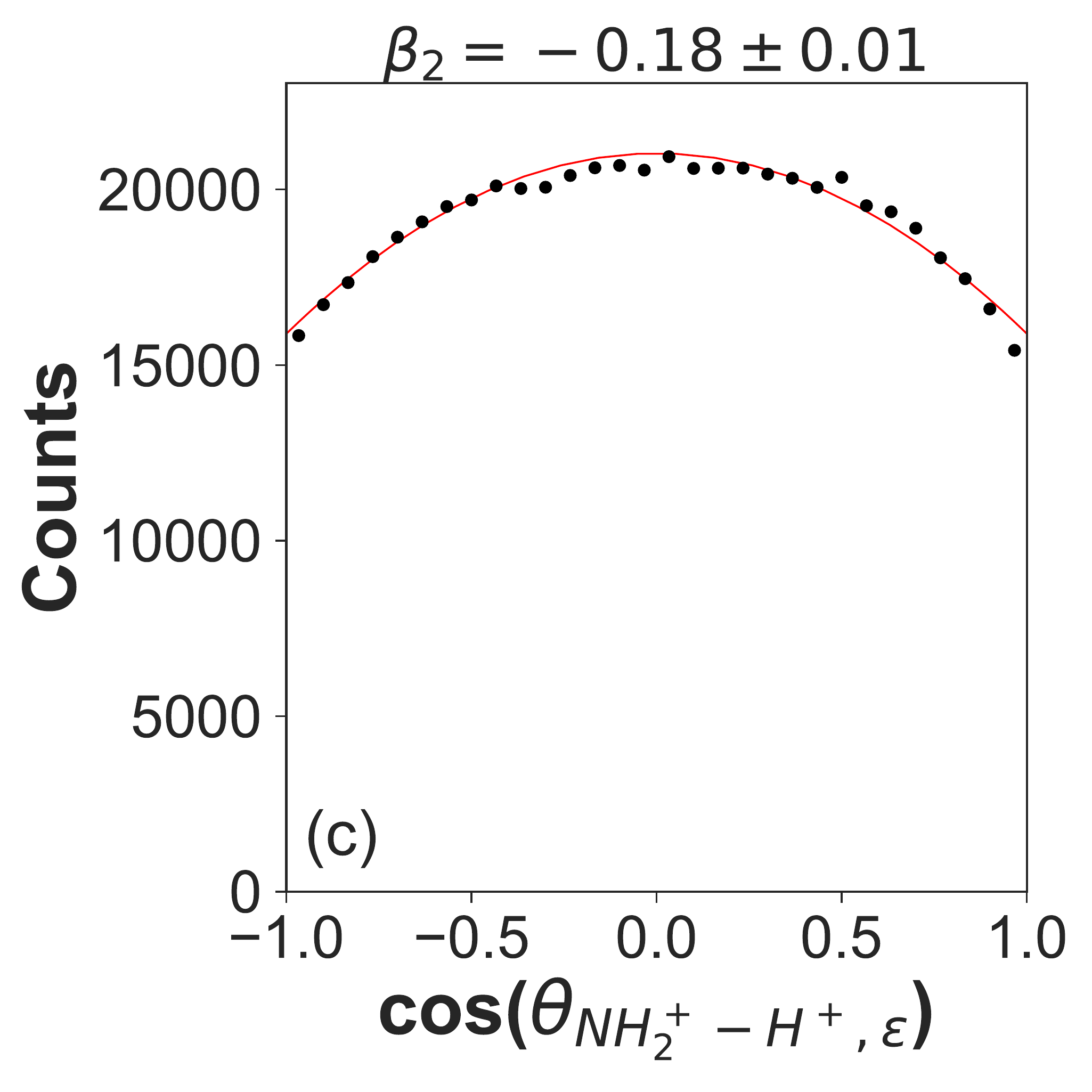}
\caption{The yield of NH$_2^+$ + H$^+$ dissociation after PDI of NH$_3$ at 61.5~eV as a function of the cosine of the measured relative angle between the NH$_2^+$-H$^+$ recoil axis and XUV polarization vector $\varepsilon$ for the (a) ($3a_1^{-2}$) $^1A_1$, (b) ($3a_1^{-1},1e^{-1}$) $^3E$, and (c) ($3a_1^{-1},1e^{-1}$) $^1E$ dication states. The fits, representing the parametrizations in terms of the anisotropy (see Eq.~\ref{AngDiff_PICS}), are shown in red, where the retrieved $\beta_2$ value is shown above each plot.}
\label{fig:recoil_angle_2body}
\end{figure}

From the PES cuts shown in Fig.~\ref{fig:PEC_NH3}(a) and the energetics presented in Fig.~\ref{fig:EeKER_2} and Table~\ref{table:2body_EeKER}, we conclude that the ($3a_1^{-2}$) $^1A_1$ state dissociates to the NH$_2^+(^3B_1)$ + H$^+$ limit, the ($3a_1^{-1},1e^{-1}$) $^3E$ state directly dissociates to this very same limit, and the ($3a_1^{-1},1e^{-1}$) $^1E$ state directly dissociates to the NH$_2^+(^1B_1)$ + H$^+$ limit. Here the ($3a_1^{-2}$) $^1A_1$ dication state must undergo an intersystem crossing preceding the dissociation, as the measured KER indicates that it does not reach its adiabatic limit (which would only produce $\sim$5.05~eV of KER). In Fig.~\ref{fig:PEC_NH3}(a) it can be seen that the ($3a_1^{-2}$) $^1A_1$ dication state is bound in the FC region and can predissociate by the $A'$ symmetry curve of the ($3a_1^{-1},1e^{-1}$) $^3E$ dication state. This enables a non-adiabatic population transfer mechanism that has been observed previously in near-threshold measurements \cite{Winkoun,Stankiewicz}. In Ref.~\cite{Winkoun} it was suggested that for excitation energies well above the PDI threshold of NH$_3$, direct dissocation would dominate over the predissociation, which is enabled by spin-orbit coupling. Our measurement, performed $\sim$27~eV above the double ionization threshold, is at odds with this proposition, as we observe the predissociation via intersystem crossing dominating over direct dissociation. Further, based on the measured KER, we conclude that this coupling mechanism outcompetes the dissociation from the population tunneling through the barrier of the ($3a_1^{-2}$) $^1A_1$ dication state along the asymmetric stretch coordinate, since dissociation via tunneling would result in a lower KER than what is measured. Comparing the measured and calculated KERs for the ($3a_1^{-2}$) $^1A_1$ state (see Table~\ref{table:2body_EeKER}) indicates that the resulting NH$_2^+$ fragment is formed in a relatively cold ro-vibrational state. In contrast to the ($3a_1^{-2}$) $^1A_1$ state, the ($3a_1^{-1},1e^{-1}$) $^3E$ state directly dissociates on the $A'$ curve, producing an NH$_2^+$ fragment with approximately 1.5~eV of ro-vibrational energy, which we infer by comparing the measured KER to the theoretical KER calculated for NH$_3^{2+}$ ground state fragments. However, the ($3a_1^{-1},1e^{-1}$) $^3E$ state also exhibits contributions from ro-vibrationally cold NH$_2^{+}$ fragments that reside in the long tail of the KER toward high energy values at constant electron sum energy. Here, a population transfer from the dissociative ($3a_1^{-1},1e^{-1}$) $^3E$ dication state to the ($3a_1^{-2}$) $^1A_1$ state is unlikely, as this would involve a non-adiabatic transition between states of different spin symmetry, and the wavepacket would only encounter the coupling region once as it dissociates. Similarly, the ($3a_1^{-1},1e^{-1}$) $^1E$ dication state also directly dissociates on its $A'$ curve, producing an NH$_2^+$ fragment with approximately 0.8~eV of ro-vibrational energy. We point out that the direct dissociation, producing a ro-vibrationally excited fragment, is consistent with the results in [I].

In order to assess if there are preferred molecular orientations at which PDI of NH$_3$ occurs for each of the three NH$_3^{2+}$ dication states, we plot in Fig.~\ref{fig:recoil_angle_2body}(a), (b), and (c) the yield of the NH$_2^+$ + H$^+$ two-body fragmentation as a function of the cosine of the relative angle between the recoil axis of the molecular breakup (NH$_2^+$-H$^+$) and the XUV polarization $\varepsilon$. For the features corresponding with the ($3a_1^{-2}$) $^1A_1$, and ($3a_1^{-1},1e^{-1}$) $^3E$, and ($3a_1^{-1},1e^{-1}$) $^1E$ dication states, we observe an enhancement in PDI for molecular orientations where the recoil axis is aligned at $\sim$90$^{\circ}$ angle with respect to the polarization vector. Here, this enhancement in PDI yield is strongest in the ($3a_1^{-2}$) $^1A_1$ state, weaker in the ($3a_1^{-1},1e^{-1}$) $^1E$ state, whereas the distribution of the ($3a_1^{-1},1e^{-1}$) $^3E$ dication state is the most isotropic of the three (its $\beta_2$ value is the closest to zero). This perpendicular orientation of the recoil axis with respect to the polarization vector roughly coincides with the C$_{3v}$ symmetry axis of the NH$_3$ molecule. In all dication states the PDI involves the $3a_1$ orbital, i.e. the lone-pair, which is aligned along the molecular C$_{3v}$ axis. This could explain the enhancement at geometries where the polarization vector of the ionizing field is directed along this orbital and the stronger effect in the ($3a_1^{-2}$) $^1A_1$ state, where both holes are created in the $3a_1$ orbital, as opposed to the ($3a_1^{-1},1e^{-1}$) $^3E$ and ($3a_1^{-1},1e^{-1}$) $^1E$ states, where a hole is created in each of the $3a_1$ and $1e$ orbitals. 

These photofragment angular distributions have been fitted (solid red line) using the familiar parameterization: 

\begin{equation}
\frac{d\sigma}{d\Omega} = \frac{\sigma_0}{4\pi} [1 + \beta_2 P_2 (\cos \theta)], 
\label{AngDiff_PICS}
\end{equation}
which describes the fragment angular distribution from the dissociation of a rigid rotor, where $\sigma_0$ is the total cross section, $\beta_2$ is the anisotropy parameter, $P_2$ is the second order Legendre polynomial, and $\theta$ is the angle between the recoil axis of the molecular two-body breakup and the polarization vector of the ionizing field \cite{Zare,Greene}. The retrieved $\beta_2$ parameter is shown above each plot, while the data is fitted using the projection method discussed in \cite{Liu}, where the error of $\beta_2$ is determined via statistical bootstrapping \cite{Efron}. We find $\beta_2$ values of $-0.30\pm0.01$, $-0.12\pm0.01$, and $-0.18\pm0.01$ for the ($3a_1^{-2}$) $^1A_1$, ($3a_1^{-1},1e^{-1}$) $^3E$, and  ($3a_1^{-1},1e^{-1}$) $^1E$ dication states, respectively. These values are also listed in Table~\ref{table:2body_EeKER}.

\begin{figure}[h!]
    {
        \includegraphics[width=4.225cm, trim=0.4cm 0.6cm 0.5cm 0.65cm, clip]{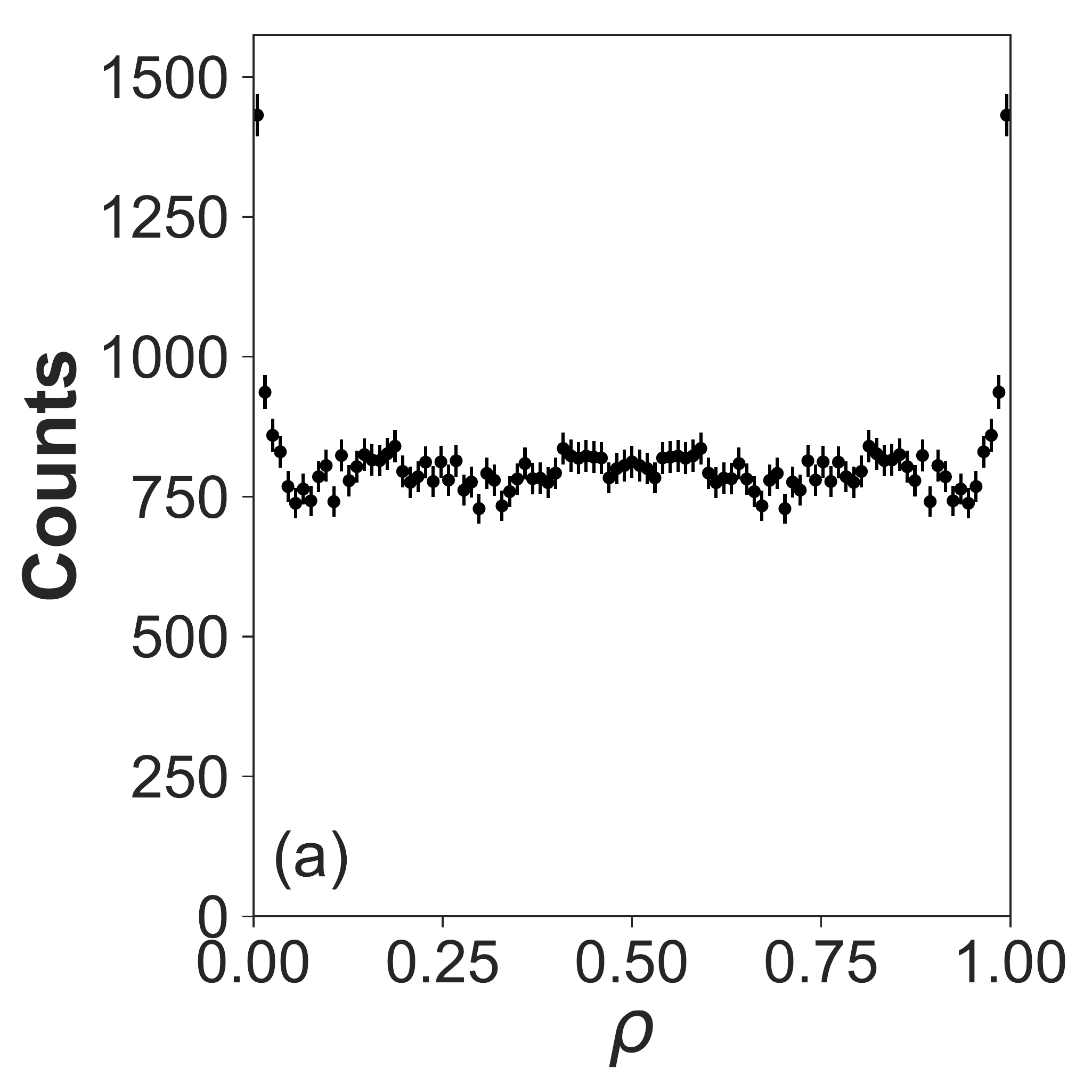}}
    {
        \includegraphics[width=4.225cm, trim=0.4cm 0.6cm 0.5cm 0.65cm, clip]{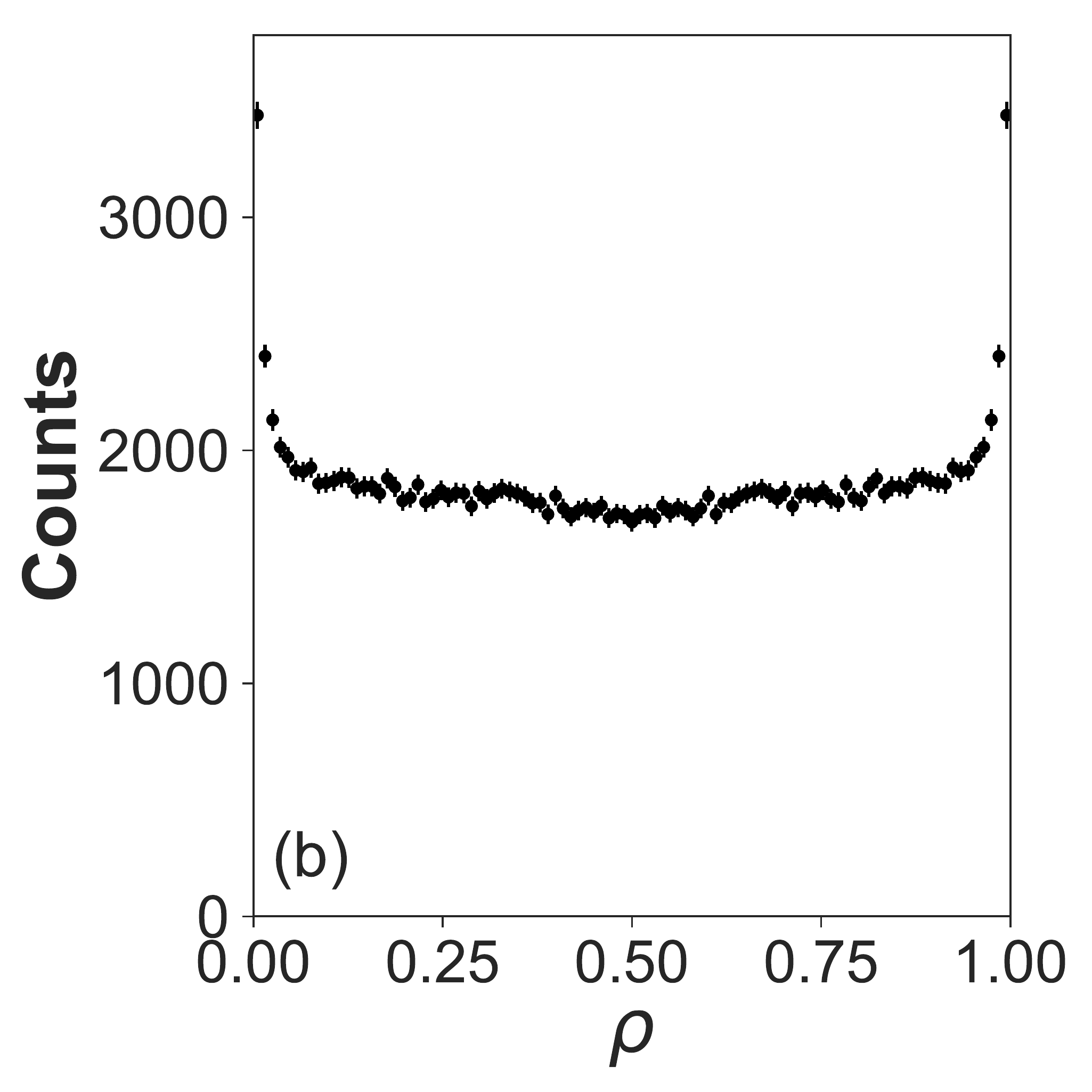}}    
    {
        \includegraphics[width=4.225cm, trim=0.4cm 0.6cm 0.5cm 0.65cm, clip]{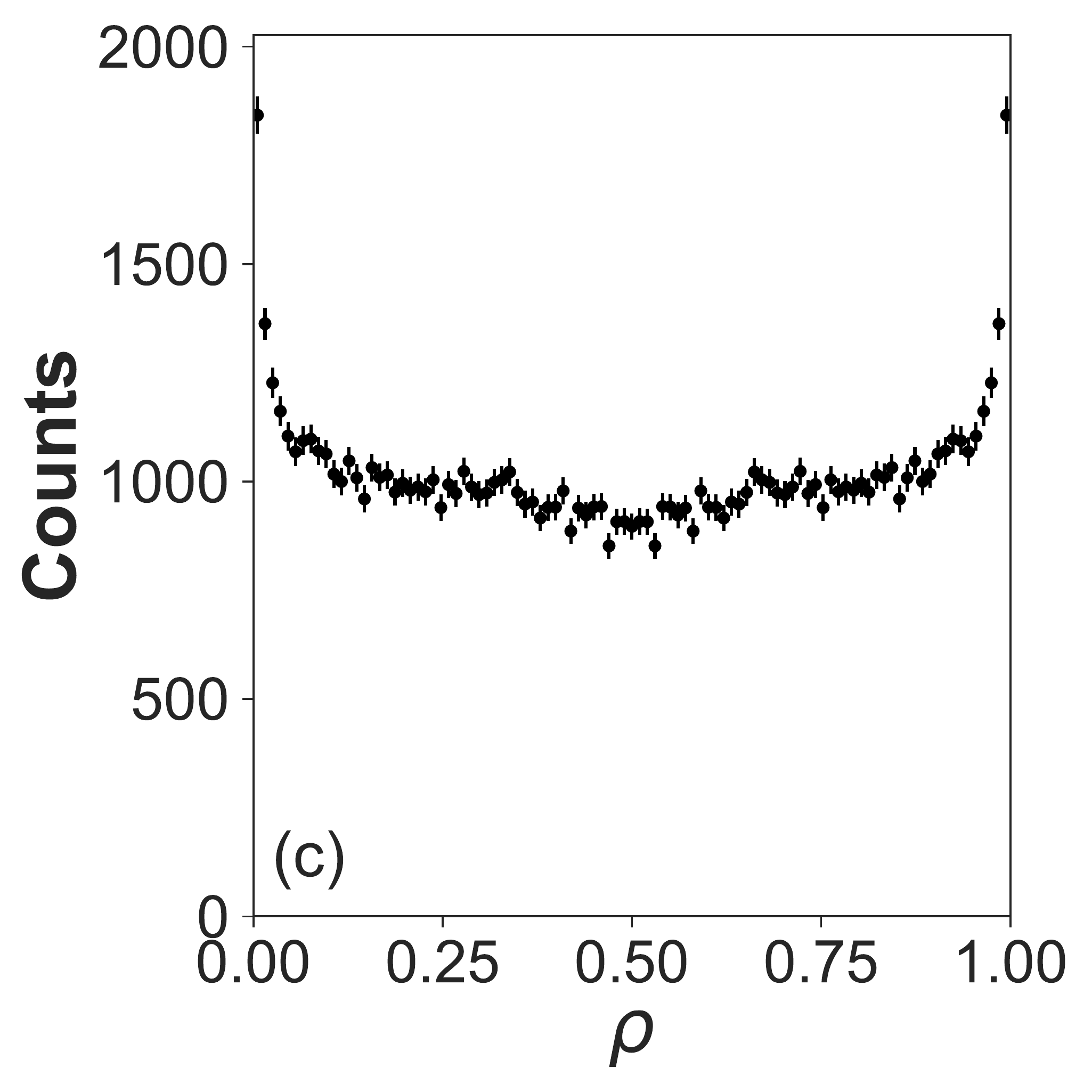}}
\caption{The yield of the NH$_2^+$ + H$^+$ two-body breakup after PDI of NH$_3$ at 61.5~eV as a function of the electron energy sharing $\rho$ (see Eq.~\ref{Eng-sharing}) for the (a) ($3a_1^{-2}$) $^1A_1$, (b) ($3a_1^{-1},1e^{-1}$) $^3E$, and (c) ($3a_1^{-1},1e^{-1}$) $^1E$ dication states.}
\label{fig:SDCS_2body}
\end{figure}

Next, we display the yield of the NH$_2^+$ + H$^+$ two-body channel as a function of the energy sharing $\rho$ between the two photoelectrons for the three features that correspond with the three NH$_3^{2+}$ states. We define the electron energy sharing as: 

\begin{equation}
\rho = \frac{E_{e_{1}}}{E_{e_{1}} + E_{e_{2}}},
\label{Eng-sharing}
\end{equation}
where $E_{e_{1}}$ and $E_{e_{2}}$ are the kinetic energies of electron 1 and 2, respectively. Values of $\rho$ near 0.5 indicate equal energy sharing between the two photoelectrons, while values near 0 or 1 indicate unequal energy sharing between the two photoelectrons. The results are shown in Fig.~\ref{fig:SDCS_2body}(a), (b), and (c). 

We attribute the sharp features near 0 and 1, observed in Fig.~\ref{fig:SDCS_2body}(a), (b), and (c), to the PDI of NH$_3$ via autoionization, corresponding with a fast photoelectron and slow electron emerging from the autoionization. The fraction of PDI via autoionization relative to direct PDI is appoximately $1.9\% \pm 1.0\%$ in the ($3a_1^{-2}$) $^1A_1$ state, $4.4\% \pm 0.7\%$ in the ($3a_1^{-1},1e^{-1}$) $^3E$ state, and $6.2\% \pm 0.9\%$ in the ($3a_1^{-1},1e^{-1}$) $^1E$ state. This is estimated by extrapolating the average number of counts in the equal energy sharing condition across all $\rho$ and then subtracting this value from the bins where $\rho$ is near 0 or 1 (the unequal energy sharing condition). Computing this residue gives an estimate on how many counts are associated with autoionization relative to the direct PDI of NH$_3$. The error in the autoionization fraction is determined from the error in the average number of counts in the equal energy sharing condition (which is extracted from Poisson statistics).

\begin{figure}[h!]
    {
        \includegraphics[width=4.225cm, trim=0.4cm 0.6cm 0.5cm 0.65cm, clip]{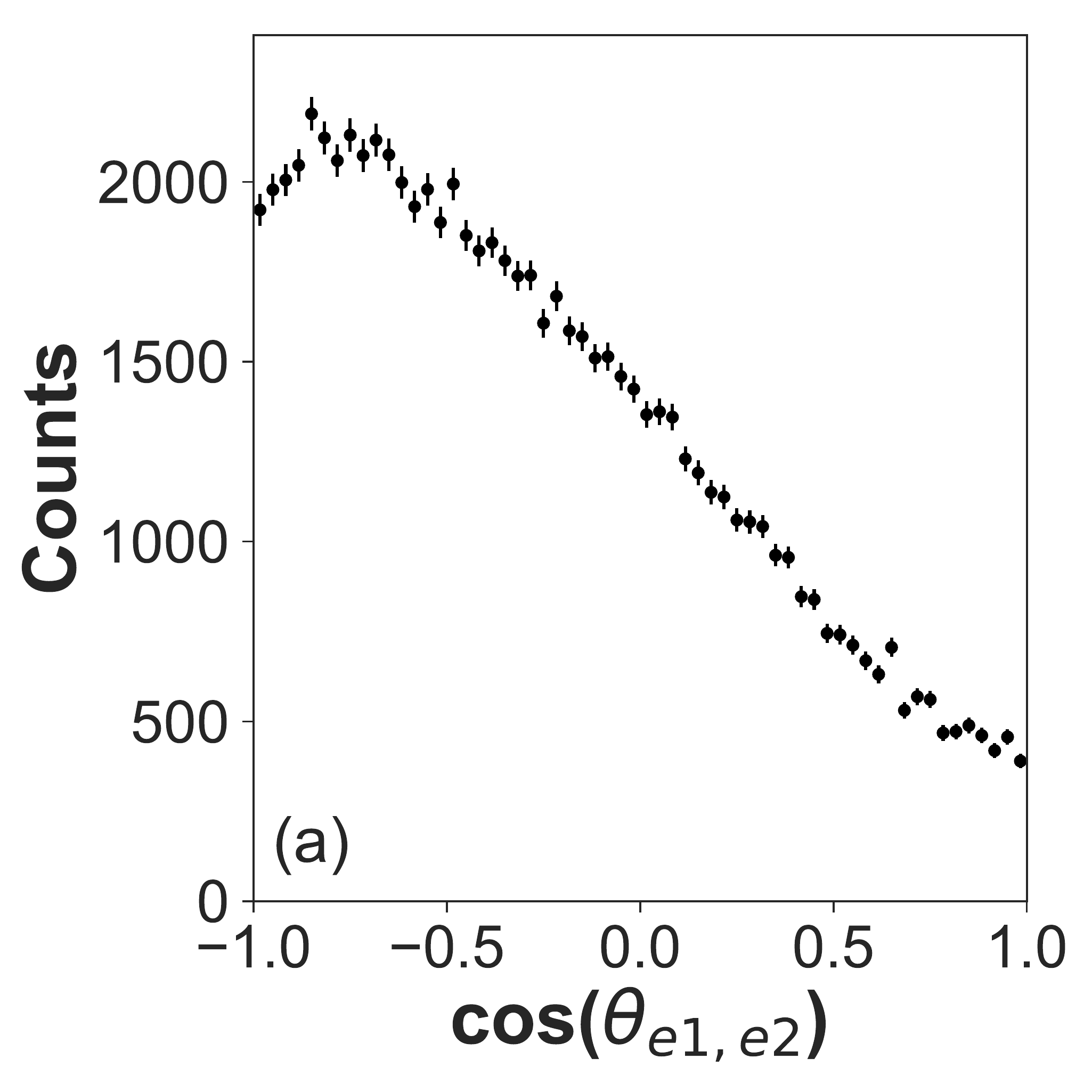}}
    {
        \includegraphics[width=4.225cm, trim=0.4cm 0.6cm 0.5cm 0.65cm, clip]{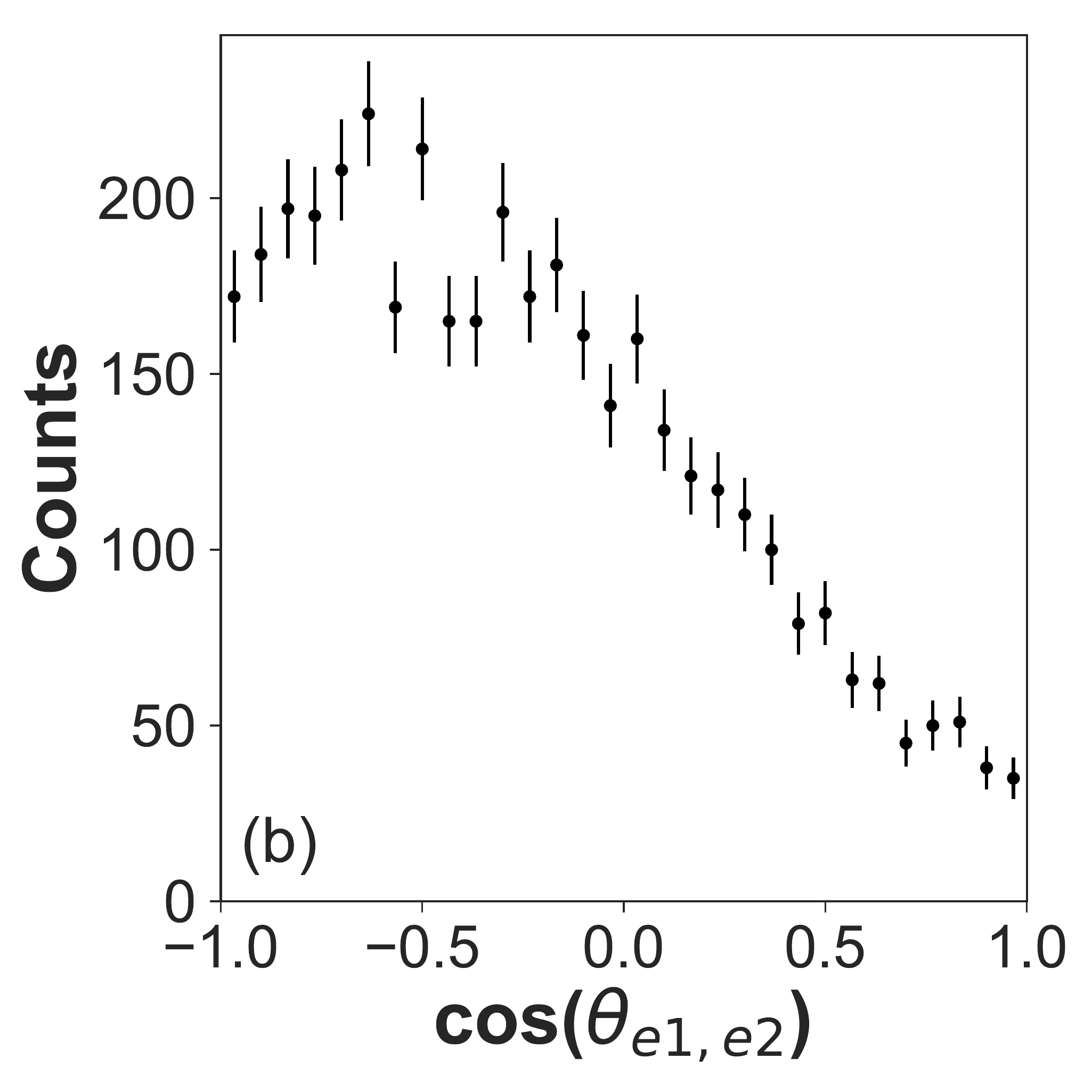}}
\caption{The yield of the NH$_2^+$ + H$^+$ two-body breakup, after PDI of NH$_3$ at 61.5~eV, as a function of the cosine of the relative emission angle between the two photoelectrons for the ($3a_1^{-2}$) $^1A_1$ dication state (a) integrated over all possible electron energy sharing and (b) for equal energy sharing ($\rho = 0.5 \pm 0.025$).}
\label{fig:X_angle_e}
\end{figure}

\begin{figure}[h!]
    {
        \includegraphics[width=4.225cm, trim=0.4cm 0.6cm 0.5cm 0.65cm, clip]{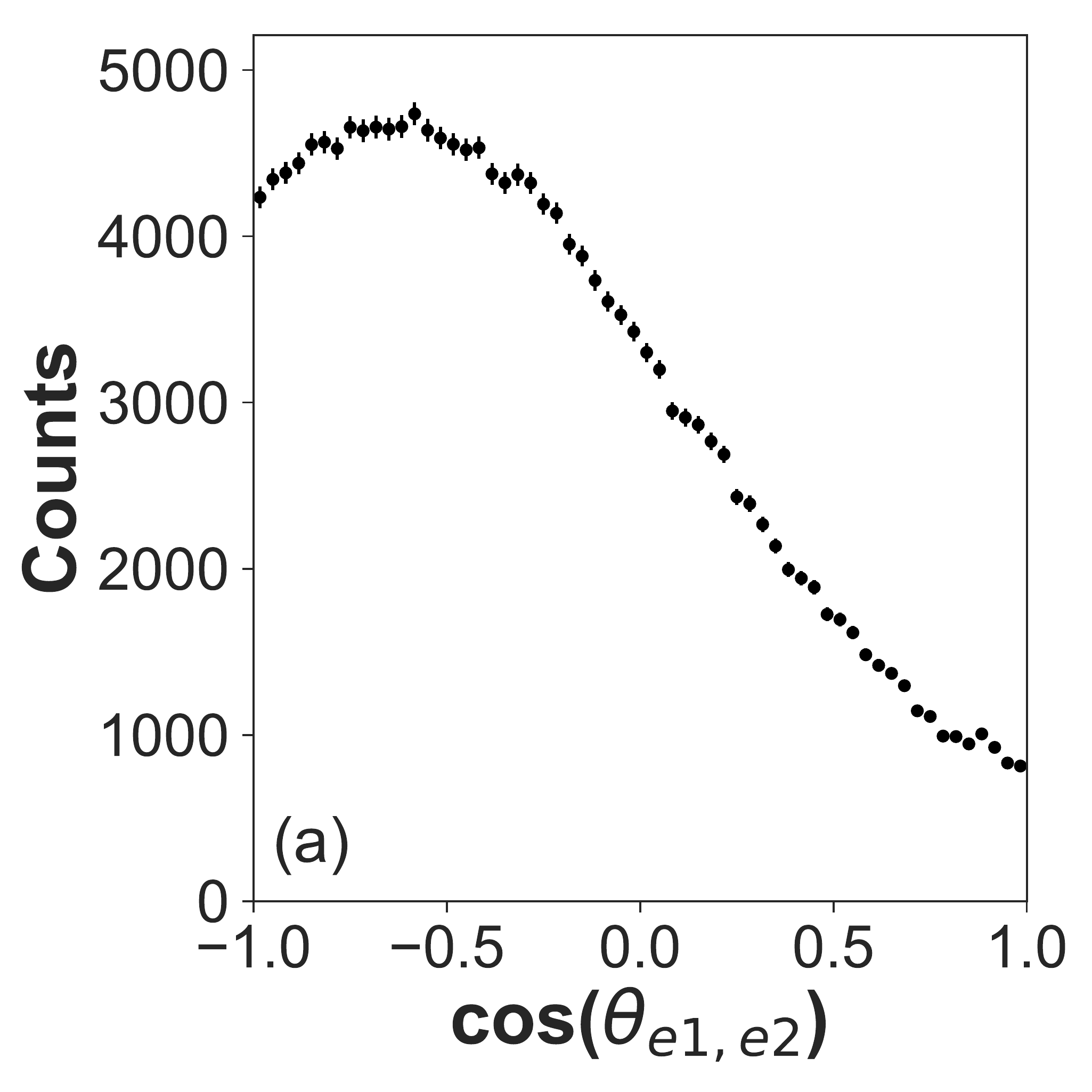}}
    {
        \includegraphics[width=4.225cm, trim=0.4cm 0.6cm 0.5cm 0.65cm, clip]{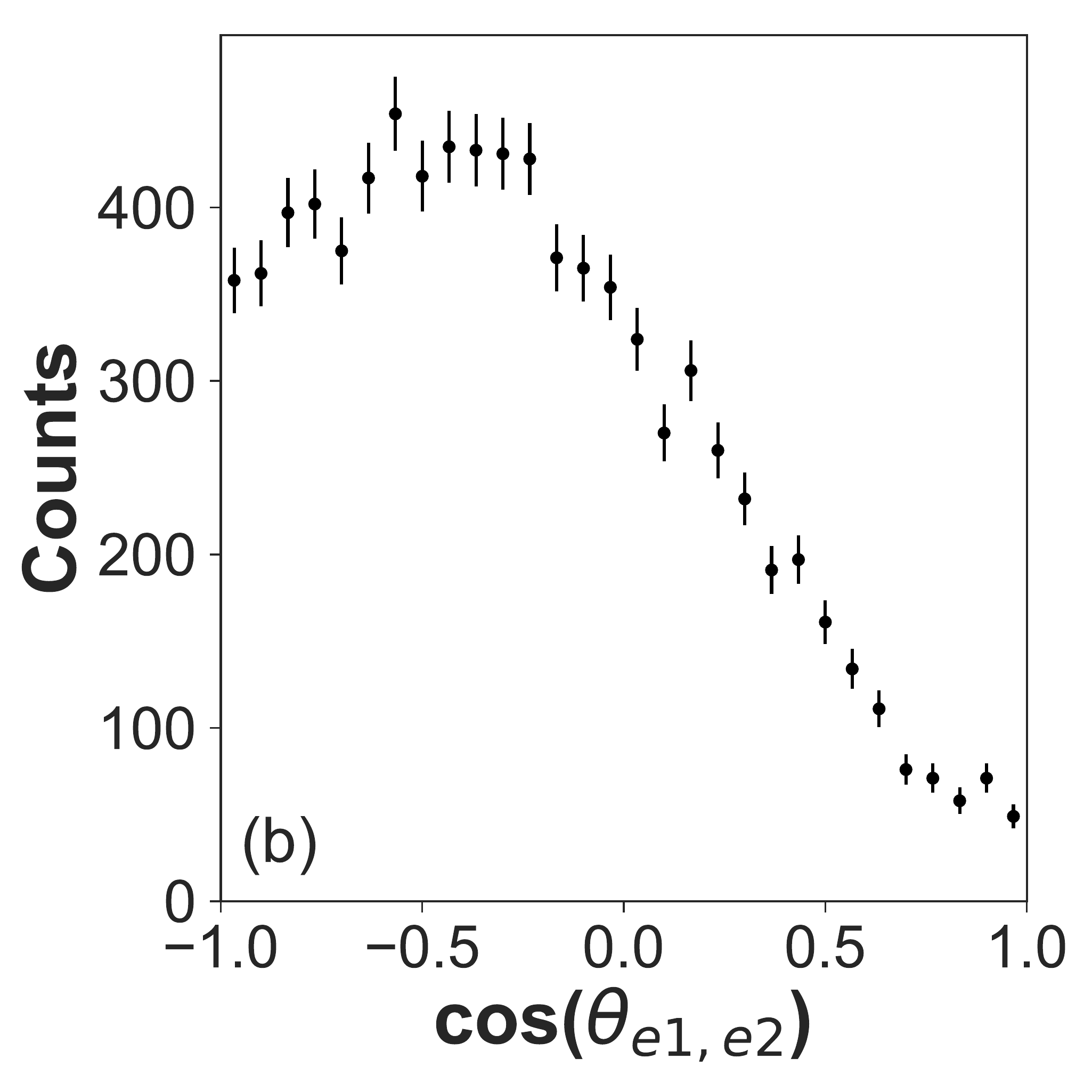}}
\caption{The yield of the NH$_2^+$ + H$^+$ two-body breakup, after PDI of NH$_3$ at 61.5~eV, as a function of the cosine of the relative emission angle between the two photoelectrons for the ($3a_1^{-1},1e^{-1}$) $^3E$ dication state (a) integrated over all possible electron energy sharing and (b) for equal energy sharing ($\rho = 0.5 \pm 0.025$).}
\label{fig:A_angle_e}
\end{figure}

\begin{figure}[h!]
    {
        \includegraphics[width=4.225cm, trim=0.4cm 0.6cm 0.5cm 0.65cm, clip]{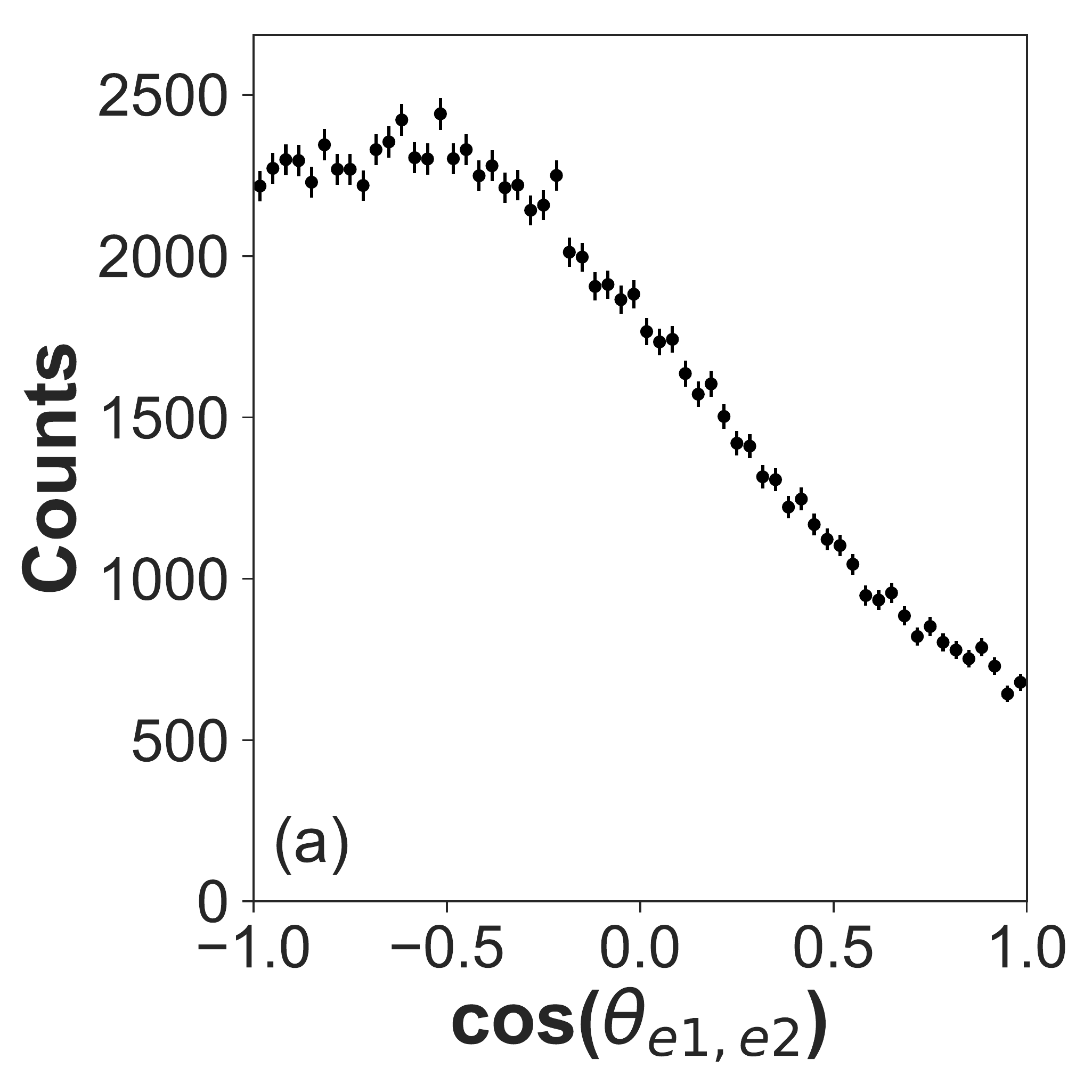}}
    {
        \includegraphics[width=4.225cm, trim=0.4cm 0.6cm 0.5cm 0.65cm, clip]{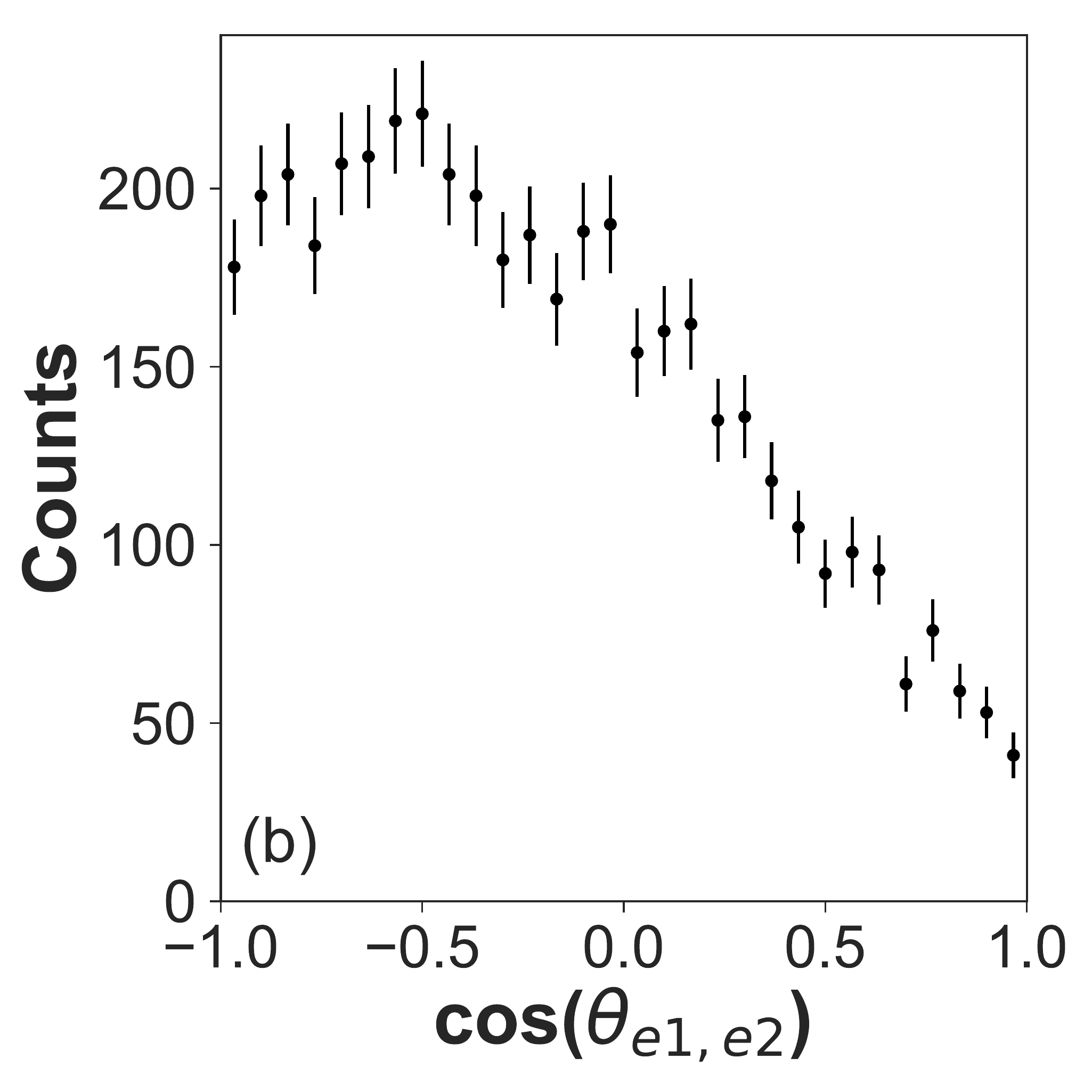}}
\caption{The yield of the NH$_2^+$ + H$^+$ two-body breakup, after PDI of NH$_3$ at 61.5~eV, as a function of the cosine of the relative emission angle between the two photoelectrons for the ($3a_1^{-1},1e^{-1}$) $^1E$ dication state (a) integrated over all possible electron energy sharing and (b) for equal energy sharing ($\rho = 0.5 \pm 0.025$).}
\label{fig:B_angle_e}
\end{figure}

Finally, we plot in Fig.~\ref{fig:X_angle_e}, Fig.~\ref{fig:A_angle_e}, and Fig.~\ref{fig:B_angle_e} the yield of the NH$_2^+$ + H$^+$ dissociation channel as a function of the cosine of the relative emission angle between the two photoelectrons, (a) integrated over all energy sharing conditions and (b) for equal energy sharing condition for the three NH$_3^{2+}$ states. In these figures, there are no conditions enforced on either the molecular orientation or the emission angle of the first detected photoelectron relative to the polarization vector of the XUV beam. In the equal energy sharing case the relative angle is plotted for $0.475 < \rho < 0.525$. We point out that our measurement suffers from some multi-hit detector dead-time effects, which influence the measured yield of photoelectrons emitted in the same direction with similar kinetic energies. In the equal energy sharing condition and for the emission into the same hemisphere $\theta_{e_1,e_2} \leq 90^\circ$, this corresponds, in worst case, to a loss of $\sim$15$\%$ of the events for the ($3a_1^{-2}$) $^1A_1$ state, $\sim$16$\%$ for the ($3a_1^{-1},1e^{-1}$) $^3E$ state, and $\sim$18$\%$ for the ($3a_1^{-1},1e^{-1}$) $^1E$ state. This worst-case scenario is simulated for an isotropic relative electron-electron emission, which very well represents autoionization processes that are sequential in nature and are subject to unequal energy sharing between the electrons. The equal energy sharing case on the other hand is dominated by knock-out processes with very few electron pairs emitted into the same hemisphere. The actual losses are hence believed to be smaller by at least a factor of 2, i.e. closer to the simulated losses for unequal electron energy sharing. We refrain from showing the measured photoelectron angular distribution in the unequal energy sharing case, which captures the autoionization feature, as there is a significant contribution due to direct PDI that pollutes the autoionization signal considerably and prevents a clear analysis of the relative angular distribution for this indirect ionization process.

The relative angles between the two electrons, integrated over all energy sharing cases, show a preferred emission of the two electrons into opposite hemispheres. The distribution for the ($3a_1^{-2}$) $^1A_1$ dication state [Fig.~\ref{fig:X_angle_e}(a)] peaks at 145$^\circ$ with a notable dip at 180$^\circ$ corresponding to a back-to-back emission. The ($3a_1^{-1},1e^{-1}$) $^3E$ state [Fig.~\ref{fig:A_angle_e}(a)] peaks at 125$^\circ$ and has a similar dip at 180$^\circ$. The ($3a_1^{-1},1e^{-1}$) $^1E$ state [Fig.~\ref{fig:B_angle_e}(a)] peaks at 120$^\circ$, with a slight increase at 180$^\circ$ compared to the two other dication states.

The photoelectron dynamics in the equal energy sharing condition ($\rho = 0.5 \pm 0.025$) reveals similar anisotropic angular distributions depicted in Fig.~\ref{fig:X_angle_e}(b), Fig.~\ref{fig:A_angle_e}(b), and Fig.~\ref{fig:B_angle_e}(b), which possess minima near 0$^\circ$ relative emission angle in all three states and peak at relative angles near 130$^\circ$ in the ($3a_1^{-2}$) $^1A_1$ state, 125$^\circ$ in the ($3a_1^{-1},1e^{-1}$) $^3E$ state, and  120$^\circ$ in the ($3a_1^{-1},1e^{-1}$) $^1E$ state. All three distributions show a dip at 180$^\circ$. This relative electron emission pattern is reminiscent of a knock-out double ionization process as found in other valence PDI investigations of atomic and molecular targets \cite{Brauning,Weber,Weber1,Weber2}. In all three cases the likelihood for emission in the same direction is roughly a factor of 10 less likely than emission at the peak angle.

\subsection{\label{}Three-body breakup channel: NH$^{+}$ + H$^{+}$ + H}

\begin{figure}[h!]
    \includegraphics[width=8.5cm]{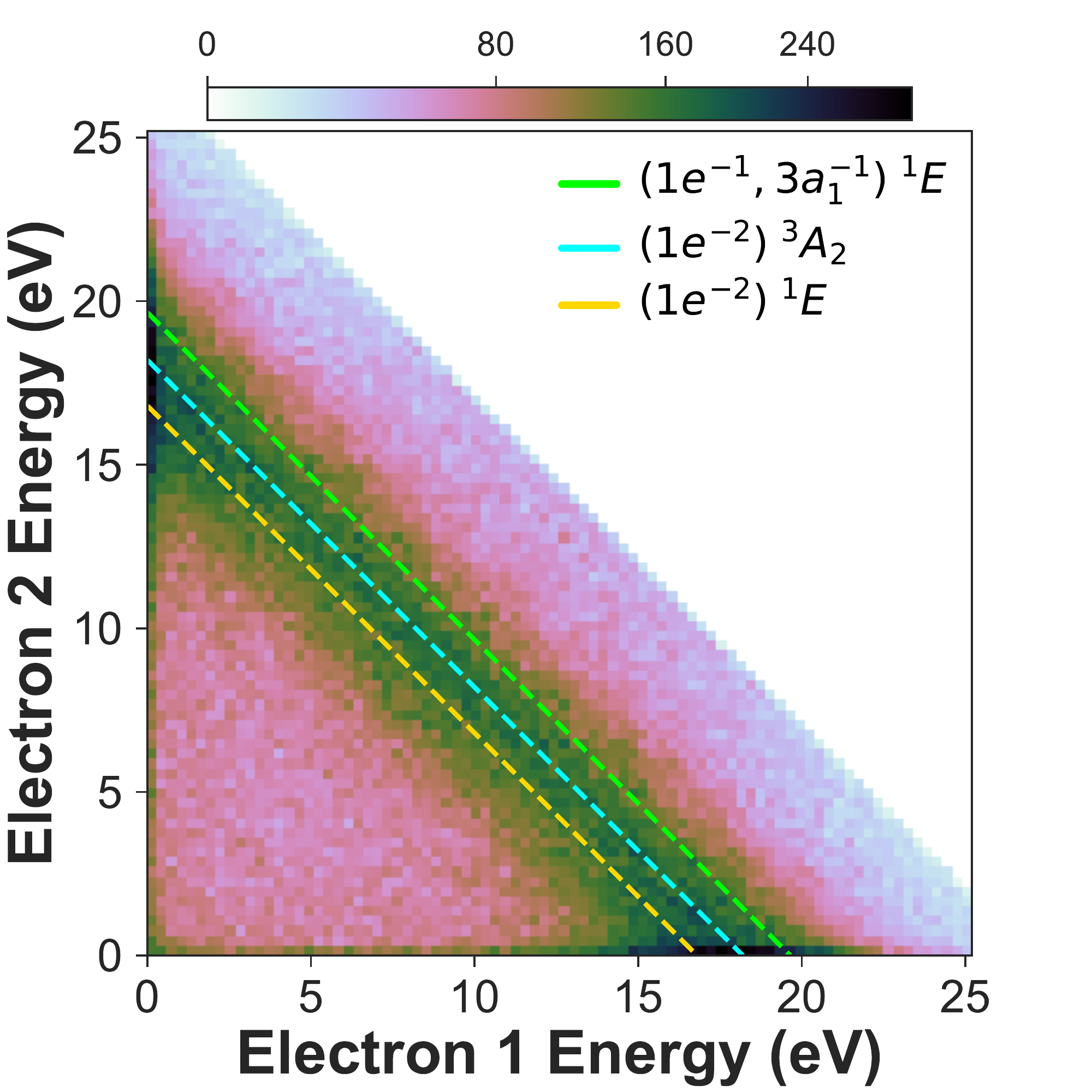}
\caption{The NH$^+$ + H$^+$ + H three-body yield, after PDI of NH$_3$ at 61.5~eV, as a function of the kinetic energy of the first and second photoelectron. The three contributing dication states are color-coded (green, cyan, and gold) and shown as diagonal lines to guide the eye. Electron sum energies beyond 27~eV are omitted for visual clarity.}
\label{fig:Ee_3}
\end{figure}

As in the previous section, we plot in Fig.~\ref{fig:Ee_3} the PDI yield of the NH$^+$ + H$^+$ + H channel as a function of the kinetic energy of the first and second detected electrons, to produce the electron-electron energy correlation map. As before, the figure has been symmetrized across the diagonal (the line E$_1$ = E$_2$) to account for the indistinguishability of the two photoelectrons. With the guidance of the calculated PES cuts, we identify three features corresponding with three NH$_3^{2+}$ dication electronic states that feed the three-body NH$^{+}$ + H$^{+}$ + H fragmentation channel. Again, although these features are difficult to visually identify here, they are better separated in different spectra that are shown below in Fig.~\ref{fig:EeKER_3}. The photoelectrons associated with these features have energy sums centered around 19.7~eV, 18.3~eV, and 16.8~eV, and a FWHM of roughly 2.5~eV, 2.3~eV, and 2.1~eV, respectively. These features, indicated by diagonal lines (taking the form E$_2$ = -E$_1$ + E$_{sum}$), are color-coded as green, cyan, and gold to guide the eye (the same green color used in the two-body breakup section is applied here because the same dication state contributes to both the two- and three-body fragmentation channels, as discussed below). Using the same protocol as described in the two-body breakup section, we choose each of these three features by selecting carefully around the center of each feature in Fig.~\ref{fig:EeKER_3}. As in the two-body breakup channel, these three states are accessed via direct PDI and also evidently through autoionization. 

Three corresponding features are present in the electron-nuclei energy correlation map, shown in Fig.~\ref{fig:EeKER_3}, which are circled by their respective color-codes to guide the eye (as before, these ellipses do not reflect the actual momentum gates). The feature circled by the cyan ellipse can be mistaken as a part of the feature circled by either the green or gold ellipse, however it is identified as a state with the assistance of the calculated PES cuts. Each feature possesses a different KER distribution centered around 6.4~eV, 7.1~eV, and 9.4~eV, each with a FWHM of roughly 1.1~eV, 1.4~eV, and 2.4~eV, respectively.

\begin{figure}[h!]
    \includegraphics[width=8.5cm]{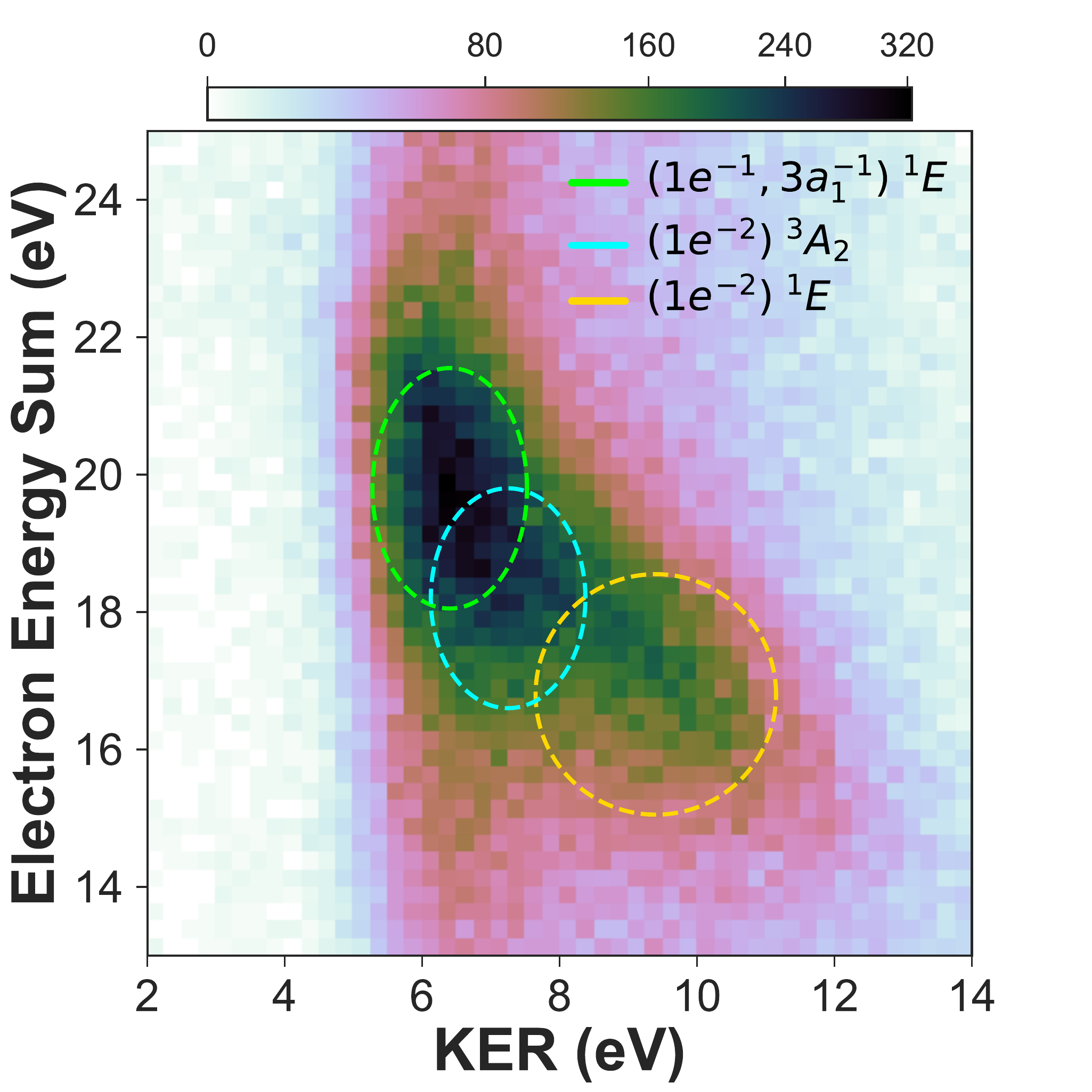}
\caption{The NH$^+$ + H$^+$ + H three-body yield, after PDI of NH$_3$ at 61.5~eV, as a function of the KER and the kinetic energy sum of the photoelectron pair. The three contributing dication states are color-coded (green, cyan, and gold) and shown as ellipses to guide the eye (they only approximately represent the software gates).}
\label{fig:EeKER_3}
\end{figure}

\begin{figure}[h!]
    \includegraphics[width=8.5cm]{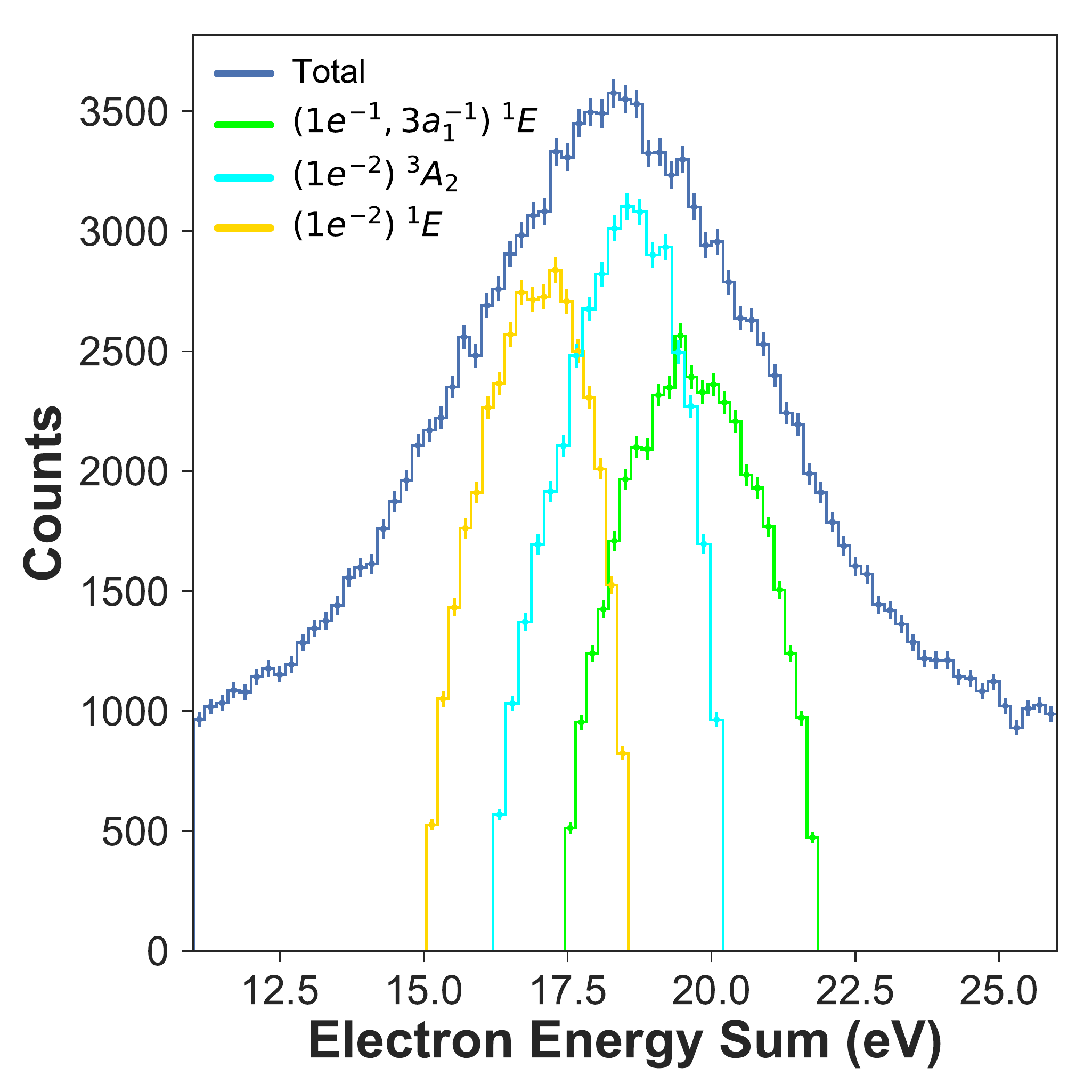}
\caption{The NH$^+$ + H$^+$ + H three-body yield, after PDI of NH$_3$ at 61.5~eV, as a function of the electron kinetic energy sum (shown in blue). The distributions for the three contributing dication states are shown in their respective color-codes (shown in green, cyan, and gold, and multiplied by a factor of 1.5 for improved visibility). Contributions from the individual dication states are extracted with gates as indicated in Fig.~\ref{fig:EeKER_3}, and hence their sum does not reflect the total (blue) yield (see text).}
\label{fig:Eesum_3}
\end{figure}

We present the NH$^+$ + H$^+$ + H three-body yield as a function of the photoelectron energy sum in Fig.~\ref{fig:Eesum_3}, where each feature we identified in Fig.~\ref{fig:EeKER_3} has been indicated by a distribution in its corresponding color. In the total yield we observe a slightly asymmetric monomodal distribution. As in the two-body channel, the wings on the distribution originate from false coincidences and background events that are challenging to completely eliminate in the analysis, causing a near-uniform background (clearly visible in Fig.~\ref{fig:Ee_3}) to underlie the spectrum, resulting in exaggerated wings.

\begin{figure}[h!]
    \includegraphics[width=8.5cm]{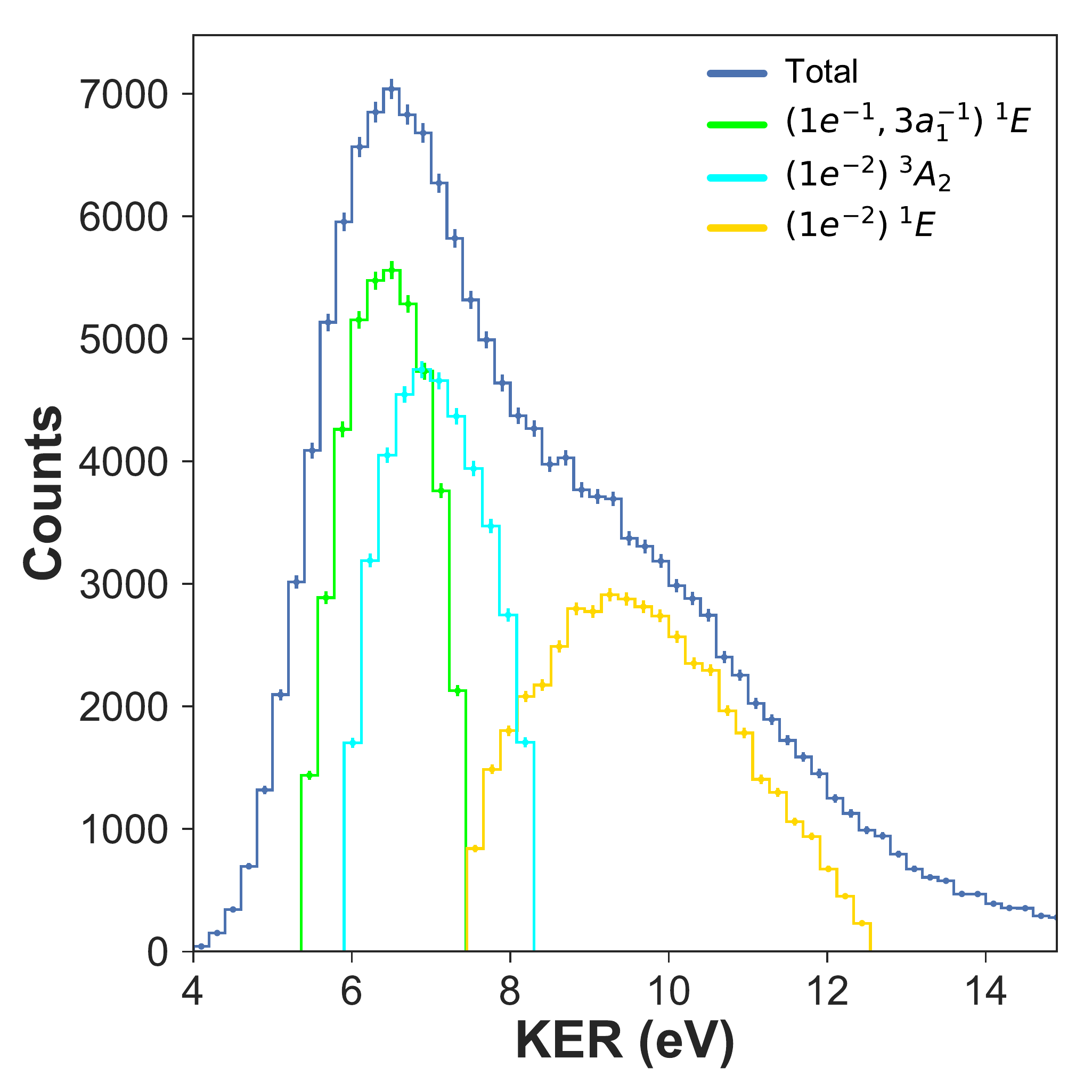}
\caption{The NH$^+$ + H$^+$ + H three-body yield, after PDI of NH$_3$ at 61.5~eV, as a function of the KER (shown in blue). The distributions for the three contributing dication states are shown in their respective color-codes (shown in green, cyan, and gold, and multiplied by a factor of 1.5 for improved visibility). Contributions from the individual dication states are extracted with gates as indicated in Fig.~\ref{fig:EeKER_3}, and hence their sum does not reflect the total (blue) yield (see text).}
\label{fig:KER_3}
\end{figure}

Next, we show the NH$^+$ + H$^+$ + H yield as a function of KER in Fig.~\ref{fig:KER_3}, following the color code of Fig.~\ref{fig:EeKER_3}. In the total yield we observe a broad asymmetric bimodal structure in the KER distribution. This bimodal distribution shows a rapid increase in yield on the low energy side of the peak and a slow decay in yield towards high KER, where the second mode is located. Both, the experimental and calculated photoelectron energy sums and KERs for each dication state are listed in Table~\ref{table:3body_EeKER}, showing good agreement for the ($1e^{-2}$) $^3A_2$ (cyan) and ($1e^{-2}$) $^1E$ (gold) dication states. In the case of the ($3a_1^{-1},1e^{-1}$) $^1E$ state (green), the agreement between experiment and theory is not as close, although the two values lie within the theoretically estimated FWHM. It is noteworthy that Fig.~\ref{fig:PEC_NH3}(b) indicates that the ($3a_1^{-1},1e^{-1}$) $^1E$ state possesses a barrier to dissociation when starting from the neutral equilibrium geometry, which is lifted when starting from geometries where the two N-H distances are compressed. This is consistent with the fact that for this state the calculated vertical photoelectron energy is greater than the measured value, whereas the calculated KER is smaller than the measured value. Consequently, PDI at geometries with contracted bond lengths are more likely to undergo three-body dissociation, as these excitations can directly fragment over the barrier. We point out that the difference in electron energy sum between the measurement and calculation is approximately the same as the difference in KER, which further supports this interpretation.

\begin{table*}
\centering
\begin{tabular}{  c  c  c  c  c  c  c  } 
 \hline\hline
 State & \multicolumn{2}{c}{Photoelectron Energy Sum (eV)} & \multicolumn{2}{c}{KER (eV)} & Branching Fraction & $\beta_2$ \\
  & Experiment & Theory$^{\rm{a}}$ & Experiment & Theory$^{\rm{a,b}}$ & & \\
 \hline
 ($3a_1^{-1},1e^{-1}$) $^1E$ (green) & 19.7 (2.5) & 21.0 (2.4) & 6.4 (1.1) & 5.2 (2.4) & 29\% $\pm$ 4\% & -0.27 $\pm$ 0.01 \\
 ($1e^{-2}$) $^3A_2$ (cyan) & 18.3 (2.3) & 18.1 (3.3) & 7.1 (1.4) & 7.7 (3.3) & 24\% $\pm$ 4\% & -0.17 $\pm$ 0.01 \\
 ($1e^{-2}$) $^1E$ (gold) & 16.8 (2.1) & 16.8 (3.4) & 9.4 (2.4) & 9.4 (3.4) & 47\% $\pm$ 4\% & 0.04 $\pm$ 0.01 \\
 \hline
\end{tabular}
\caption{The measured and calculated photoelectron energy sum and KER for each of the three active dication states leading to NH$^{+}$ + H$^{+}$ + H fragmentation following PDI of NH$_3$ at 61.5~eV, as well as the estimated branching fractions and $\beta_2$ anisotropy parameter (see text). $^{\rm{a}}$Theoretical FWHM values estimated from square of symmetric stretch vibrational wavefunction of NH$_3$ projected onto dication state  (see text). $^{\rm{b}}$Theoretical KER values are calculated assuming ro-vibrational ground state fragments, i.e. assuming maximum KER with no energy channeled into internal excitations.}
\label{table:3body_EeKER}
\end{table*}

The three contributing dication states were also identified using MRCI calculations, as highlighted in the theory section, and are consistently color-coded throughout the paper, where the green state corresponds with the ($3a_1^{-1},1e^{-1}$) $^1E$ state (the same as in the two-body breakup section), the cyan state corresponds with the ($1e^{-2}$) $^3A_2$ state, and the gold state with the ($1e^{-2}$) $^1E$ state. The ion yield measurements estimate the branching ratios for these three dication states, listed in Table~\ref{table:3body_EeKER}, to be approximately $29\% \pm 4\%$ for the ($3a_1^{-1},1e^{-1}$) $^1E$ state, $24\% \pm 4\%$ for the ($1e^{-2}$) $^3A_2$ state, and $47\% \pm 4\%$ for the ($1e^{-2}$) $^1E$ dication state. These branching ratios are derived in the same manner as described in the two-body breakup section. As before, the main contribution to the uncertainty of the branching ratio is rooted in the aforementioned electron pair dead-time, which influences the detection yield of the electron-ion coincidences for each dication state as a function of the electron sum energy. Applying the simulation mentioned above, we estimate the total possible loss in PDI yield for electron sum energies of 19.7~eV (($3a_1^{-1},1e^{-1}$) $^1E$), 18.3~eV (($1e^{-2}$) $^3A_2$), and 16.8~eV (($1e^{-2}$) $^1E$) to be 6.5\%, 7.1\%, and 7.9\%, respectively. This translates to an error of up to 4\% in the branching ratio. Errors due to deviations from the assumed Gaussian shape of each feature in the fitting process and the quality of the fit are estimated to be small (both $<$1\%). 

The energetics observed in the PES cuts of Fig.~\ref{fig:PEC_NH3}(b), Fig.~\ref{fig:EeKER_3}, and Table~\ref{table:3body_EeKER} indicate that the ($3a_1^{-1},1e^{-1}$) $^1E$ dication state dissociates to the NH$^+(^2\Pi)$ + H$^+$ + H($^2S$) limit, with the ($1e^{-2}$) $^3A_2$ and ($1e^{-2}$) $^1E$ states dissociating to this very same limit. This finding suggests that following PDI of NH$_3$ to either of these $E$ symmetry dication states, i.e. the ($3a_1^{-1},1e^{-1}$) $^1E$ and ($1e^{-2}$) $^1E$ states, three-body dissociation ensues on the $A'$ symmetry PESs, shown in Fig.~\ref{fig:PEC_NH3}(b). In the case of the ($3a_1^{-1},1e^{-1}$) $^1E$ dication state, the $A''$ symmetry curve possesses a large barrier to dissociation, whereas fragmentation on the $A'$ symmetry curve is much more favorable. As for the ($1e^{-2}$) $^1E$ state, we do not observe fragmentation on the $A''$ symmetry curve, although the PES cuts shown in Fig.~\ref{fig:PEC_NH3}(b) indicate that the excitation lies above the shallow barrier of this curve. However, similar to the case of the ($3a_1^{-1},1e^{-1}$) $^1E$ dication state the fragmentation on the $A'$ symmetry curve is favored in the ($1e^{-2}$) $^1E$ state. Populating the ($1e^{-2}$) $^3A_2$ dication state results in direct three-body fragmentation on the PES (which is not doubly degenerate in the FC region like the $E$ symmetry states), reaching the NH$^+(^2\Pi)$ + H$^+$ + H($^2S$) limit. We point out that at large internuclear separations ($>$18~Bohr) a charge-exchange mechanism was observed between the NH$^+$ and H fragments in this dication state, which produces the fragments NH + H$^+$ + H$^+$ and is discussed in detail in [I]. An analogous asymptotic electron transfer mechanism has also been observed in dissociative electron attachment to NH$_3$ \cite{Rescigno16}.

\begin{figure}[h!]
    \includegraphics[width=4.225cm, trim=0.4cm 0.6cm 0.5cm 0.45cm, clip]{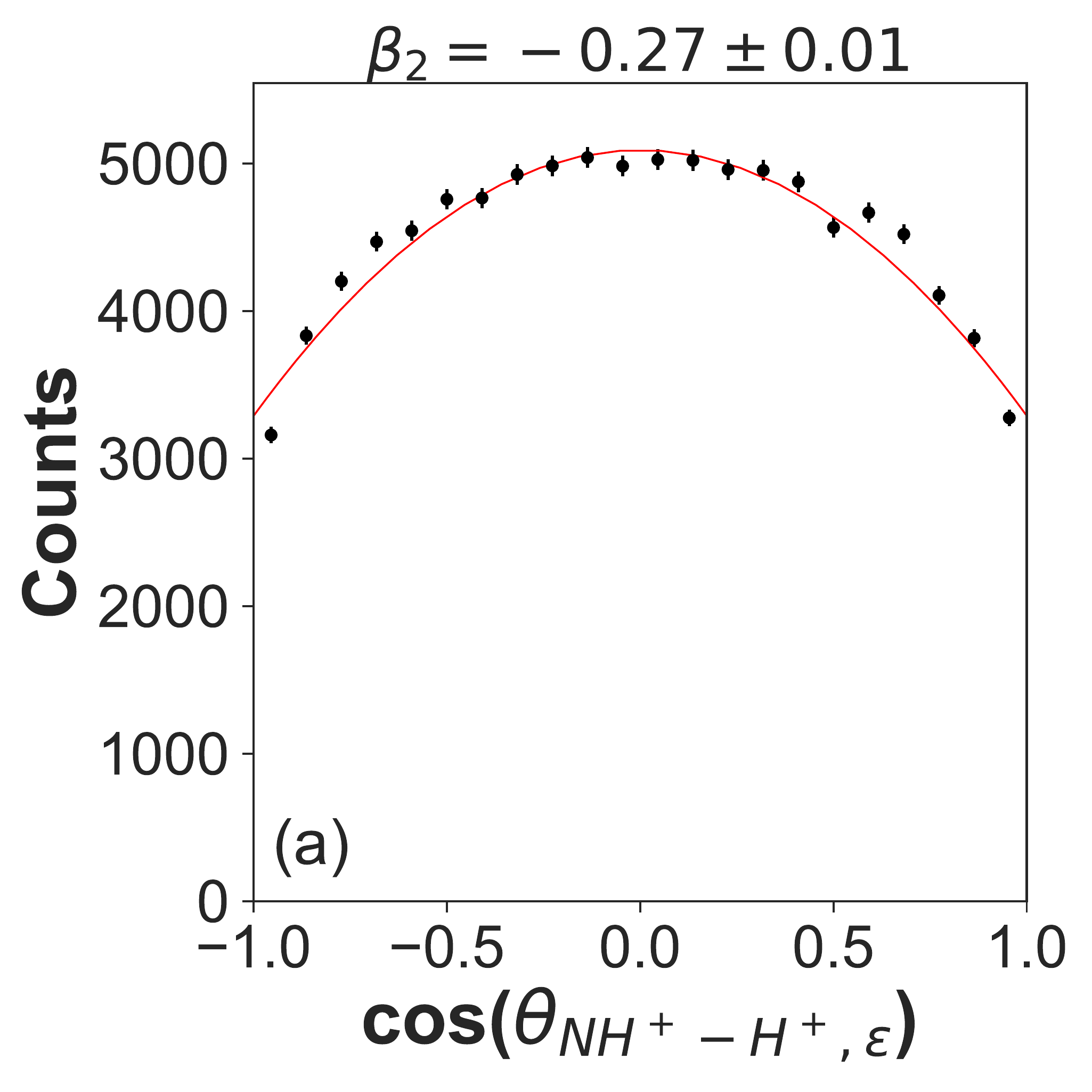}
    \includegraphics[width=4.225cm, trim=0.4cm 0.6cm 0.5cm 0.45cm, clip]{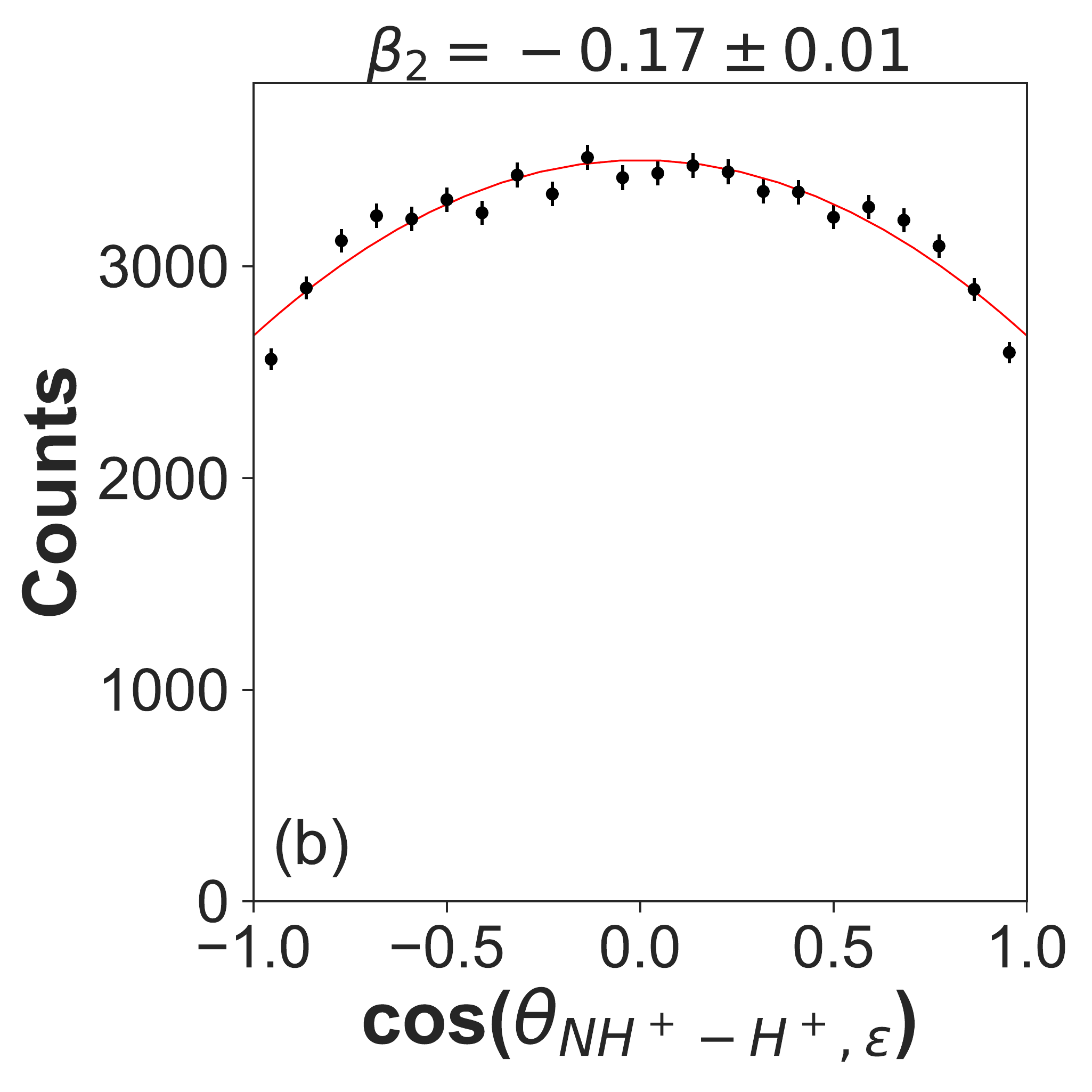}
    \includegraphics[width=4.225cm, trim=0.4cm 0.6cm 0.5cm 0.45cm, clip]{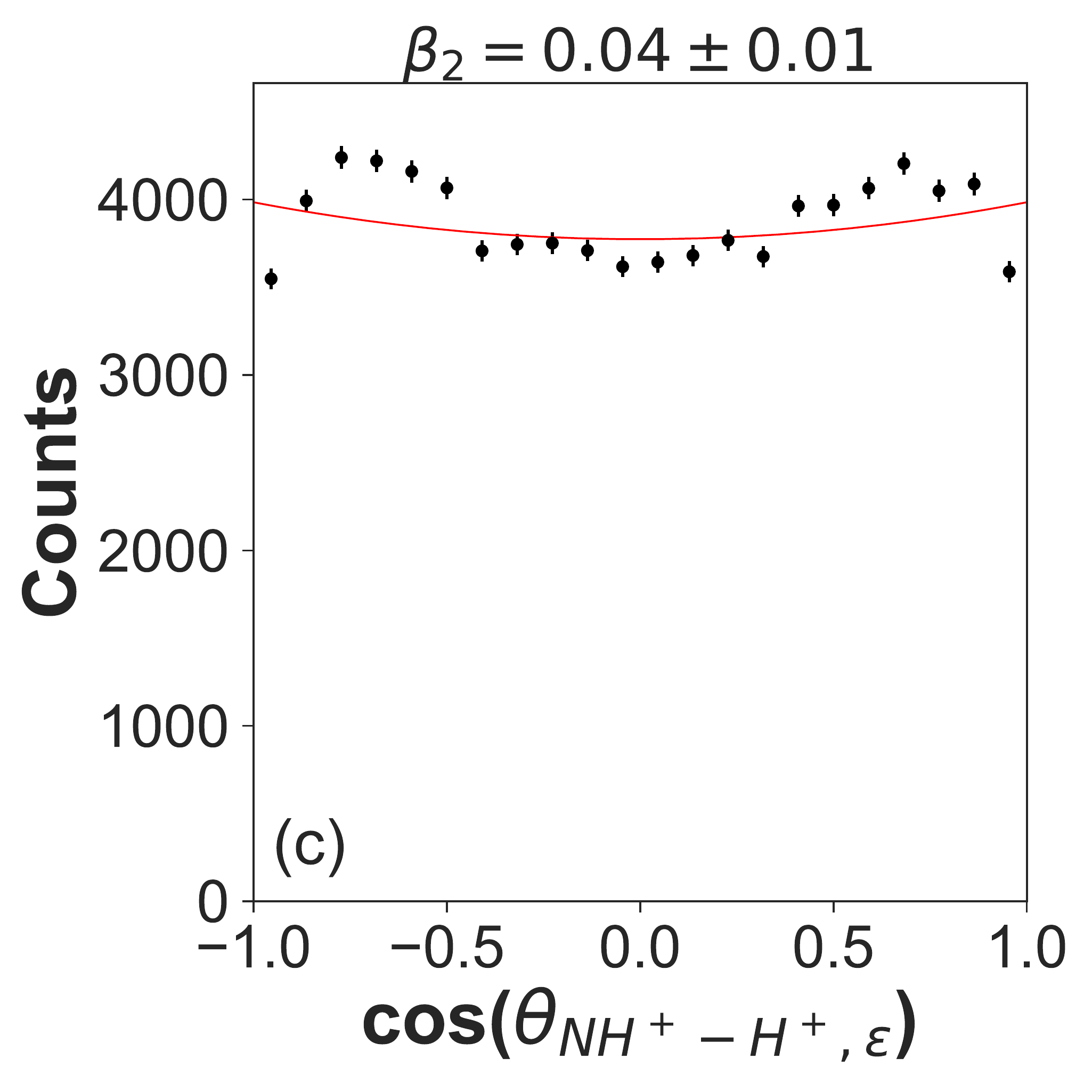}
\caption{The NH$^+$ + H$^+$ + H three-body fragmentation yield, after PDI of NH$_3$ at 61.5~eV, as a function of the cosine of the measured relative angle between the NH$^+$-H$^+$ recoil axis and the XUV polarization vector $\varepsilon$ for the (a) ($3a_1^{-1},1e^{-1}$) $^1E$, (b) the ($1e^{-2}$) $^3A_2$, and (c) the ($1e^{-2}$) $^1E$ dication states. The fits, representing the parametrizations in terms of the anisotropy (see Eq.~\ref{AngDiff_PICS}), are shown in red, where the retrieved $\beta_2$ value is shown above each plot.}
\label{fig:recoil_angle_3body}
\end{figure}

\begin{figure}[h!]
    \includegraphics[width=4.225cm, trim=0.4cm 0.6cm 0.5cm 0.2cm, clip]{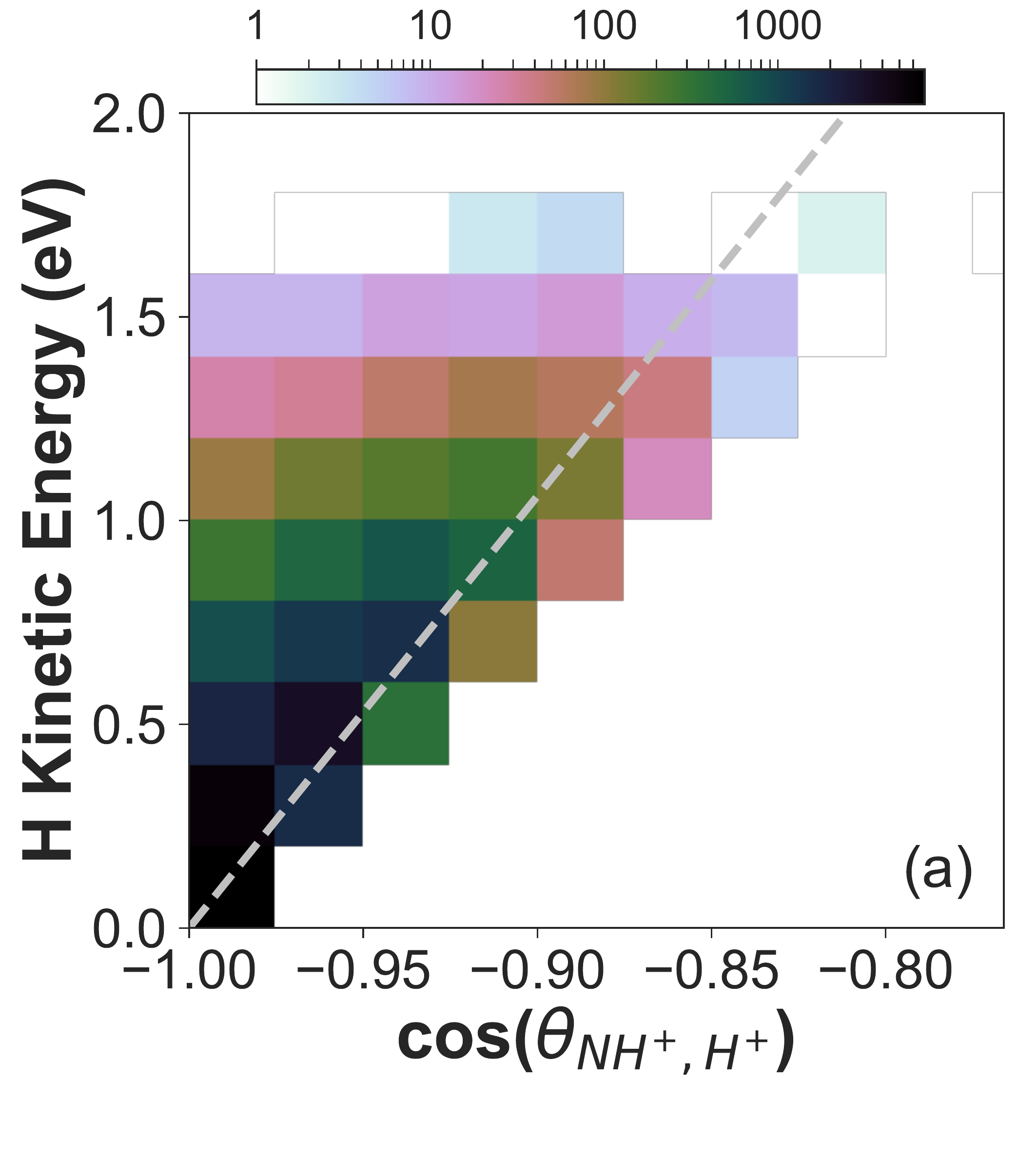}
    \includegraphics[width=4.225cm, trim=0.4cm 0.6cm 0.5cm 0.2cm, clip]{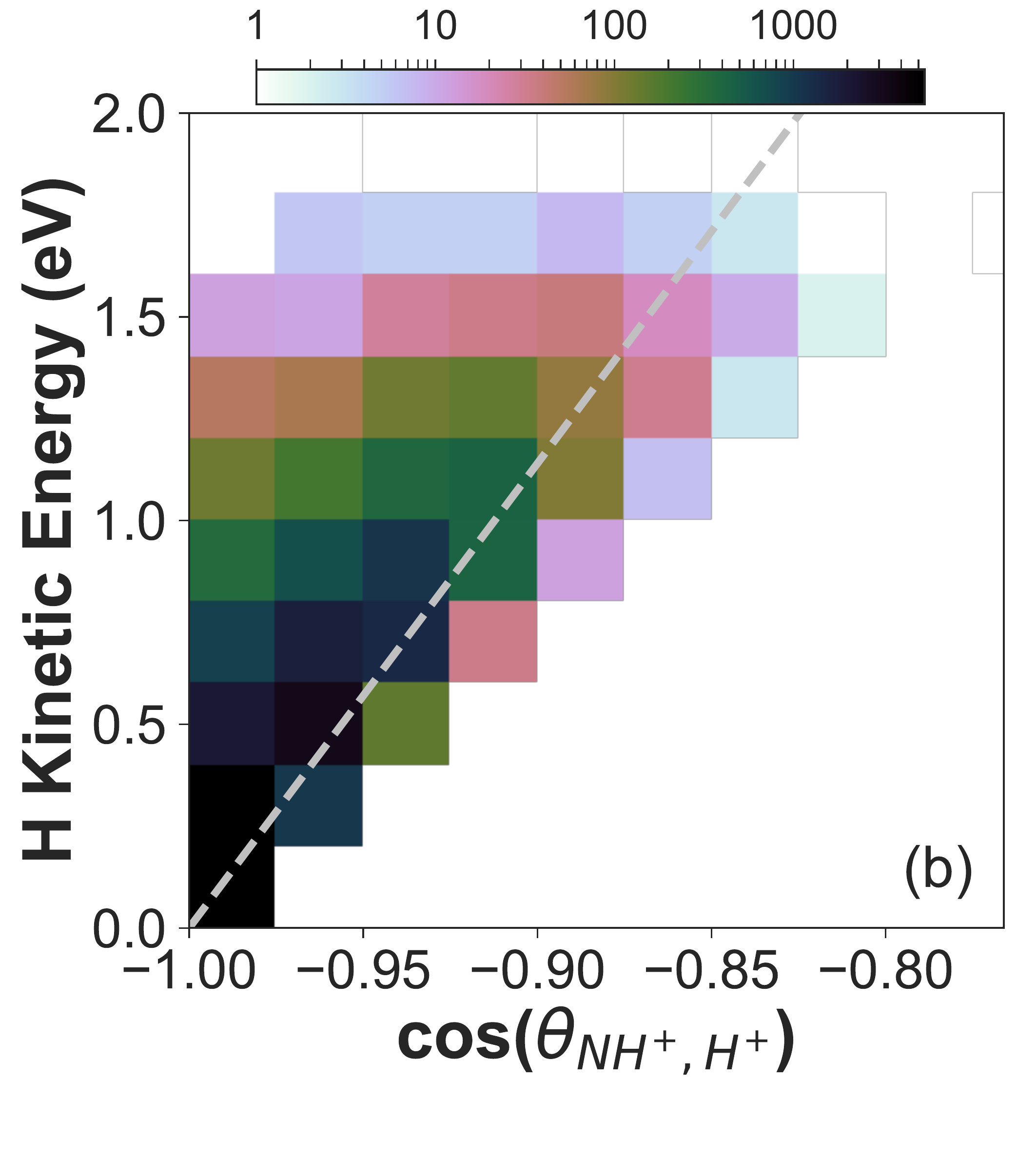}
    \includegraphics[width=4.225cm, trim=0.4cm 0.6cm 0.5cm 0.2cm, clip]{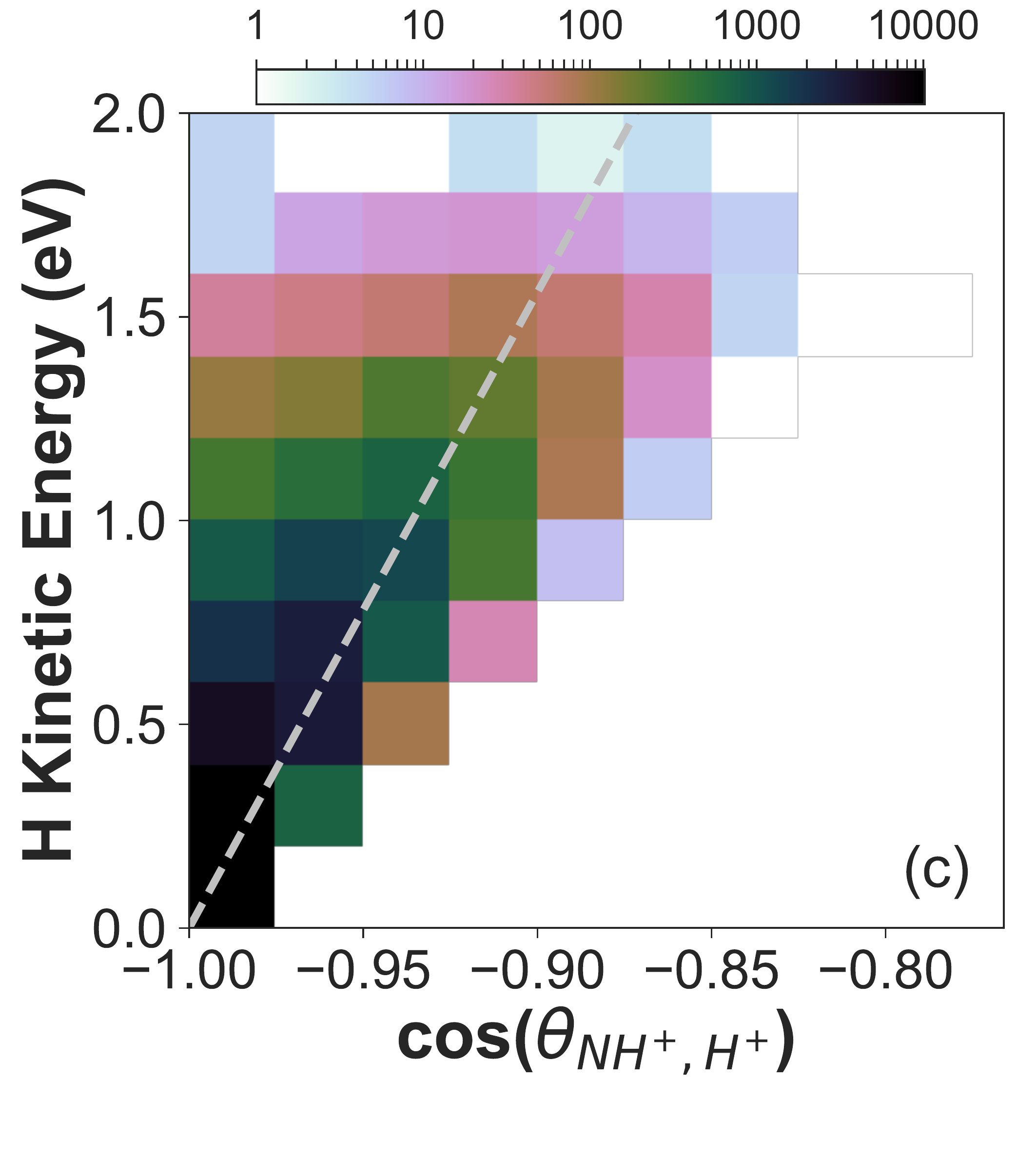}
\caption{The NH$^+$ + H$^+$ + H three-body fragmentation yield after PDI of NH$_3$ at 61.5~eV as a function of the cosine of the measured relative angle between the NH$^+$ and H$^+$ fragment momentum vectors and the kinetic energy of the neutral H fragment, for the (a) the ($3a_1^{-1},1e^{-1}$) $^1E$ , (b) the ($1e^{-2}$) $^3A_2$, and (c) the ($1e^{-2}$) $^1E$ dication states. The dashed silver line in each figure serves to guide to the eye towards the slope of the feature.}
\label{fig:ion_angle}
\end{figure}

To determine if there are preferred molecular orientations where PDI of NH$_3$ occurs for each of the three dication states, we plot in Fig.~\ref{fig:recoil_angle_3body}(a), (b), and (c) the yield of the NH$^+$ + H$^+$ + H three-body fragmentation as a function of the cosine of the relative angle between the recoil axis of the charged fragments of the molecular breakup (NH$^+$-H$^+$) and the XUV polarization ($\varepsilon$). While in a three-body breakup all particles carry away momentum, we can deduce from the PIPICO spectrum, presented in Fig.~\ref{fig:PIPICO}, that for the most part the charged fragments solely compensate their momenta, while the third particle (the neutral H atom) takes on a spectator role. This becomes apparent by examining the TOF correlation of the NH$^+$ + H$^+$ + H channel, which is almost as wide in the TOF difference and as narrow in the TOF sum as the NH$_2^+$ + H$^+$ two-body fragmentation. This underlines that the charged fragments of the three-body breakup repel each other and interact with each other via the Coulomb force over a long range of internuclear distances. Thus, their recoil axis, which is calculated via the difference of the measured momenta, appears to be an appropriate choice for a distinguished axis. We will see later that the relative angle between the charged fragments almost exclusively peaks at 180$^{\circ}$ and the neutral H fragment carries away rather little kinetic energy (Fig.~\ref{fig:ion_angle}), further supporting this selection.

Eq.~\ref{AngDiff_PICS} is valid for any vectorial quantity arising from single-photon ionization of a randomly oriented sample by linearly polarized light. Consequently it can be applied to the recoil axis of the charged fragments H$^+$ + NH$^+$ of the three-body breakup channel as well. For the ($3a_1^{-1},1e^{-1}$) $^1E$ and ($1e^{-2}$) $^3A_2$ dication states, we observe an enhancement in PDI for molecular orientations where the NH$^+$ + H$^+$ recoil axis is orientated at a $\sim90^{\circ}$ angle with respect to the polarization vector. As in the two-body breakup case, this roughly coincides with the C$_{3v}$ symmetry axis of the molecule. We find a $\beta_2$ value of $-0.27\pm0.01$. In the ($3a_1^{-1},1e^{-1}$) $^1E$ dication state, the PDI involves the $3a_1$ orbital, which is aligned along the molecular C$_{3v}$ axis. This could explain the enhancement in PDI at geometries where the polarization vector of the ionizing field is directed along this orbital, and why this enhancement is reduced in the ($1e^{-2}$) $^3A_2$, where the ionization no longer involves the $3a_1$ orbital. The $\beta_2$ value was determined to be $-0.17\pm0.01$. The anisotropy parameter for both states are also listed in Table~\ref{table:3body_EeKER}.
 
In contrast, the ($1e^{-2}$) $^1E$ dication state appears to exhibit a small enhancement in PDI for orientations where the recoil axis of the charged NH$^+$ and H$^+$ fragments is orientated at a $\sim35^{\circ}$ and $\sim145^{\circ}$ angle with the polarization vector [see Fig.~\ref{fig:recoil_angle_3body}(c)]. The angular distribution is nearly isotropic, appearing almost flat, meaning that compared to the other two dication states of the three-body breakup, we observe a lower likelihood of PDI where the recoil axis of the charged fragments NH$^+$ and H$^+$ is orientated perpendicular to the polarization vector in the ($1e^{-2}$) $^1E$ state. Here, since the PDI of NH$_3$ no longer involves the $3a_1$ orbital, enhancement near the C$_{3v}$ axis is suppressed. Because the PDI only involves the $1e$ orbital, which lies along the N-H bonds of the molecule, we observe an enhancement in PDI for molecular orientations near angles where the recoil axis of the charged fragments NH$^+$ and H$^+$ lies along the XUV polarization vector. We point out that this interpretation is qualitative and does not explain the aforementioned differences between the ($1e^{-2}$) $^3A_2$ and ($1e^{-2}$) $^1E$ states, which both invoke ionization from solely the $1e$ orbital. We find a $\beta_2$ value of $0.04\pm0.01$.

The angular distribution of Fig.~\ref{fig:recoil_angle_3body}(c) may be influenced by a breakdown of the axial recoil approximation during the dissociation \cite{Rakitzis}, which can occur for select dication states and has been identified in the PDI of water molecules for a similar photon energy recently \cite{Streeter}. Given the purely statistical error bars, the measured angular distribution for the ($1e^{-2}$) $^1E$ state is somewhat inconsistent with the $\beta_2$ functional form, while for the ($3a_1^{-1},1e^{-1}$) $^1E$ and ($1e^{-2}$) $^3A_2$ states, the angular distributions can be accurately fitted by  Eq.~\ref{AngDiff_PICS}. The inconsistency (Fig.~\ref{fig:recoil_angle_3body}(c)) is rather insensitive to fine details of the momentum calibration, and does not appear to be subject to multi-hit problems of the ion and electron detectors, all of which were inspected thoroughly. We could not identify any problems in these analysis domains that could explain the small deviation of the measured angular distribution in Fig.~\ref{fig:recoil_angle_3body}(c) from the flat or shallow parabolic form of Eq.~\ref{AngDiff_PICS}, but we need to point out that the KER and electron sum energy distributions for the weak ($1e^{-2}$) $^1E$ dication state are quite broad and may overlap with the distributions of the ($1e^{-2}$) $^3A_2$ dication state, as apparent in Fig.~\ref{fig:EeKER_3}.

The diffuse features corresponding to the weak ($1e^{-2}$) $^1E$ dication state may also contain background contributions from false coincidences with parent NH$_3^+$ ions and background H$^+$ as well as H$_2^+$ ions (observed as the horizontal and vertical features in Fig.~\ref{fig:PIPICO}). Moreover, a small number of events from the NH$_2^+$ + H$^+$ two-body breakup channel, which lies adjacent to the NH$^+$ + H$^+$ + H channel in the PIPICO spectrum (seen in Fig.~\ref{fig:PIPICO}), can be falsely assigned and may contribute to the two small peaks. The spread in TOF of the NH$_2^+$ + H$^+$ coincidence feature in Fig.~\ref{fig:PIPICO} may result in a false assignment of some NH$_2^+$ + H$^+$ fragment pairs to any of the 3-body channels. Both of the latter two sources of background, although largely eliminated in the calibration and analysis, can be challenging to completely remove in some cases, which can result in select ionization channels being contaminated by a few percent of erroneously assigned events. Since the two small peaks in Fig.~\ref{fig:recoil_angle_3body}(c) lie only a few ($\sim 5\% \pm 1\%$) percent above the isotropic distribution, we speculate that these features arise from any of the three forms of pollution mentioned above, and do not point towards some unusual dissociation mechanism or previously unobserved photodissociation dynamics of the ($1e^{-2}$) $^1E$ dication state of NH$_3$.

In Fig.~\ref{fig:ion_angle} we plot on a logarithmic scale the NH$^+$ + H$^+$ + H three-body fragmentation yield, following the PDI of NH$_3$ by 61.5~eV photons, as a function of the cosine of the measured relative angle between the NH$^+$ and H$^+$ ion momenta, and the kinetic energy of the neutral H fragment. The triangular shape of the distribution is governed by momentum conservation of the three fragments, and we use it to elucidate the interaction between the heavy particles during the dissociation as a function of the NH$_3^{2+}$ state. All three distributions are peaked at 180$^\circ$, indicating a preferred back-to-back emission between the photoions with the neutral H largely spectating. However, as the H fragment receives more kinetic energy from the dissociation, the relative angle between the two NH$^+$ and H$^+$ ionic fragments correspondingly opens up, which becomes nicely apparent in this 2D-spectrum. We observe that the range of angles spanned between the two photoions, following the PDI to the ($3a_1^{-1},1e^{-1}$) $^1E$ and ($1e^{-2}$) $^3A_2$ dication states [Fig.~\ref{fig:ion_angle}(a) and (b)], is broader than in the ($1e^{-2}$) $^1E$ state [Fig.~\ref{fig:ion_angle}(c)]. The range of kinetic energies spanned by the neutral H fragment is broader in the ($1e^{-2}$) $^1E$ dication state [Fig.~\ref{fig:ion_angle}(c)] compared with the ($3a_1^{-1},1e^{-1}$) $^1E$ and ($1e^{-2}$) $^3A_2$ dication states [Fig.~\ref{fig:ion_angle}(a) and (b)]. We also observe a difference in the correlation between the kinetic energy of the neutral H and the measured relative angle between the NH$^+$ and H$^+$ photoions. The dashed silver lines in Figs.~\ref{fig:ion_angle}(a), (b), and (c) are intended to guide the eye towards the slope of the features and improve the visibility of the energy-angle correlations. In the ($3a_1^{-1},1e^{-1}$) $^1E$ and ($1e^{-2}$) $^3A_2$ dication states [Fig.~\ref{fig:ion_angle}(a) and (b)] the angle between the NH$^+$ and H$^+$ fragments opens up more rapidly than in the ($1e^{-2}$) $^1E$ state [Fig.~\ref{fig:ion_angle}(c)], as the neutral H fragment takes away more kinetic energy. This suggests that the neutral H fragment acts more like a passive spectator in the dissociation that ensues following PDI to the ($1e^{-2}$) $^1E$ dication state as compared to the dissociation from the ($3a_1^{-1},1e^{-1}$) $^1E$ and ($1e^{-2}$) $^3A_2$ dication states, which show a stronger influence of the kinetic energy of the neutral H fragment on the relative angle between the photoions.

\begin{figure}[h!]
    \includegraphics[width=4.225cm, trim=0.4cm 0.6cm 0.5cm 0.65cm, clip]{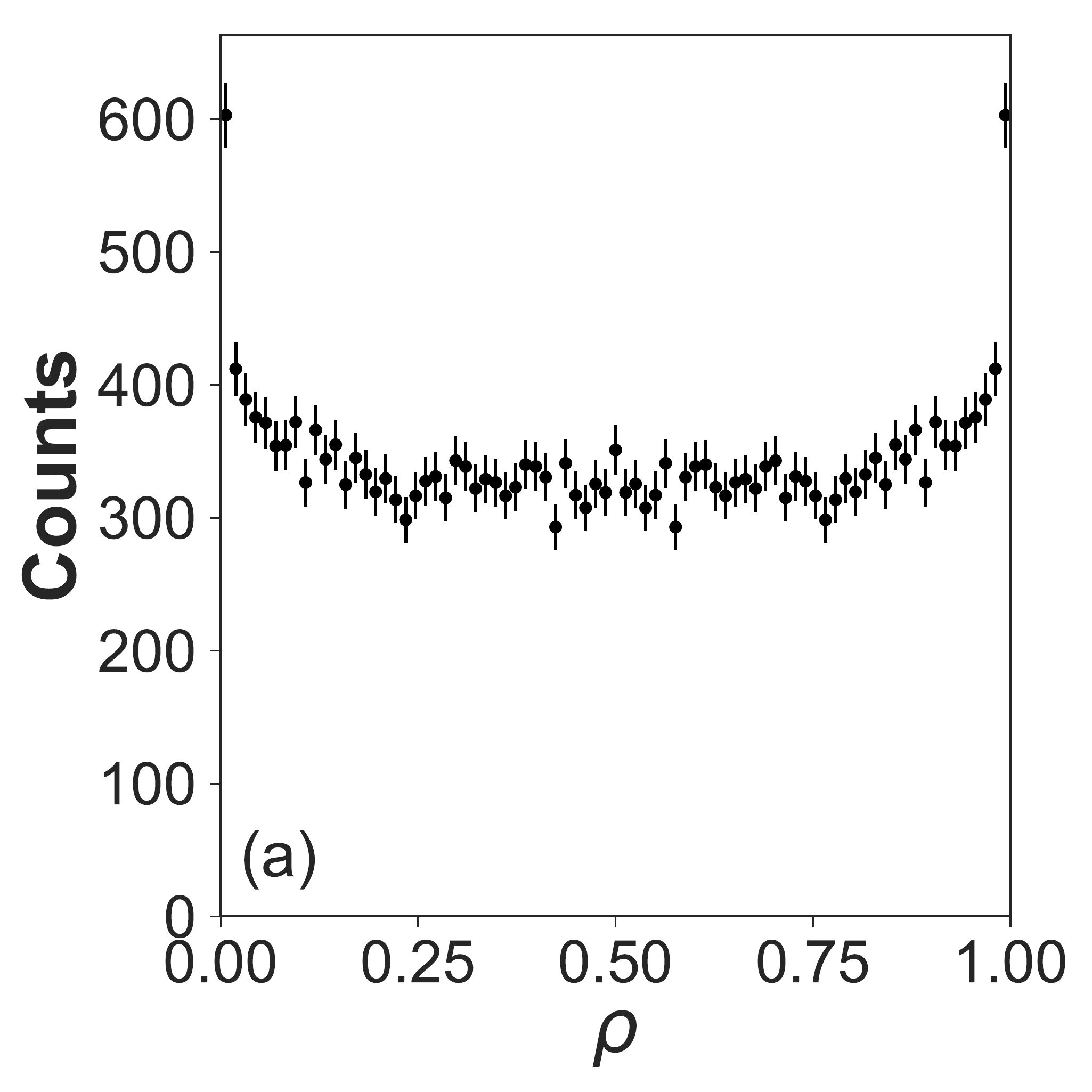}
    \includegraphics[width=4.225cm, trim=0.4cm 0.6cm 0.5cm 0.65cm, clip]{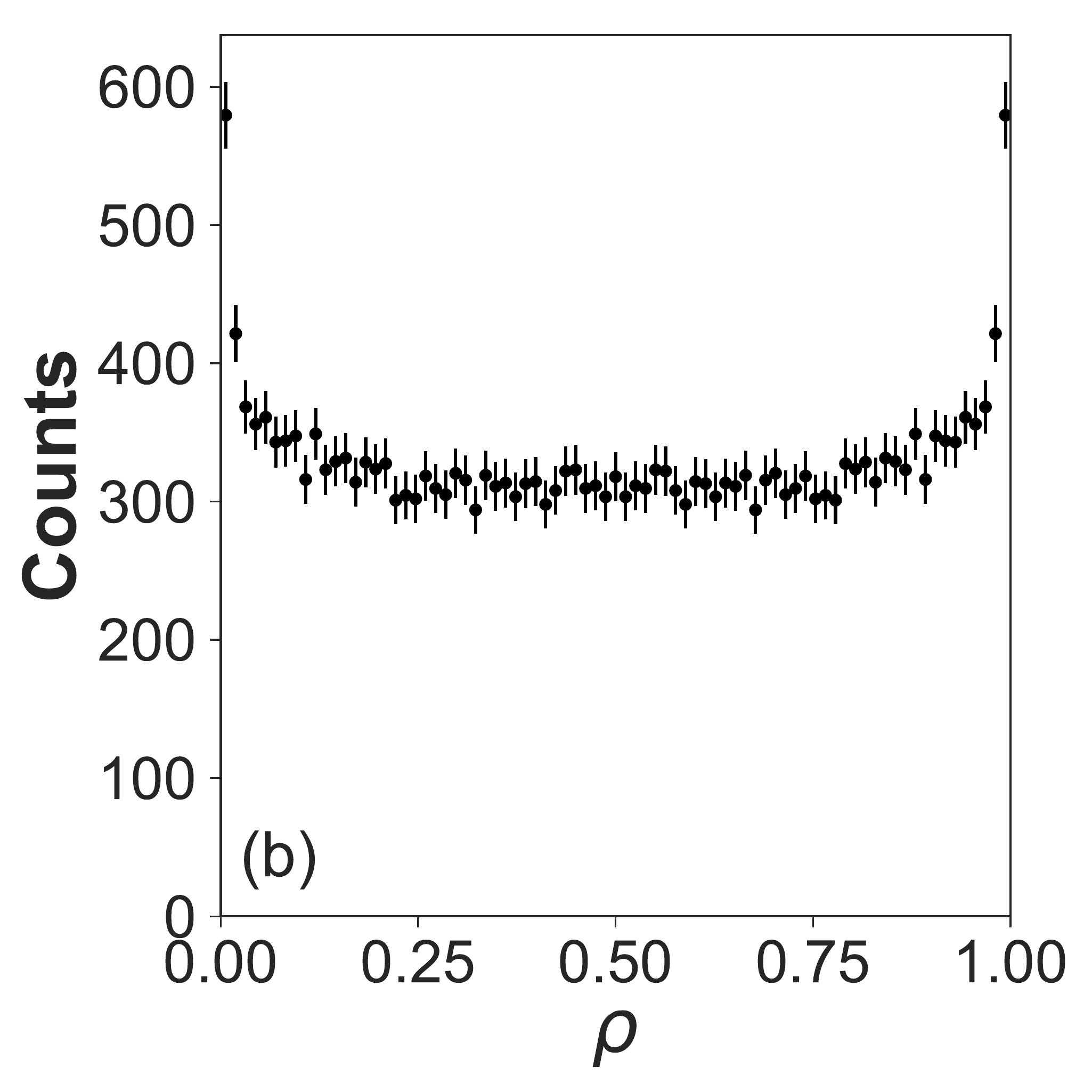}
    \includegraphics[width=4.225cm, trim=0.4cm 0.6cm 0.5cm 0.65cm, clip]{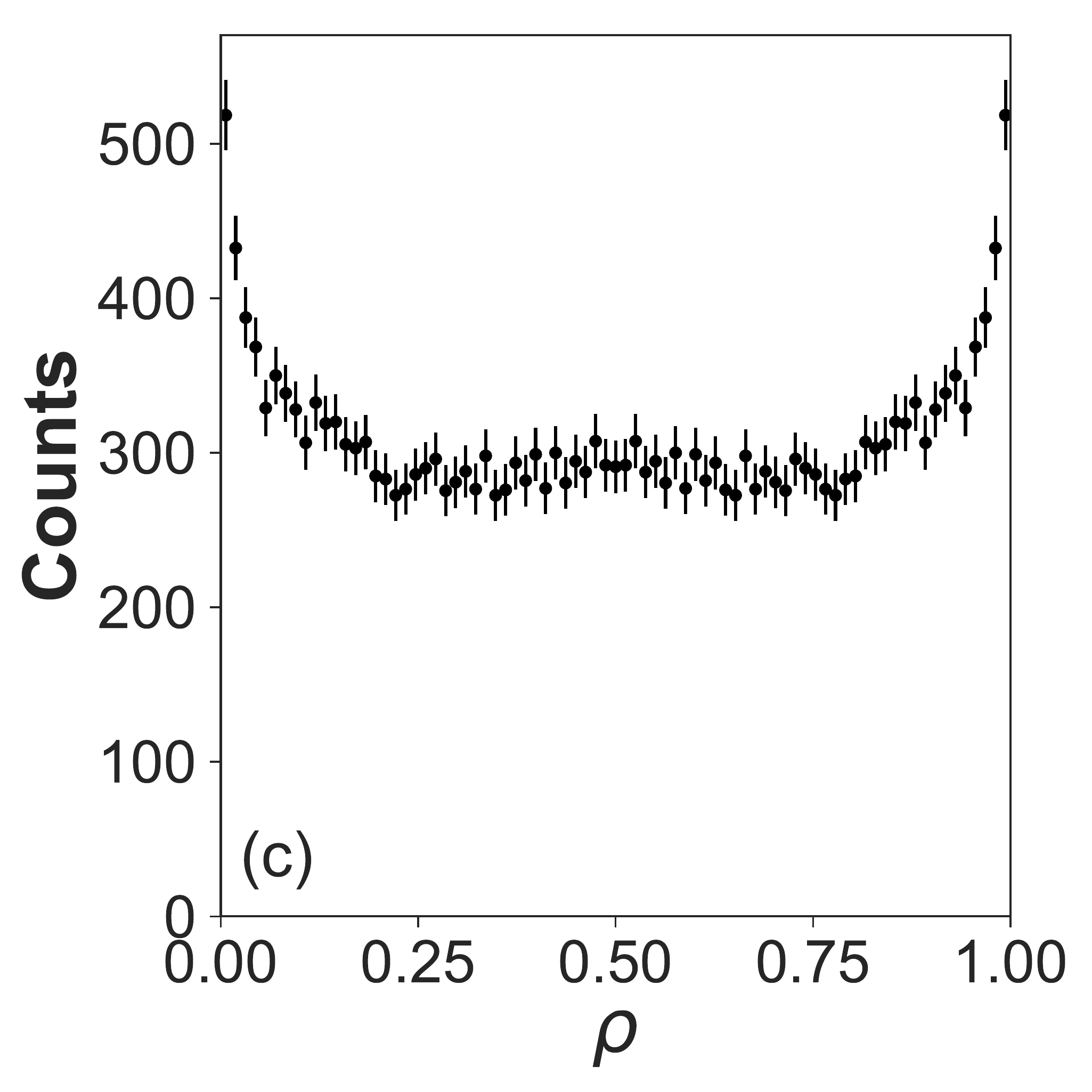}
\caption{The NH$^+$ + H$^+$ + H three-body fragmentation yield, after PDI of NH$_3$ at 61.5~eV, as a function of the electron energy sharing $\rho$ (Eq.~\ref{Eng-sharing}) for the (a) ($3a_1^{-1}, 1e^{-1}$) $^1E$, (b) the ($1e^{-2}$) $^3A_2$, and (c) the ($1e^{-2}$) $^1E$ dication states.}
\label{fig:SDCS_3body}
\end{figure}

Next, we display the NH$^+$ + H$^+$ + H three-body fragmentation yield as a function of the energy sharing between the two photoelectrons for the three features that correspond with the three dication states. These results are shown in Fig.~\ref{fig:SDCS_3body}(a), (b), and (c). As before, we attribute the sharp features near 0 and 1, observed in Fig.~\ref{fig:SDCS_3body}(a), (b), and (c), to an autoionization process, corresponding with a fast photoelectron and slow electron emerging from autoionization. The fraction of PDI via autoionization relative to direct PDI is $\sim$5.6$\% \pm 2.0\%$ in the ($3a_1^{-1},1e^{-1}$) $^1E$ state, $\sim$7.6$\% \pm 1.1\%$ in the ($1e^{-2}$) $^3A_2$ state, and $\sim$7.7$\% \pm 2.1\%$ in the ($1e^{-2}$) $^1E$ dication state. This fraction is estimated using the same protocol described in the two-body breakup section.

\begin{figure}[h!]
    \includegraphics[width=4.225cm, trim=0.4cm 0.6cm 0.5cm 0.6cm, clip]{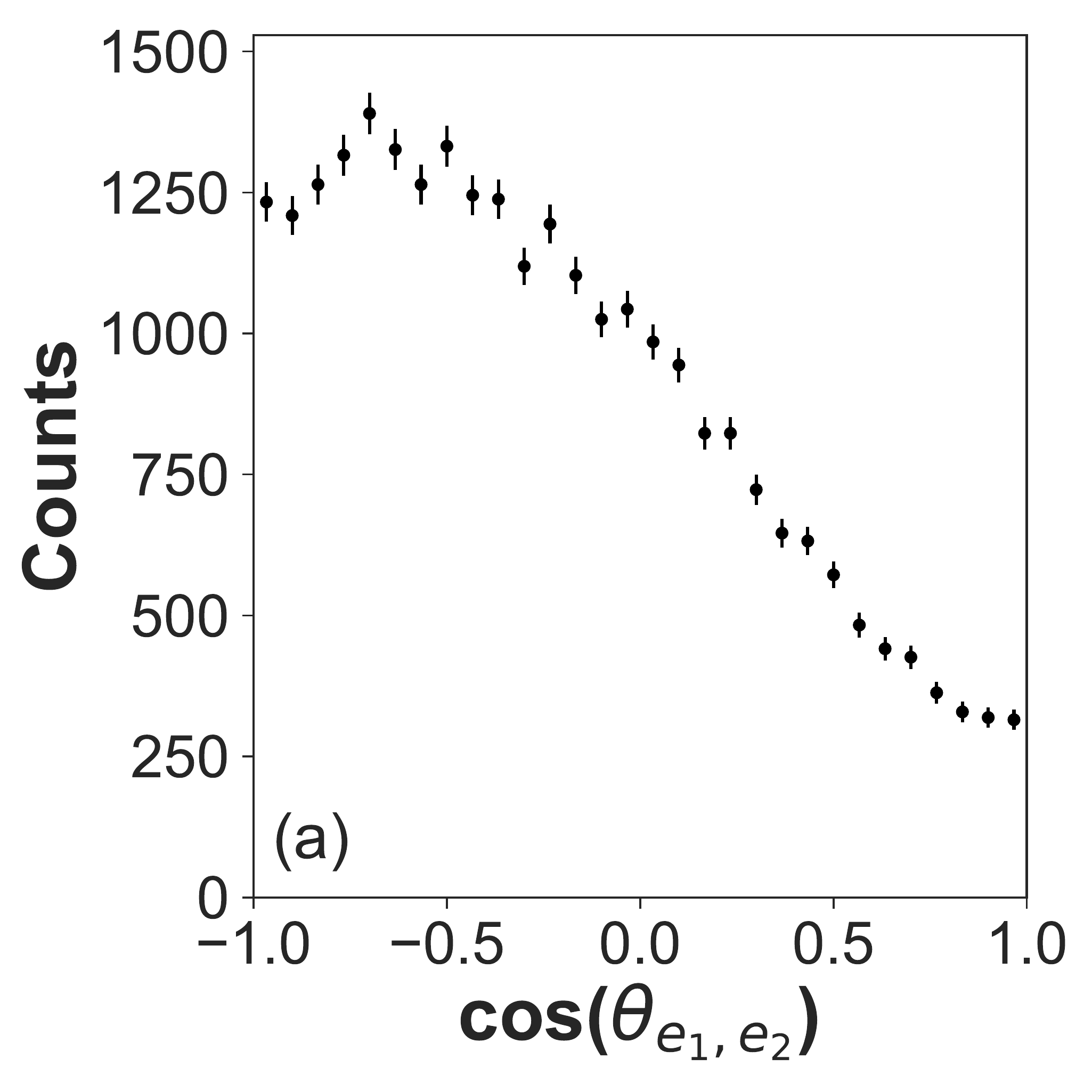}
    \includegraphics[width=4.225cm, trim=0.4cm 0.6cm 0.5cm 0.6cm, clip]{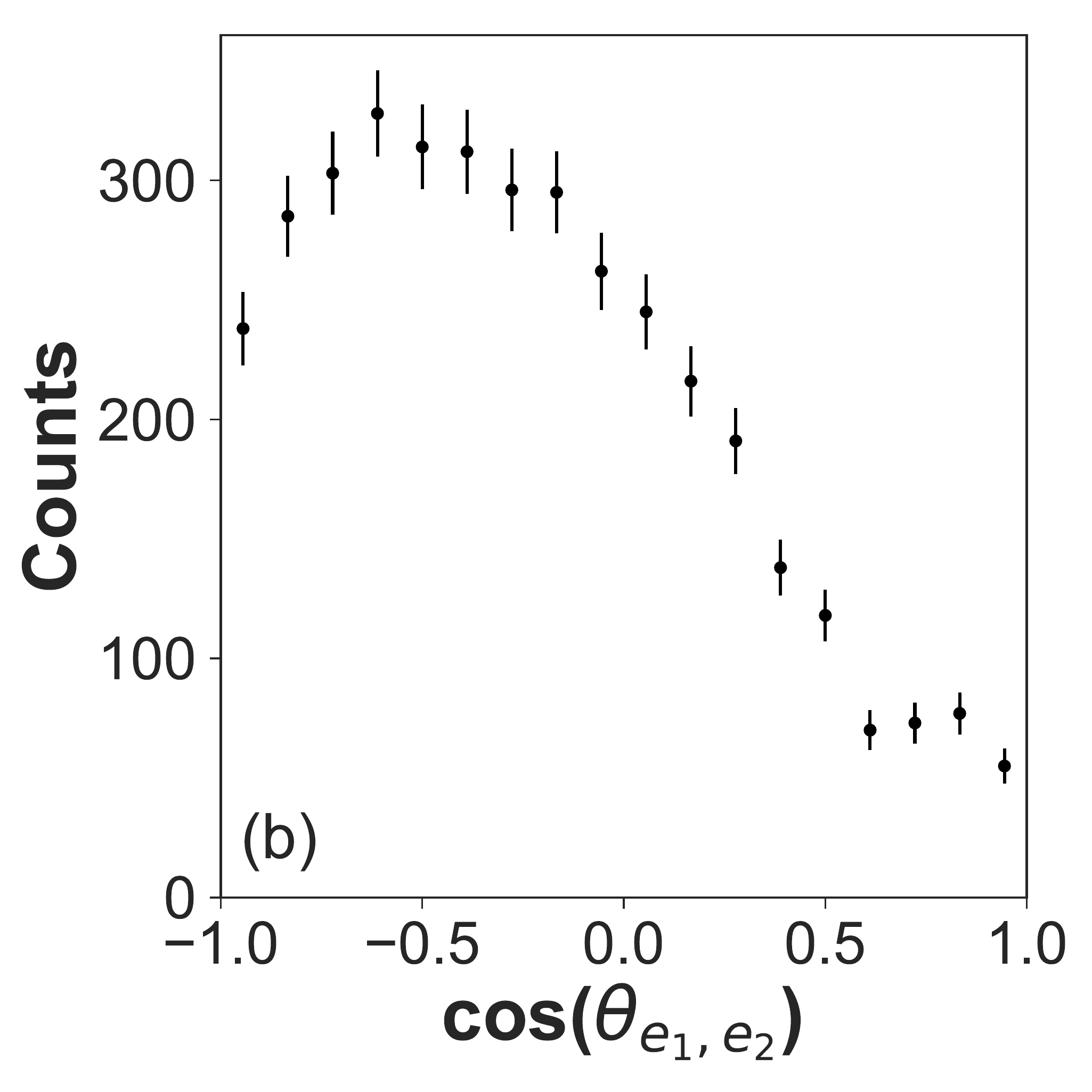}
\caption{The NH$^+$ + H$^+$ + H three-body fragmentation yield, after PDI of NH$_3$ at 61.5~eV, as a function of cosine of the relative emission angle between the two photoelectrons for the ($3a_1^{-1},1e^{-1}$) $^1E$ dication state, (a) integrated over all possible electron energy sharing and (b) for equal energy sharing ($\rho = 0.5 \pm 0.075$).}
\label{fig:B_angle_e_3}
\end{figure}

\begin{figure}[h!]
    \includegraphics[width=4.225cm, trim=0.4cm 0.6cm 0.5cm 0.25cm, clip]{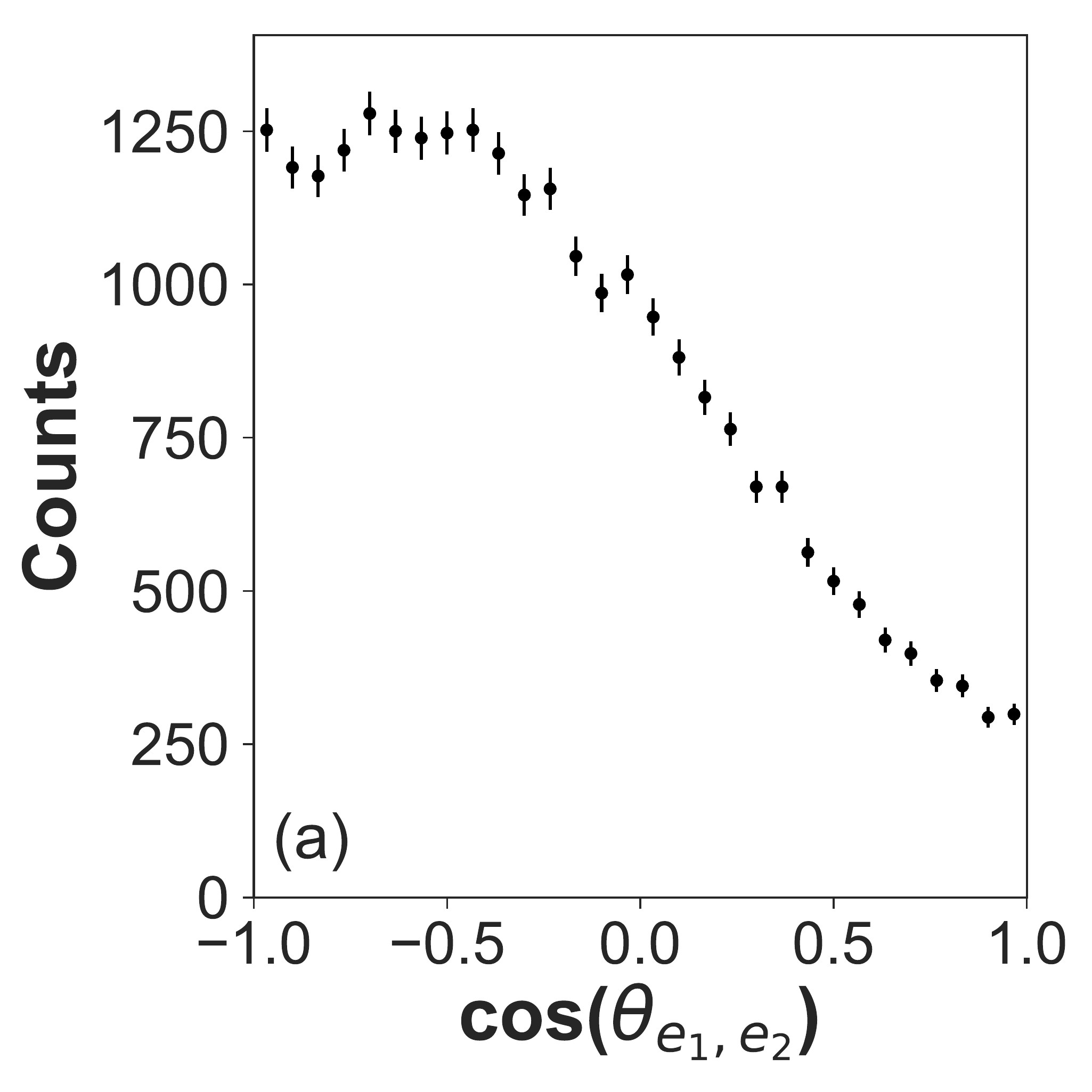}
    \includegraphics[width=4.225cm, trim=0.4cm 0.6cm 0.5cm 0.25cm, clip]{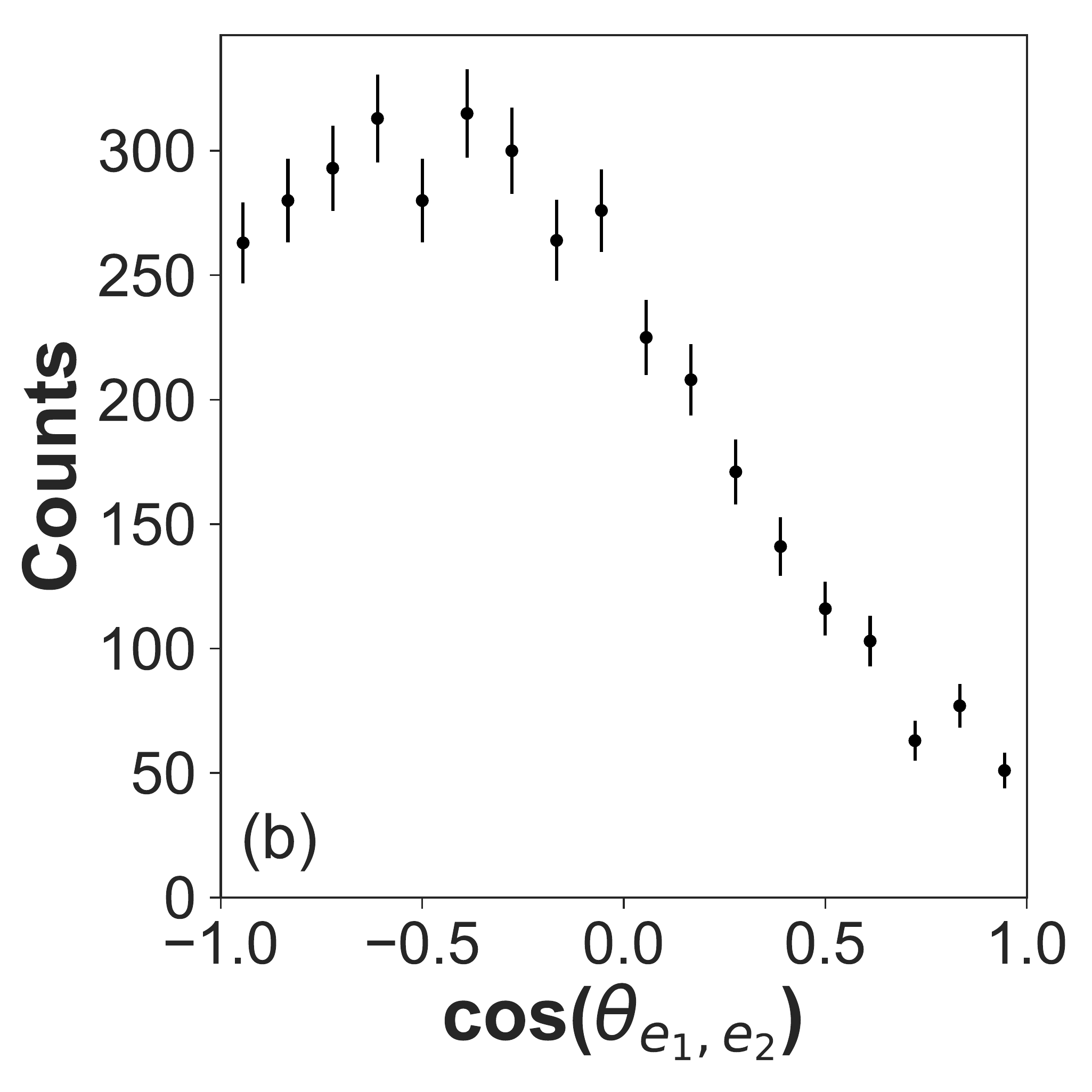}
\caption{The NH$^+$ + H$^+$ + H three-body fragmentation yield, after PDI of NH$_3$ at 61.5~eV, as a function of the cosine of the relative emission angle between the two photoelectrons in two different energy sharing conditions for the ($1e^{-2}$) $^3A_2$ dication state, (a) integrated over all possible electron energy sharing and (b) for equal energy sharing ($\rho = 0.5 \pm 0.075$).}
\label{fig:D_angle_e}
\end{figure}

\begin{figure}[h!]
    \includegraphics[width=4.225cm, trim=0.4cm 0.6cm 0.5cm 0.25cm, clip]{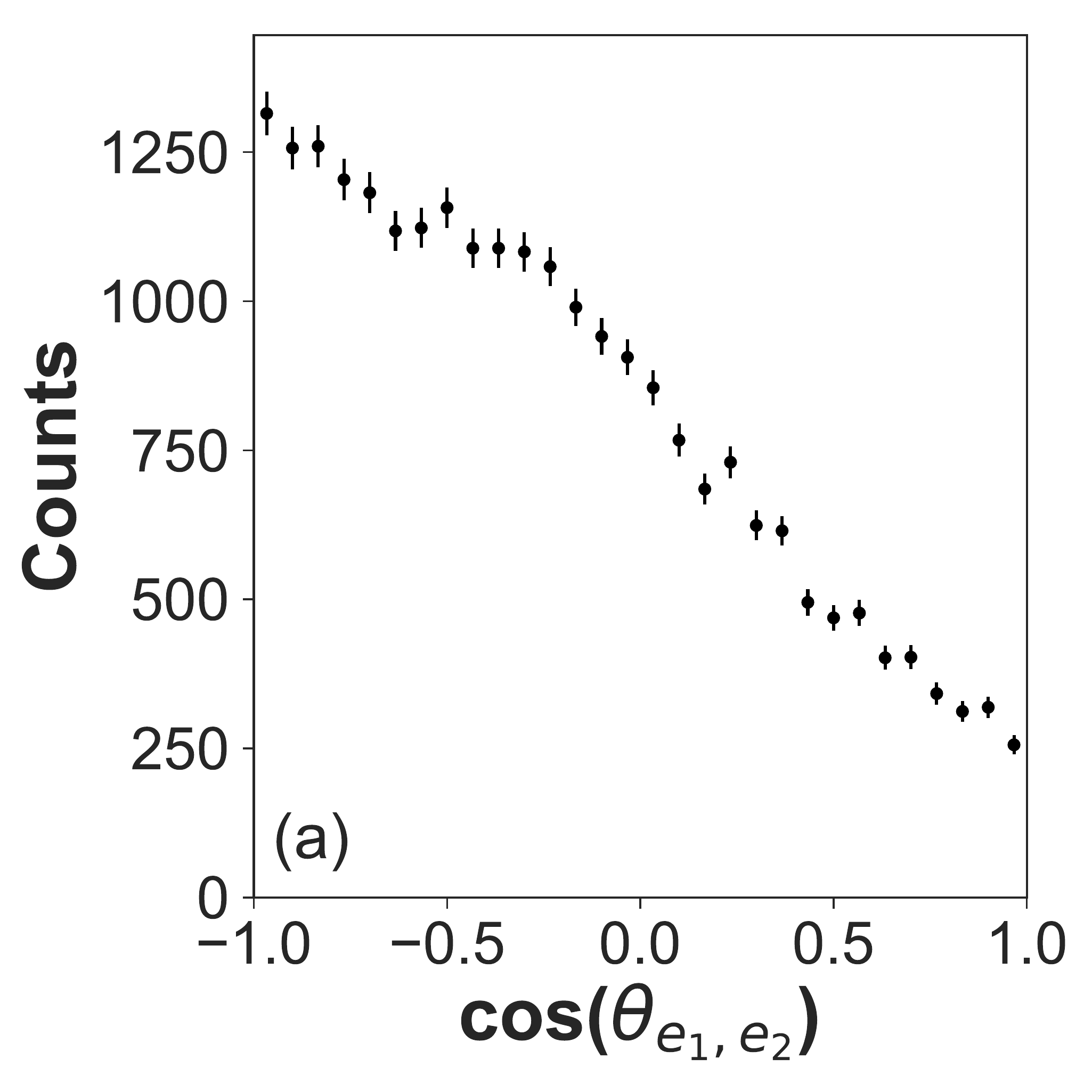}
    \includegraphics[width=4.225cm, trim=0.4cm 0.6cm 0.5cm 0.25cm, clip]{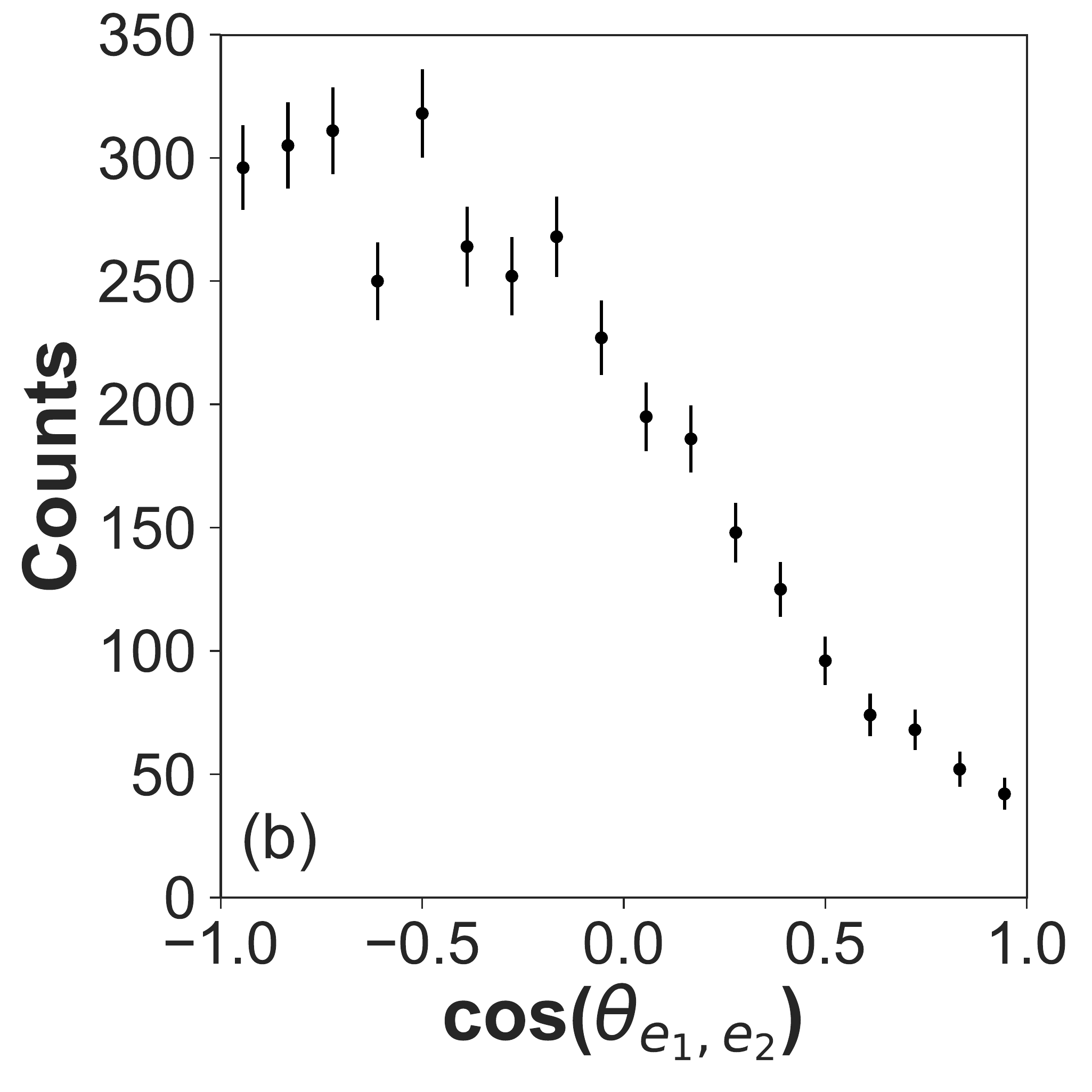}
\caption{The NH$^+$ + H$^+$ + H three-body fragmentation yield, after PDI of NH$_3$ at 61.5~eV, as a function of the cosine of the relative emission angle between the two photoelectrons in two different energy sharing conditions for the ($1e^{-2}$) $^1E$ dication state, (a) integrated over all possible electron energy sharing and (b) for equal energy sharing ($\rho = 0.5 \pm 0.075$).}
\label{fig:C_angle_e}
\end{figure}

Lastly, we plot in Fig.~\ref{fig:B_angle_e_3}, Fig.~\ref{fig:D_angle_e}, and Fig.~\ref{fig:C_angle_e} the NH$^+$ + H$^+$ + H three-body fragmentation yield as a function of the cosine of the relative emission angle between the two photoelectrons, (a) integrated over all energy sharing conditions and (b) in an equal energy sharing condition for the three dication states. There are no conditions enforced on either the molecular orientation nor the emission angle of the first detected photoelectron relative to the polarization vector of the XUV beam. In the equal energy sharing case the relative angle is plotted for $0.425 < \rho < 0.575$. As pointed out in the two-body breakup section, our measurement suffers from some multi-hit detector dead-time effects, influencing the measured yield of photoelectrons emitted in the same direction with similar kinetic energies. In the equal energy sharing condition and for $\theta_{e_1,e_2} \leq 90^\circ$ this corresponds with a maximum loss of $\sim$20$\%$ of the events for the ($3a_1^{-1},1e^{-1}$) $^1E$ state, $\sim$21$\%$ for the ($1e^{-2}$) $^3A_2$ state, and $\sim$23$\%$ for the ($1e^{-2}$) $^1E$ state. Like in the two-body breakup case we believe that the actual loss is at least a factor of two smaller than the worst-case scenario listed above, i.e. closer to our simulated results for the unequal electron energy sharing case. As in the two-body breakup section, we refrain from showing the photoelectron angular distribution of the unequal energy sharing case, which captures the autoionization feature, as there is a significant contribution due to direct PDI that significantly pollutes the autoionization signal and prevents a clear analysis of this relative angular distribution.

The relative angles between the two electrons, integrated over all energy sharing cases, show a preferred emission of the two particles into opposite hemispheres. The distribution for the ($3a_1^{-1},1e^{-1}$) $^1E$ dication state [Fig.~\ref{fig:B_angle_e_3}(a)] peaks at 135$^\circ$ with a notable dip at 180$^\circ$ corresponding to a back-to-back emission. The ($1e^{-2}$) $^3A_2$ state [Fig.~\ref{fig:D_angle_e}(a)] peaks at 130$^\circ$ with a slight increase at 180$^\circ$ compared to the ($3a_1^{-1},1e^{-1}$) $^1E$ dication state. The ($1e^{-2}$) $^1E$ state [Fig.~\ref{fig:C_angle_e}(a)] has a peak at 120$^\circ$, but now exhibits dominating back-to-back emission between the two electrons.

The photoelectron dynamics in the equal energy sharing condition [Fig.~\ref{fig:B_angle_e_3}(b), Fig.~\ref{fig:D_angle_e}(b), and Fig.~\ref{fig:C_angle_e}(b)], again, reveals similar anisotropic angular distributions, which possess node-like features near 0$^\circ$ relative electron-electron emission angle and a peak at approximately 125$^\circ$ for the ($3a_1^{-1},1e^{-1}$) $^1E$ state, 115$^\circ$ for the ($1e^{-2}$) $^3A_2$ state, and 120$^\circ$ for the ($1e^{-2}$) $^1E$ state. A dip at 180$^\circ$ remains visible. All three dication states resemble the dynamics of a knock-out double ionization process. In all three cases, the likelihood for an emission of the two electrons in the same direction is roughly a factor of 10 less likely than the emission at peak angle.

\section{\label{sec:level5}Conclusion}

In this experiment we performed state-selective measurements on the two-body NH$_2^{+}$ + H$^{+}$ and three-body NH$^{+}$ + H$^{+}$ + H dissociation channels of neutral NH$_{3}$, following PDI at 61.5~eV, where the two photoelectrons and two cations were detected in coincidence on an event-by-event basis using charged particle 3-D momentum imaging. With the help of theory, five dication states could be identified as active in this photon energy range and assigned to the two different breakup channels. Three of these PDI channels produce ro-vibrationally excited ionic fragments, where the ($3a_1^{-1},1e^{-1}$) $^1E$ and ($3a_1^{-1},1e^{-1}$) $^3E$ dication states lead to hot NH$_2^+$ fragments, while the ($1e^{-2}$) $^3A_2$ dication state leads to a hot NH$^+$ fragment.

Our measurement identifies three dication electronic states that dissociate to NH$_2^{+}$ + H$^{+}$ fragments, which are populated via direct PDI as well as through autoionization. We observe that the initial excitations in one of these dication states, the ($3a_1^{-2}$) $^1A_1$ state, undergoes intersystem crossing preceding dissociation. This effect has been observed before, but only close to the PDI threshold. By plotting the relative angle between the recoil axis of the molecular breakup and polarization vector of the XUV beam, we see an anisotropic PDI yield that illustrates the connection between the molecular orbitals participating in the PDI and the molecular orientations experiencing enhanced PDI. 

In the three-body dissociation channel of NH$_3^{2+}$ we identify three contributing dication states, two of which are different from the states of the two-body breakup channel, that dissociate to NH$^{+}$ + H$^{+}$ + H fragments. However, these state are also populated via the same ionization mechanisms, i.e. direct PDI and autoionization. In contrast to the two-body fragmentation channel, in this three-body breakup channel we observe that the three contributing dication states directly dissociate without any non-adiabatic transitions preceding the fragmentation. Plotting the relative angle between the recoil axis of the charged fragments of the breakup and the polarization vector of the XUV beam again demonstrates the connection between the molecular orbitals participating in the PDI and the molecular orientations experiencing enhanced PDI. Moreover, the dissociation of the two dication states of the three-body fragmentation result in different spectator roles of the neutral H atom.

Lastly, we presented the relative electron-electron angular distribution of all dication electronic states for all and equal electron-electron energy sharing. The distributions indicate the dominance of a knock-out PDI mechanism in all cases, and they are similar to the kinematics reported for the PDI of atoms (He) and small molecules (D$_2$ and H$_2$O) for comparable excess energies \cite{Randazzo,Brauning,Weber,Knapp}. 

\section{\label{sec:level6}Acknowledgments}

Work at LBNL was supported by the U.S. Department of Energy, Office of Science, Office of Basic Energy Sciences, Division of Chemical Sciences, Biosciences, and Geosciences under contract No. DE-AC02-05CH11231 and used the Advanced Light Source and National Energy Research Computing Center. JRML personnel were supported by the same US DOE funding source under Award No. DE-FG02-86ER13491. A.G. was supported by the ALS through a Doctoral Fellowship in Residence. Personnel from the University of Nevada, Reno was supported by the National Science Foundation Grant No. NSF-PHY-1807017. The Frankfurt group acknowledges the support of the Deutsche Akademische Austausch Dienst (DAAD) and the Deutsche Forschungsgemeinschaft (DFG). We thank the staff at the Advanced Light Source for operating beamline 10.0.1.3 and providing the photon beam. Moreover, we thank the RoentDek GmBH for longtime support with detector hardware and software.

\bibliography{Refs}

\end{document}